\documentclass[epj,nopacs]{svjour}
\usepackage{epsfig}
\usepackage{amssymb}
\usepackage{appendix}
\usepackage{color}
\usepackage{bm}
\usepackage{amsmath}
\begin{document}
\hugehead

\title{
  Ratios of Helicity Amplitudes for Exclusive \bm{$\rho^0$} Electroproduction 
}

\author{ 
The HERMES Collaboration \medskip \\
A.~Airapetian,$^{12,15}$
N.~Akopov,$^{26}$
Z.~Akopov,$^{5}$
E.C.~Aschenauer,$^{6,}$\footnote{Now at: Brookhaven National Laboratory, Upton, New York 11772-5000, USA}
W.~Augustyniak,$^{25}$
R.~Avakian,$^{26}$
A.~Avetissian,$^{26}$
E.~Avetisyan,$^{5}$
%B.~Ball,$^{15}$
S.~Belostotski,$^{18}$
N.~Bianchi,$^{10}$
% ...
}

\institute{
$^1$Physics Division, Argonne National Laboratory, Argonne, Illinois 60439-4843, USA\\
$^2$Istituto Nazionale di Fisica Nucleare, Sezione di Bari, 70124 Bari, Italy\\
$^3$School of Physics, Peking University, Beijing 100871, China\\
$^4$Nuclear Physics Laboratory, University of Colorado, Boulder, Colorado 80309-0390, USA\\
$^5$DESY, 22603 Hamburg, Germany\\
$^6$DESY, 15738 Zeuthen, Germany\\
$^7$Joint Institute for Nuclear Research, 141980 Dubna, Russia\\
$^8$Physikalisches Institut, Universit\"at Erlangen-N\"urnberg, 91058 Erlangen, Germany\\
$^9$Istituto Nazionale di Fisica Nucleare, Sezione di Ferrara and Dipartimento di Fisica, Universit\`a di Ferrara, 44100 Ferrara, Italy\\
$^{10}$Istituto Nazionale di Fisica Nucleare, Laboratori Nazionali di Frascati, 00044 Frascati, Italy\\
$^{11}$Department of Subatomic and Radiation Physics, University of Gent, 9000 Gent, Belgium\\
$^{12}$Physikalisches Institut, Universit\"at Gie{\ss}en, 35392 Gie{\ss}en, Germany\\
$^{13}$SUPA, School of Physics and Astronomy, University of Glasgow, Glasgow G12 8QQ, United Kingdom\\
$^{14}$Department of Physics, University of Illinois, Urbana, Illinois 61801-3080, USA\\
$^{15}$Randall Laboratory of Physics, University of Michigan, Ann Arbor, Michigan 48109-1040, USA \\
$^{16}$Lebedev Physical Institute, 117924 Moscow, Russia\\
$^{17}$National Institute for Subatomic Physics (Nikhef), 1009 DB Amsterdam, The Netherlands\\
$^{18}$Petersburg Nuclear Physics Institute, Gatchina, 188300 Leningrad region, Russia\\
$^{19}$Institute for High Energy Physics, Protvino, 142281 Moscow region, Russia\\
$^{20}$Institut f\"ur Theoretische Physik, Universit\"at Regensburg, 93040 Regensburg, Germany\\
$^{21}$Istituto Nazionale di Fisica Nucleare, Sezione Roma 1, Gruppo Sanit\`a and Physics Laboratory, Istituto Superiore di Sanit\`a, 00161 Roma, Italy\\
$^{22}$TRIUMF, Vancouver, British Columbia V6T 2A3, Canada\\
$^{23}$Department of Physics, Tokyo Institute of Technology, Tokyo 152, Japan\\
$^{24}$Department of Physics and Astronomy, VU University, 1081 HV Amsterdam, The Netherlands\\
$^{25}$Andrzej Soltan Institute for Nuclear Studies, 00-689 Warsaw, Poland\\
$^{26}$Yerevan Physics Institute, 375036 Yerevan, Armenia }
%$^{27}$Department of Physics and Astronomy 666 W.Hancock,Wayne State University, Detroit MI 48201, USA} 

\date{Received: \today / Revised version:}

\titlerunning{Paper Tag: HELAMPRHO}
\authorrunning{The HERMES Collaboration}

\abstract{
Exclusive $\rho^0$-meson electroproduction is studied in the HERMES experiment, using a $27.6$~GeV longitudinally polarized electron/positron beam and unpolarized hydrogen and deuterium  targets in the kinematic region $0.5$~GeV$^2 <Q^2<7.0$~GeV$^2$, $3.0$~GeV $<W<6.3$~GeV, and $-t^\prime<0.4$~GeV$^2$. Real and imaginary parts of the ratios of the natural-parity-exchange helicity amplitudes $T_{11}$ ($\gamma^*_T \rightarrow \rho_T$),  $T_{01}$ ($\gamma^*_T \rightarrow \rho_L$), $T_{10}$ ($\gamma^*_L \rightarrow \rho_T$), and $T_{1-1}$ ($\gamma^*_{-T} \rightarrow \rho_T$)  to $T_{00}$ ($\gamma^*_L \rightarrow \rho_L$)  are extracted from the data. For the unnatural-parity-exchange amplitude $U_{11}$,  the  ratio $|U_{11}/T_{00}|$ is obtained. The $Q^2$ and $t'$ dependences of these ratios are presented and compared with perturbative QCD predictions.
}

%\PACS{
%{13.60.-r,13.60.Le,13.88.+e}{}
%}

%\PACS{
%{13.60.-r,13.60.Le,13.88.+e}{}
%}

\maketitle

%%%%%%%%%%%%%%%%%%%%%%%%%%%%%%%%%%%%%%%%%
\section{Introduction}
Exclusive electroproduction of vector mesons, $e + N \rightarrow e' + V + N'$, has been the focus of investigation for decades.  Not only is the reaction mechanism of intrinsic interest, but  this process also offers the possibility of studying, in a model-dependent way, the structure of hadrons involved in the process~\cite{Bauer,INS}. Using the one-photon-exchange approximation, all the measurable observables in electroproduction can be expressed
in terms of the virtual photon spin-density matrix and  the helicity amplitudes $F_{\lambda_V \lambda '_N \lambda_{\gamma} \lambda_N}$ of the
process
\begin{equation}
\gamma^*(\lambda_{\gamma})+N(\lambda_N) \rightarrow V(\lambda_V)+N^{\prime}(\lambda '_N) \; .
\label{react01}
\end{equation}
Here, $\gamma ^*$ denotes the virtual photon exchanged between the lepton and the target nucleon,  $V$ denotes the produced vector meson, and $N(N^{\prime})$ the initial (final) nucleon. The helicities of the particles are given in parentheses in Eq.~(\ref{react01}). The helicity amplitudes are defined in the virtual-photon-nucleon center-of-mass (CM) system. In order to  make the discussion of the transition $\gamma^* \rightarrow V$ more transparent, we shall henceforth often  omit the  nucleon helicity indices using the notation $F_{\lambda_V \lambda_{\gamma}}$ instead of $F_{\lambda_V \lambda '_N \lambda_{\gamma} \lambda_N}$. The properties of helicity amplitudes can be studied  in detail because the spin-density matrix of the virtual photon is well known from quantum electrodynamics and the spin-density matrix of the produced vector meson is experimentally accessible.

For unpolarized targets, the formalism of the spin-density matrix elements (SDMEs) of the produced vector meson 
was first presented in Ref.~\cite{SW}, where expressions of SDMEs in terms 
of  helicity amplitudes were established. The formalism was then extended to the case of polarized targets 
in Ref.~\cite{Fraas}. Recently, a new general formalism for the description of the process in
Eq.~(\ref{react01}) through SDMEs was  presented in Ref.~\cite{Diehl}.

In order to determine SDMEs from experimental data, the SDMEs are considered as being
independent free parameters in the fitting of the  production and decay angular distribution of the vector meson. 
 This is referred to as the ``SDME method" in the rest of this work.

SDMEs are dimensionless quantities and therefore depend on ratios of amplitudes rather than on amplitudes themselves.
The exact expressions for SDMEs given in Refs.~\cite{SW,Fraas,Diehl} can be rewritten in terms of amplitude ratios. In an alternative
method of fitting the angular distribution, these ratios are considered as being independent free parameters. This method is referred
 to as the ``amplitude method" in the rest of this work.

In order to extract the helicity amplitudes themselves, experimental data on the 
differential cross section with respect to the Mandelstam $t$ variable,
$d \sigma/dt$ (which is proportional to the sum of squared moduli of all the amplitudes) are required in addition to the experimentally determined
amplitude ratios. An analysis of these combined data  would allow
the extraction of the moduli of all amplitudes and of the phase differences between 
them with the common phase remaining undetermined. 
However, the requisite information on nucleon spin-flip amplitudes is not available for data taken with unpolarized 
targets.

Exclusive meson production in hard lepton scattering has been shown to offer the possibility of constraining 
 generalized parton distributions (GPDs), which provide correlated information on transverse spatial 
and longitudinal momentum distributions of partons in the 
nucleon~\cite{gpd1,gpd2,gpd3,MD,BR,golos1,golos2,golos3,VGG,GPV,GPRV}. 
 Vector-meson production amplitudes contain various linear combinations of process-independent GPDs for quarks 
 of various flavors and gluons.  Access to  GPDs  relies on the factorization property of the process amplitude,  
 i.e.,  the amplitude can be written as  convolution of  ``non-perturbative" GPDs with amplitudes of hard 
 partonic subprocesses calculated within the framework of perturbative quantum chromodynamics (pQCD) and  
quantum electrodynamics.

The amplitudes $F_{0\frac{1}{2}0 \pm \frac{1}{2}}$  are the most interesting because, only for these 
and quantities constructed from them, was factorization proven~\cite{strikman}. Factorization is not 
proven for the other amplitudes nor for SDMEs, which depend on all helicity amplitudes. The amplitudes 
$F_{0\frac{1}{2}0 \pm \frac{1}{2}}$ correspond to the transition of a longitudinally polarized (L) virtual 
photon to a longitudinally polarized vector meson, $\gamma^*_L \to V_L$, and dominate at large
photon virtuality $Q^2$. 
Nevertheless, an application of the ``modified perturbative approach"
 \cite{golos2,golos3} assumes that the factorization property also  holds for the amplitudes 
$F_{11}$ and $F_{01}$. The agreement found between certain calculated SDMEs and those extracted 
from  HERMES~\cite{DC-24}, ZEUS~\cite{ZEUS2}, and  H1~\cite{H1-amp} data supports this 
assumption. 
%Currently, the only available  GPD-based calculations  of SDMEs for electroproduction 
%of vector mesons are performed in the 
%model developed in Refs.~\cite{golos1,golos2,golos3}.
The differential and total cross sections for longitudinal $\rho^0$-meson 
production by longitudinal photons are reasonably well described
in the GPD-based approach of Refs.~\cite{golos1,golos2,VGG,GPV,GPRV} not
only at the high energies of the HERA collider 
experiments~\cite{ZEUS3,ZEUS4,H1-1,H1,H1-2} but also at intermediate energies covered by the
fixed-target experiments E665~\cite{E665}, HERMES~\cite{rho-xsec}, and
CLAS~\cite{CLAS1,CLAS2}.

Amplitude ratios are useful not only as a tool for obtaining $F_{00}$. They can be used
 to study the general properties~\cite{Diehl,IK,KNZ,RK} of the vector-meson-production
 amplitudes  at very small and very  large values of $Q^2$ and $t$. They can also be used to test theoretical 
 models. For these purposes, an amplitude ratio is more 
 convenient than an SDME,  which constitutes a more complicated object since any SDME  depends on all the 
 amplitude ratios. The amplitudes of the non-diagonal $\gamma^* \rightarrow V$ transitions $F_{01}$ and $F_{10}$ 
 are known~\cite{KNZ} to be  zero if any valence quark in the vector meson carries half of the longitudinal
 momentum of the meson in the infinite momentum frame.
 These amplitudes are the most sensitive objects for the study of the quark motion in 
 vector mesons. The double-spin-flip amplitude $F_{1-1}$ contains information on the gluon transversity 
generalized parton distribution~\cite{MD,BR,IK,KNZ,RK,Kivel} in the nucleon,  which cannot be obtained from
 inclusive deep-inelastic lepton-nucleon scattering.

Amplitude ratios in $\rho^0$ and $\phi$ meson production on the proton were first studied by the H1 experiment~\cite{H1-amp} at the HERA
 collider. The results of the analysis of $\rho^0$-meson production at HERMES using the SDME method were published in Ref.~\cite{DC-24},
 where also the description of the HERMES spectrometer and details of the data treatment etc. can be found.
The present work is a continuation of that analysis. In the analysis presented in this paper,  ratios of helicity amplitudes  to the
amplitude $F_{0\frac{1}{2}0\frac{1}{2}}$   are extracted  separately for data taken with unpolarized hydrogen and deuterium
 targets first and then also for the combined datasets in a single analysis.  A comparison of the proton and deuteron results 
allows the study of the degree of interference between $I=1$ exchanges of $q\bar{q}$ pairs and $I=0$
exchanges of $q\bar{q}$ pairs and two gluons.

 %%%%%%%%%%%%%%%%%%%%%%%%%%%%%%%%%%%
\section{Kinematics}

In accordance with the notation of Ref.~\cite{DC-24}, the kinematic variables of the  process
\begin{equation}\label{rhoprod}
eN\rightarrow e' \gamma^* N\rightarrow e'\rho^o N'   \rightarrow e' \pi^+ \pi^- N'
\end{equation}
are defined as follows. The four-momenta of the incident and outgoing leptons are denoted by $k$ and
$k^{\prime}$, the difference of which defines the four-momentum $q=k-k'$
of the virtual photon $\gamma^*$. In the target rest frame (which is also called the laboratory or lab frame in this work),
 $\vartheta$ is the scattering angle between the incident and outgoing leptons,  the energies of the leptons are denoted by $E$
  and $E^{\prime}$.
\begin{figure}[hbtc!]
  \begin{center}
\epsfig{file=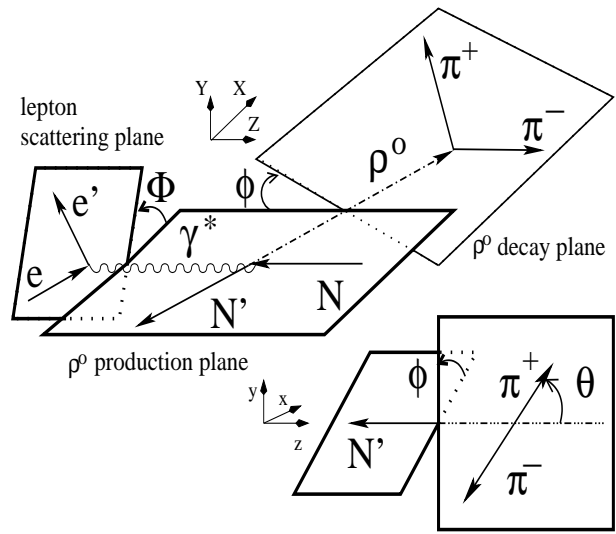,width=8.0cm,height=7.cm}
\vspace{0.5cm}
\caption{ {\small
Definition of angles in the process $eN\rightarrow e' \gamma^* N\rightarrow e'\rho^o N'   \rightarrow e' \pi^+ \pi^- N'$. Here $\Phi$ is
the angle between
 the $\rho^0$ production plane and the lepton scattering plane in the CM system of virtual photon and target nucleon.
 See Ref.~\cite{DESY2} for details. The variables $  \theta$ and $\phi$ are respectively the polar and azimuthal angles of the decay
 $\pi^+$ in the $\rho^0$-meson rest frame, with the $Z$ axis being anti-parallel to the outgoing nucleon momentum.
}}
\label{hellen}
\end{center}
\end{figure}
 The photon virtuality, given by
\begin{equation}
Q^2 = -q^2=-(k-k^{\prime})^2\stackrel{lab}{\approx}4\,E\,
E^{\prime}\sin^2 \frac{\vartheta}{2},
\end{equation}
is positive in leptoproduction. In this equation, the electron rest mass is neglected. The four-momentum of the incident (recoiling)  nucleon is denoted by $p$ ($p^{\prime}$). The Bjorken scaling variable $x_B$ is defined as
\begin{equation}
x_B = \frac{Q^2}{2\,p\cdot q}  =\frac{Q^2}{2\,M\,\nu},
\end{equation}
with
\begin{equation}
\nu = \frac{p\cdot q}{M}\stackrel{lab}{=}E-E^{\prime},
\end{equation}
so that $\nu$ represents the energy transfer from the incoming lepton to the virtual photon in the laboratory frame. The squared  invariant mass of the photon-nucleon system is given by
\begin{equation} \label{wdef}
W^2=(p+q)^2
=M^2+2\,M\,\nu-Q^2.
\end{equation}
The Mandelstam variable $t$ is defined by the relation
\begin{equation}\label{eqt}
t=(q-v)^2,
\end{equation}
where $v$ is the four-momentum of the $\rho^0$ meson being equal to $p_{\pi^+}+p_{\pi^-}$, the sum of $\pi^+$ and $\pi^-$  four-momenta. The variables $t$, $t_0$, and $t^\prime=t-t_0$ are always negative, where  $-t_{0}$ is the minimal value of $-t$ for given values of $Q^2$, $W$, and the $\rho^0$-meson mass $M_V$. At small values of $-t^\prime$, the approximation $-t^\prime \approx v_T^2$ holds, where $v_T$ is the  transverse momentum of the $\rho^0$  meson with respect to the direction of the virtual photon in the CM system.

The variable $\epsilon$ represents the  ratio of fluxes of longitudinally and transversely polarized virtual  photons and is given by
\begin{eqnarray}\label{expreps}
\epsilon=\frac{1-y - \frac{Q^2}{4E^2} }{1-y+ \frac{y^2}{2} + \frac{Q^2}{4E^2}} \!\!
\stackrel{lab}{\approx}\left(1+2\left(1+\frac{\nu^2}{Q^2}\right)\tan^2\frac{\vartheta}{2}\right)^{-1}
\end{eqnarray}
with $y = p\cdot q / p\cdot k   \stackrel{lab}{=} \nu / E$.

The ``exclusivity'' of $\rho^0$ production in  the process in Eq.~(\ref{rhoprod}) is characterized by the variable
\begin{equation}
\label{deltae}
\Delta E = \frac{M_{X}^{2} -M^{2}}{2 M } \stackrel{lab} = E_V - (E_{\pi^+}+E_{\pi^-}),
\end{equation}
where $M_X = \sqrt{ (k - k' + p - p_{\pi^+}-p_{\pi^-})^2}$ is the reconstructed invariant mass of  the undetected hadronic  system (missing mass), $E_V= \nu + t/(2 M )$ is the energy of the exclusively produced $\rho^o$ meson, and $(E_{\pi^+}+E_{\pi^-})$ is the sum of the energies
 of the two detected pions in the laboratory system. For  exclusive $\rho^0$ electroproduction $M_X =M$ and therefore $\Delta E=0$.

 The  angles used for the description of the process in Eq.~(\ref{rhoprod}) are defined according to Ref.~\cite{DESY2} (see also
 Ref.~\cite{DC-24}) and presented in  Fig.~\ref{hellen}.
%%%%%%%%%%%%%%%%%%%%%%%%%%%%%%%%%%%%%%%%%
\section{Formalism}
\subsection{Natural and Unnatural-Parity Exchange Helicity Amplitudes}

In Ref.~\cite{SW}, exclusive $\rho^0$-meson leptoproduction is described by helicity amplitudes 
$F_{\lambda_{V}\lambda '_N\lambda_{\gamma}\lambda _N}$
defined in the right-handed CM   system of the virtual photon and target nucleon. In this system, the $Z$-axis  is aligned along
 the virtual photon three-momentum $\vec{q}$ and the $Y$-axis is parallel to $\vec{q} \times \vec{v}$ where $\vec{v}$ is the $\rho^0$ meson
 three-momentum as shown in Fig.~\ref{hellen}. The helicity amplitude can be expressed as the scalar product of the matrix element of
 the electromagnetic current vector $J^{\kappa}$ and the virtual-photon polarization vector $e_{\kappa}^{(\lambda_{\gamma})}$
\begin{eqnarray}
F_{\lambda_{V} \lambda '_{N} \lambda_{\gamma}  \lambda_{N} } = (-1)^{\lambda_{\gamma}}
\langle v \lambda_{V} p' \lambda '_{N} | J ^{\kappa} | p \lambda_{N} \rangle
e_{\kappa}^{(\lambda_{\gamma})},
 \label{jacobwick}
\end{eqnarray}
 where a summation over the Lorentz index $\kappa$ is performed. Here $e_{\kappa}^{(\pm 1)}$ and 
 $e_{\kappa}^{(0)}$ indicate respectively transverse and longitudinal polarization of the virtual photon in the CM 
system:
\begin{eqnarray}
\label{transvphot}
e^{(\pm 1)}& =& (e_0,e_X,e_Y,e_Z)=(0,\mp \frac{1}{\sqrt{2}},-\frac{i}{\sqrt{2}},0)\;,\\
\nonumber
e^{(0)}&=&\frac{1}{Q}(q_Z,0,0,q_0) \\
&=&\frac{1}{WQ}(M\sqrt{\nu^2+Q^2},0,0,M\nu-Q^2)\;.
 \label{longphot}
\end{eqnarray}
The ket vector $| p \lambda_{N} \rangle$ corresponds to the  initial  nucleon and the bra vector $\langle v \lambda_{V} p' \lambda '_{N} |$ represents the final state  consisting of a $\rho ^0$ meson and the scattered nucleon.

Any helicity amplitude $F_{\lambda_{V} \lambda '_{N} \lambda_{\gamma}  \lambda_{N} }$ can be decomposed into the 
sum of an amplitude $T_{\lambda_{V} \lambda '_{N} \lambda_{\gamma}  \lambda_{N} }$ for natural-parity exchange 
(NPE) and an amplitude $U_{\lambda_{V} \lambda '_{N} \lambda_{\gamma}  \lambda_{N} }$ 
for unnatural-parity exchange (UPE)~\cite{SW,Fraas,Diehl}
\begin{equation}
F_{\lambda_{V} \lambda '_{N} \lambda_{\gamma}  \lambda_{N} } =
T_{\lambda_{V} \lambda '_{N} \lambda_{\gamma}  \lambda_{N} }+
U_{\lambda_{V} \lambda '_{N} \lambda_{\gamma}  \lambda_{N} },
\label{decomp}
\end{equation}
which  obey the  following  symmetry relations
\begin{eqnarray}
T_{\lambda_{V} \lambda '_{N}\lambda_{\gamma} \lambda_{N}}=
(-1)^{-\lambda_{V}+\lambda_{\gamma}}
T_{-\lambda_{V} \lambda '_{N}-\lambda_{\gamma} \lambda_{N}} \nonumber \\
=(-1)^{\lambda '_{N}-\lambda_{N}}
T_{\lambda_{V} -\lambda '_{N}\lambda_{\gamma} -\lambda_{N}},
 \label{symmnat}
\end{eqnarray}
\begin{eqnarray}
U_{\lambda_{V} \lambda '_{N}\lambda_{\gamma} \lambda _{N}}=
-(-1)^{-\lambda_{V}+\lambda_{\gamma}}
U_{-\lambda_{V} \lambda '_{N}-\lambda_{\gamma} \lambda _{N}} \nonumber \\
= -(-1)^{\lambda '_{N}-\lambda_{N}}
U_{\lambda_{V} -\lambda '_{N}\lambda_{\gamma} -\lambda _{N}}\;.
  \label{symmunn}
\end{eqnarray}
 There are three  important consequences of the symmetry relations (\ref{symmnat}) and (\ref{symmunn})~\cite{SW,Diehl}:
\newline
i)~The number of linearly independent NPE amplitudes is equal to 10 while only 
8 independent UPE amplitudes describe the process in Eq.~(\ref{react01}),
\newline
ii)~No UPE amplitude exists for the transition $\gamma_L \to \rho^0_L$, so that in particular
$F_{0 \frac{1}{2} 0 \frac{1}{2} } \equiv T_{0 \frac{1}{2} 0 \frac{1}{2} } \equiv T_{00}$,
\newline
iii)~For unpolarized targets there is no interference between NPE and UPE amplitudes~\cite{SW,Diehl}.

In Regge phenomenology~\cite{IW,KS}, an exchange of a single natural-parity  reggeon  [$P=(-1)^J$ (pomeron, secondary reggeons $\rho$, $f_2$,
 $a_2$, ...)] contributes to the NPE amplitudes, while an exchange of a single unnatural-parity reggeon
  [$P=-(-1)^J$ ($\pi$, $a_1$, $b_1$,...)]  contributes to the UPE amplitudes~\cite{CSM}. It is worth noting that for a multi-reggeon-exchange 
 contribution, there is no such one-to-one correspondence. For example, an exchange of two reggeons of  
``unnatural'' parity contributes to the NPE amplitudes. 

 For convenience, we introduced in Ref.~\cite{DC-24} the abbreviation $\widetilde {\sum} \equiv 
\frac{1}{2}{\sum}_{\lambda '_N  \lambda_N}$ for the summation  over the  final nucleon  helicity indices and for 
averaging over the initial spin states of the nucleon. For NPE amplitudes, transitions that are diagonal in 
nucleon helicity ($\lambda '_N = \lambda _N $) are dominant.
 In this case, neglecting the small nucleon helicity-flip amplitudes $T_{\lambda _V \pm \frac{1}{2} \lambda 
_{\gamma} \mp \frac{1}{2}}$
and using  Eq.~(\ref{symmnat}), the summation and averaging operation $\widetilde  {\sum}$  reduces to one term:
\begin{eqnarray} \label{tsum}
\nonumber
\widetilde  {\sum}T_{\lambda_V \lambda_{\gamma}}
T^*_{\lambda '_V \lambda '_{\gamma}}
&\equiv&  \frac{1}{2} \sum_{\lambda_N \lambda '_N}
T_{\lambda_V \lambda '_N\lambda_{\gamma} \lambda_N}
T^*_{\lambda '_V \lambda '_N \lambda '_{\gamma}\lambda_N}  \\
&\hspace{-3.cm}=&\hspace{-1.5cm} T_{\lambda_V \frac{1}{2}\lambda_{\gamma}\frac{1}{2}}
                                 T^*_{\lambda '_V \frac{1}{2}\lambda '_{\gamma}\frac{1}{2}}
                                +T_{\lambda_V -\frac{1}{2}\lambda_{\gamma}\frac{1}{2}}
                                 T^*_{\lambda '_V -\frac{1}{2}\lambda '_{\gamma}\frac{1}{2}} \nonumber \\ [0.2cm] 
&\hspace{-3.cm}\approx&\hspace{-1.5cm} T_{\lambda_V \frac{1}{2}\lambda_{\gamma}\frac{1}{2}}
                                       T^*_{\lambda '_V \frac{1}{2}\lambda '_{\gamma}\frac{1}{2}}
                               \equiv  T_{\lambda_V \lambda_{\gamma}}T^*_{\lambda '_V \lambda '_{\gamma}},
\end{eqnarray}
  where $T^*_{\lambda '_V \lambda '_N \lambda '_{\gamma}\lambda_N}$ represents the complex conjugate of the  
amplitude $T_{\lambda '_V \lambda '_N \lambda '_{\gamma}\lambda_N}$. Due to symmetry properties (see 
 Eq.~(\ref{symmnat})), the abbreviated notation $T_{\lambda_{V}\lambda_{\gamma}}$ is used for the amplitudes 
$T_{\lambda_{V}\frac{1}{2}\lambda_{\gamma}\frac{1}{2}}=T_{\lambda_{V}-\frac{1}{2}\lambda_{\gamma}-\frac{1}{2}}$.

 In general, no dominance for UPE amplitudes can be proven either for diagonal transitions ($\lambda_N = \lambda 
'_N$) or for transitions with nucleon helicity flip. Therefore,  no relation analogous to Eq.~(\ref{tsum}) can be derived.

%%%%%%%%%%%%%%%
\subsection{Spin Density Matrix Elements} 

The photon-spin-density matrix normalized to unit flux of transversely polarized virtual photons
embodies the matrices $\varrho^U_{\lambda_{\gamma}\mu_{\gamma}}$ and
$\varrho^L_{\lambda_{\gamma}\mu_{\gamma}}$ corresponding to 
unpolarized (U) and longitudinally (L) polarized  lepton beams:
\begin{equation} \label{matr}
\varrho^{U+L}_{\lambda_{\gamma} \mu_{\gamma }} =
\varrho_{\lambda_{\gamma} \mu_{\gamma }}^{U} +
P_B \; \varrho_{\lambda_{\gamma} \mu_{\gamma}}^{L},
\end{equation}
where $P_B$ is the longitudinal polarization of the lepton beam and $\varrho_{\lambda_{\gamma} \mu_{\gamma 
}}^{U}$
and $\varrho_{\lambda_{\gamma} \mu_{\gamma}}^{L}$ are
defined in Ref.~\cite{DC-24}.
The spin density matrix $\rho_{\lambda_{V} \mu_{V}}$
of the produced $\rho^0$  meson  is 
related to that of the virtual photon, 
$\varrho^{U+L}_{\lambda_{\gamma} \mu_{\gamma }}$, 
through the von~Neumann formula~\cite{SW} : 
\begin{eqnarray} \label{neumann}
\rho_{\lambda_{V} \mu_{V}} =   \frac{1}{2 {\mathcal{N}} } 
\sum_{\lambda_{\gamma} 
\mu_{\gamma}\lambda_N \lambda '_N}
\!\!\!\!\!\!\!\!  F_{\lambda_{V}\lambda '_N\lambda_{\gamma}\lambda _N}\;
 \varrho^{U+L}_{\lambda_{\gamma} \mu_{\gamma }}\;
  F_{\mu_{V} \lambda '_N \mu_{\gamma}\lambda _N}^{*}.  \hspace*{0.40cm}
 \end{eqnarray}
Note that the phase factor $(-1)^{\lambda_{\gamma}}$
in Eq.~(\ref{jacobwick}) is important  for  usual matrix summation
over $\lambda_{\gamma}$ and $\mu_{\gamma}$ in Eq.~(\ref{neumann}).
The normalization factor is given by 
\begin{eqnarray}
{\mathcal{N}} =   {\mathcal{N}}_T + \epsilon {\mathcal{N}}_L,
\label{ntotal}
\end{eqnarray}
with
\begin{eqnarray} 
 {\mathcal{N}}_{T} &=&  \widetilde{\sum}
 \Big( |T_{11}|^2+|T_{01}|^2+|T_{1-1}|^2  \nonumber \\
& & + |U_{11} |^2+|U_{01}|^2+|U_{1-1}|^2 \Big), \\ [0.2cm]
\label{sigmatrans}
{\mathcal{N}}_L &=& \widetilde{\sum} \Big( |T_{00}|^2+2|T_{10}|^2+2|U_{10}|^2 \Big).
\label{sigmalong}
\end{eqnarray}
Equation~(\ref{sigmalong}) is obtained by using the symmetry relations~(\ref{symmnat}) and (\ref{symmunn}).

If the spin-density matrix of the virtual photon is decomposed into the standard set (see Ref.~\cite{SW}) of nine hermitian matrices $\Sigma ^{\eta}$ ($\eta=0,\;1,\;...,\;8$), for the  produced $\rho^0$  meson a set of nine matrices
$r^{\eta}_{\lambda_{V} \mu_{V}}$ is  obtained:
\begin{eqnarray}  \label{rhomatr}
r_{\lambda_{V} \mu_{V}}^{\eta} \!\!\!
= \frac{1}{2 {\mathcal{N}}}  \!\!\!  \sum_{\lambda_{\gamma} \mu_{\gamma} \lambda '_N \lambda _N} 
\!\!\!\!\!\!\!\!\!
  F_{\lambda_{V}\lambda '_N \lambda_{\gamma}\lambda _N}
 \Sigma_{\lambda_{\gamma} \mu_{\gamma }}^{\eta} 
  F_{\mu_{V} \lambda ' _N\mu_{\gamma}\lambda _N}^{*}.
\end{eqnarray}

If an experiment is performed at  fixed  beam energy,  a (Rosenbluth~\cite{rosenbluth}) decomposition
 into contributions from longitudinally and transversely polarized virtual photons is
 impossible. In this case, the contributions from the matrix elements  $r_{\lambda_{V} \mu_{V}}^{0}$
 and $r_{\lambda_{V} \mu_{V}}^{4}$ cannot be disentangled and the only measurable combination is
\begin{eqnarray} 
r^{04}_{\lambda_{V} \mu_{V}} \equiv r^{0}_{\lambda_{V} \mu_{V}}
+\epsilon r^{4}_{\lambda_{V} \mu_{V}}\;.
\label{r04}
\end{eqnarray}

\subsection{Three-dimensional Angular Distribution}
\label{app:app1}

 The formula for the angular distribution of the scattered electrons/positrons and $\pi^+$ mesons from the decay 
 $\rho^0 \rightarrow \pi^+\pi^-$ for the case of a  longitudinally polarized beam and an unpolarized 
 target~\cite{SW} is given by
\begin{eqnarray}
 {\mathcal{W}} ^{U+L} (\Phi,\phi,\cos{\theta}) =
\nonumber \\
{\mathcal{W}}^{U}(\Phi,\phi,\cos{\theta}) +
P_B {\mathcal{W}}^{L}(\Phi,\phi,\cos{\theta})\;.
\label{eqang1}
\end{eqnarray}
 Here $\mathcal{W}^{U}(\Phi,\phi,\cos{\theta})$ represents the angular distribution when both  beam and target 
 are unpolarized  (see Eq.~(\ref{eqang2})) and $\mathcal{W}^{L}(\Phi,\phi,\cos{\theta})$ is the additional term 
 arising for longitudinally polarized beam (see Eq.~(\ref{eqang3})).
 \begin{figure*}[hbtc!]
%\begin{widetext}
\begin{eqnarray}
\mathcal{W}^{U}(\Phi,\phi,\cos{\theta})  &=& \frac{3} {8 \pi^{2}} \Bigg[
         \frac{1}{2} (1 - r^{04}_{00}) + \frac{1}{2} (3 r^{04}_{00}-1) \cos^2{\theta}
- \sqrt{2} \mathrm{Re} \{ r^{04}_{10} \} \sin 2\theta
\cos \phi - r^{04}_{1-1}  \sin ^{2} \theta \cos 2 \phi \hspace*{1.0cm}
\nonumber \\
&-& \epsilon \cos 2 \Phi \Big( r^{1}_{11} \sin ^{2} \theta  + r^{1}_{00} \cos^{2}{\theta}
  - \sqrt{ 2}  \mathrm{Re} \{r^{1}_{10}\} \sin 2  \theta  \cos  \phi
    - r^{1}_{1-1} \sin ^{2} \theta \cos 2 \phi   \Big)   \nonumber  \\
&-& \epsilon \sin 2 \Phi \Big( \sqrt{2} \mathrm{Im} \{r^{2}_{10}\} \sin 2 \theta \sin \phi +
       \mathrm{Im} \{ r^{2}_{1-1} \} \sin ^{2} \theta \sin 2 \phi  \Big)  \nonumber \\
&+& \sqrt{ 2 \epsilon (1+ \epsilon)}  \cos \Phi
\Big(  r^{5}_{11} \sin ^2 {\theta} +
 r^{5}_{00} \cos ^{2} \theta - \sqrt{2} \mathrm{Re} \{r^{5}_{10}\}
\sin 2 \theta \cos \phi -
 r^{5}_{1-1} \sin ^{2} \theta \cos 2 \phi  \Big)  \nonumber \\
&+& \sqrt{ 2 \epsilon (1+ \epsilon)}  \sin \Phi
\Big( \sqrt{ 2} \mathrm{Im} \{ r^{6}_{10} \} \sin 2 \theta \sin \phi
+ \mathrm{Im} \{r^{6}_{1-1} \} \sin ^{2} \theta \sin 2 \phi \Big) \Bigg],
\label{eqang2} \\
\mathcal{W}^{L}(\Phi,\phi,\cos \theta)  &=& \frac{3}{8 \pi^{2}} \Bigg[
  \sqrt{ 1 - \epsilon ^{2} }  \Big(  \sqrt{ 2}  \mathrm{Im} \{ r^{3}_{10} \}
\sin 2 \theta \sin \phi +
   \mathrm{Im} \{ r^{3}_{1-1}\} \sin ^{2} \theta \sin 2 \phi  \Big)  \nonumber  \\
&+& \sqrt{ 2 \epsilon (1 - \epsilon)} \cos \Phi
\Big( \sqrt{2} \mathrm{Im} \{r^{7}_{10}\} \sin 2 \theta \sin \phi
+  \mathrm{Im} \{ r^{7}_{1-1} \}  \sin ^{2} \theta \sin 2 \phi   \Big)  \nonumber \\
&+& \sqrt{ 2 \epsilon (1 - \epsilon)} \sin \Phi
\Big( r^{8}_{11} \sin ^{2} \theta + r^{8}_{00} \cos ^{2}
\theta -  \sqrt{2} \mathrm{Re}\{ r^{8}_{10}\} \sin 2 \theta \cos \phi
- r^{8}_{1-1} \sin ^{2} \theta \cos 2\phi \Big)  \Bigg].
\label{eqang3}
\end{eqnarray}
%\end{widetext}
\end{figure*}
%%%%%%%%%%%%%%%%%%%%%%%%%%%%%%%
\section{The Amplitude Method}
\subsection{Comparison of the SDME and Amplitude Methods}

 The SDME  method for analyzing experimental data has the disadvantage that, in general, numerators of SDMEs
depend on several ratios of  helicity amplitudes (see Refs.~\cite{SW,DC-24}) while the denominator common for
all SDMEs depends on all the amplitude ratios.  Therefore, if an SMDE extracted from data differs from model
calculations, the source of the discrepancy is difficult to identify.

 In total there are  18 independent amplitudes~\cite{SW,Fraas,Diehl}. Since SDMEs depend on ratios of these
complex amplitudes  to $F_{00}$, any SDME comprises 34 real, independent functions. However, the number of SDMEs
to be extracted from  experimental data  in polarized particle scattering is larger. For instance, there are 47
SDMEs when both beam and target are longitudinally polarized,  while the angular distribution for longitudinal 
beam polarization and transverse target polarization depends on 71  SDMEs~\cite{Diehl}. 
While the SDMEs can be completely expressed in terms of amplitude ratios, 
the fact that there are many more 
SDMEs than amplitude ratios implies that the SDMEs obey some relations. Due to the 
complicated  inter-dependence on amplitude ratios, SDMEs cannot be considered as independent quantities  and 
amplitude ratios  can  provide a more economical basis for fitting angular distributions of decay particles. 
This implies that the   SDME values calculated from  extracted amplitude ratios can be expected to be more 
accurate than those from direct SDME fits.

 SDMEs calculated using the extracted helicity amplitude ratios may differ from those obtained with the {SDME 
method. The SDMEs calculated
 from amplitude ratios are more constrained than those obtained by the SDME method. 
 In general, any set of SDMEs obtained with the SDME method is physical only if it can be described in terms of 
amplitude ratios. Certain conditions, called positivity constraints, that must be satisfied by SDMEs in order to 
be expressed through amplitude ratios were considered in Ref.~\cite{Diehl}. However,  the full set of conditions
  that must  be  satisfied by SDMEs in order to be expressible  in terms of helicity amplitude ratios is 
currently unknown.
%%%%%%%%%%%%%%%%%%%%%%%%%%%%%%%%%%%%%%%
\subsection{Hierarchy of  Helicity Amplitudes}
  The number of amplitude ratios to be extracted from data can be reduced if  there exists  a hierarchy for the 
moduli of helicity amplitudes. For large photon virtuality
$Q^2$ and small $|t^{\prime}|$, such a hierarchy among  helicity amplitudes was predicted
theoretically~\cite{IK,KNZ} and confirmed experimentally~\cite{H1,DC-24,ZEUS1}.  
According to this hierarchy, there are  a few  significant amplitudes while 
the contributions of other amplitudes may be neglected within the present experimental accuracy.
In a given kinematic region, the hierarchy of the amplitudes is governed by one or more ``small kinematic  
parameters"  that determine the contribution of various amplitudes to the process. 
%%%%%%%%%%%%%%%%%%%%%%
\subsubsection{$s$-Channel Helicity Conservation}
\label{sec:schc}
It was observed~\cite{Bauer,INS} that, at small $|t^{\prime}|$,  the amplitudes with a helicity flip in the 
transition  $\gamma^* \to V$ ($\lambda _V \neq \lambda _{\gamma}$) are much smaller than those for diagonal 
transitions where the helicity of the vector  meson is equal to that of the virtual photon ($\lambda _V=\lambda 
_{\gamma}$). This behavior is controlled by the small parameter  
\begin{eqnarray}
\alpha =\frac{\sqrt{-t^{\prime}}}{M},
\label{alpha}
\end{eqnarray}
where $M$ is a typical hadronic mass, usually taken to be the nucleon mass. 
This dominance of the $\gamma^*\rightarrow V$ transition that is diagonal with respect to helicity is called the
$s$-channel helicity conservation (SCHC) approximation. Furthermore,  it was shown~\cite{MSI} that for small 
 $|t^\prime|$ and $Q^2 > 2M^2$,  NPE amplitudes with  nucleon helicity flip are suppressed compared to  amplitudes 
describing diagonal transitions with $\lambda _N=\lambda '_N$. The same small parameter given by Eq.~(\ref{alpha})
  controls this suppression.

As only an unpolarized target is considered here,  there is no linear contribution of nucleon-helicity-flip amplitudes to  the relevant  
SDMEs~\cite{SW,Diehl}. 
The fractional contribution of  NPE amplitudes with  nucleon helicity flip is suppressed by a factor $\alpha^2$,
 so that we need to consider only NPE amplitudes with $\lambda_{N}=\lambda'_{N}$ if we neglect terms of order $\alpha^2$.

 The SCHC NPE amplitudes are $T_{00}$ and $T_{11}$. The helicity-flip amplitudes $T_{01}$ and $T_{10}$ are  
proportional to the small factor $\alpha$, while the double-helicity-flip amplitude $T_{1-1}$ is proportional to 
 $\alpha^2$.
%%%%%%%%%%%%%%%%%%%%%
\subsubsection{Twist decomposition}
\label{sec:twistdec}
 At asymptotically large photon virtuality, all amplitudes can be decomposed into power series of another small 
parameter
\begin{eqnarray}
\beta =\frac{M_V}{Q}.\;
\label{beta}
\end{eqnarray}
 An expansion in $\beta$ corresponds to the twist decomposition. A theoretical analysis  in the framework of 
pQCD  was performed~\cite{IK,KNZ}
  for the amplitudes of the process in Eq.~(\ref{react01}). It was shown~\cite{strikman}  that only the 
dominant amplitude  $T_{00}$  receives twist-2  contribution, while all other amplitudes contain only 
higher-twist contributions. The amplitudes $T_{11}$, $T_{01}$, $T_{1-1}$ are suppressed
 by $\beta$  while the amplitude $T_{10}$ is suppressed by $\beta^2$. SCHC and the twist decomposition  lead
 to the following hierarchy of NPE amplitudes at small $|t^\prime|$ and  large $Q^2$:
\begin{eqnarray}
|T_{00}|^2 \gg |T_{11}|^2\gg |T_{01}|^2 
\gg |T_{10}|^2 \sim  |T_{1-1}|^2\;.
\label{T-hier}
\end{eqnarray}
Such a hierarchy was observed  in the HERA collider experiments~\cite{ZEUS2,H1-amp,ZEUS1,H1}. 

%%%%%%%%%%%%%%%%%%%%%%%%
\subsubsection{Asymptotic Behavior of Amplitudes at High Energy}
At high energy,  a third small parameter can be defined:
\begin{eqnarray}
\gamma =\frac{M}{W}\;.
\label{gamma}
\end{eqnarray}
The asymptotic behavior of amplitudes at large $W$ and small $|t|$ was studied experimentally, providing 
information that led to the development of Regge phenomenology~\cite{IW,KS}. The fractional contribution of 
 various amplitudes to SDMEs can be estimated by applying the  formula for the amplitude of the exchange of a 
 reggeon $R$ at small $|t|$ and large $W$, $F \propto (W^2/M^2)^{\alpha_R(t)}$, where $\alpha_R(t)$  is the 
Regge  trajectory (the spin of the exchanged reggeon). As a rough estimate, we assume that the pomeron intercept 
 $\alpha_P(0) \approx 1$. Hence the amplitude of pomeron exchange is proportional to the factor $W^2/M^2$. 
 Secondary reggeons  with natural parity have $\alpha_R(0) \approx 0.5$ which results in  $F \propto W/M$. 
 Intercepts for reggeons with  unnatural parity  are negative in accordance with the results of a Regge 
 phenomenology analysis~\cite{IW,KS} of experimental data on soft scattering of hadrons. Therefore, the Regge
 factor is less than $(W^2/M^2)^0=1$. In QCD, exchanges of secondary reggeons and reggeons with unnatural parity 
 correspond to quark-antiquark exchanges, while pomeron exchange corresponds to two-gluon exchange. In other 
 words, the ratio of amplitudes of quark-antiquark exchanges  with natural parity to the amplitude of pomeron 
 exchange is proportional to the parameter $\gamma$, while the ratio of amplitudes of quark-antiquark exchanges  
 with unnatural parity to the amplitude of pomeron exchange is proportional to $\gamma^2$. Since there is no 
 interference between contributions of NPE and UPE amplitudes to SDMEs measured on unpolarized targets, the 
 fractional contribution of UPE amplitudes is suppressed by a factor $\gamma^4$. This explains why only NPE 
 amplitudes  survive at HERA collider energies.

In the context of single-reggeon exchange, the isospin $I=1$  reggeon ($\rho$, $a_2$, ...) contribution 
to the amplitude of vector-meson production
on the proton is of opposite sign to that on the neutron. In contrast, the $I=0$  reggeon (pomeron, $\omega$, $f_2$, ...)
contribution to the amplitude is of the same sign for both proton and neutron. 
Hence, the results on  SDMEs and  amplitude ratios should be different for proton and deuteron
 if $I=1$ and $I=0$ contributions to the amplitude  interfere.
By comparing proton and deuteron results, 
the fractional contribution of $I=1$ reggeons  to the extracted amplitude ratios can be estimated.
%%%%%%%%%%%%%%%%%%%%%%%%%%
\subsubsection{Hierarchy of Amplitudes in the Kinematic Region of HERMES}
\label{sec:herm-twist}
In the HERMES kinematic region, the parameter $\beta$ is  larger than 0.3  and the relative sizes of SDMEs measured in 
exclusive $\rho ^0$ production  can be explained by the following hierarchy~\cite{DC-24}:
\begin{eqnarray}
\nonumber
|T_{00}|^2 &\sim& |T_{11}|^2 \gg |U_{11}|^2 > |T_{01}|^2  
\gg |T_{10}|^2 \\ &\sim&  |T_{1-1}|^2 
 > |U_{01}|^2 \sim |U_{10}|^2 \sim |U_{1-1}|^2, 
\label{HERMES-hier}
\end{eqnarray}
with the two largest amplitudes being $T_{00}$ and $T_{11}$. The abbreviated notation $U_{\lambda _{V} \lambda _{\gamma}}$, where
\begin{eqnarray}
|U_{\lambda _{V} \lambda _{\gamma}}|^2 \equiv
|U_{\lambda _{V}\frac{1}{2}\lambda _{\gamma}-\frac{1}{2}}|^2
+|U_{\lambda _{V}\frac{1}{2}\lambda _{\gamma}\frac{1}{2}}|^2 
\label{U-short}
\end{eqnarray}
 was introduced because, for UPE amplitudes, it is impossible to prove the dominance of those without 
nucleon helicity flip over those with helicity flip,
  in contrast to the NPE amplitudes. As shown in Eq.~(\ref{HERMES-hier}), the moduli of all the 
 UPE amplitudes except $U_{11}$ are smaller than those of the NPE amplitudes.  The numerical estimate
  $|U_{11}|^2/(|T_{11}|^2+\epsilon |T_{00}|^2) \approx 0.11 \pm 0.04$ obtained in Ref.~\cite{DC-24} shows that 
the modulus of $|U_{11}|$  is even larger than that of $|T_{01}|$ (and the result of the fit done in the present 
work confirms this result) and the contribution of $|U_{11}|$ to the SDMEs cannot be neglected. This 
contribution is suppressed by a factor $\gamma ^4$ which is smaller than $0.01$ in the HERMES kinematic region 
$3.0$~GeV $\leq W \leq 6.5$~GeV.  If the UPE amplitude $U_{11}$ is due to pion exchange, its contribution may be 
significant in the HERMES kinematic  region because of the large pion-nucleon coupling constant $g_{\pi NN}$.

Contributions from small amplitudes can be appreciable if they are multiplied by the largest amplitudes $T_{00}$ or $T_{11}$ in the numerators
 of SDME formulas. As the small amplitudes $T_{01}$, $T_{10}$, and $T_{1-1}$ are multiplied by the largest amplitudes (see Ref.~\cite{SW}
 and Appendix A of Ref.~\cite{DC-24}), five complex NPE amplitudes have to be considered in total, i.e. the four ratios 
$T_{11}/T_{00}$,
 $T_{01}/T_{00}$, $T_{10}/T_{00}$, and $T_{1-1}/T_{00}$. Concerning the ratio $U_{11}/T_{00}$, only $|U_{11}|/|T_{00}|$ can be determined
 as there is no interference between UPE and NPE amplitudes for  the case of  an unpolarized target. 
The contribution of all other UPE amplitudes can be neglected~\cite{DC-24}.
In total, there are nine independent free parameters to be determined when fitting the angular distribution using the amplitude method.
%%%%%%%%%%%%%%%
\subsection{Basic Formulas of the Amplitude Method}
The exact formulas for SDMEs expressed in terms of  helicity amplitudes
are presented in Appendix A of Ref.~\cite{DC-24}.
In order to rewrite  SDMEs in terms of the ratios  of 
significant helicity amplitudes, we 
neglect the contributions
from NPE nucleon helicity-flip amplitudes (using Eq.~(\ref{tsum}))
and from all UPE amplitudes except $U_{11}$.
Then, dividing both the numerators and the denominator in the  exact formulas
for SDMEs by $|T_{00}|^2$, we arrive  at approximate expressions for SDMEs
in terms of certain amplitude ratios: 
\begin{align}
              r_{00}^{04}    & \simeq \{\epsilon +|t_{01}|^2\}/N                                \label{a1} \,,  \\
\mathrm{Re} \{r_{10}^{04} \} & \simeq \mathrm{Re} \{\epsilon t_{10} +\frac {1}{2}t_{01}(t_{11}-t_{1-1})^*\}/N  \label{a2} \,,  \displaybreak[2] \\
              r_{1-1}^{04}   & \simeq \mathrm{Re} \{ -\epsilon |t_{10}|^2+t_{1-1}t_{11}^* \}/N  \label{a3}\,,\displaybreak[2] \\
              r_{11}^{1}     & \simeq \mathrm{Re} \{ t_{1-1}t_{11}^* \}/N                       \label{a4} \,, \displaybreak[2]\\
              r_{00}^1       & \simeq - |t_{01}|^2/ N   \label{a5} \,, \displaybreak[2]\\
\mathrm{Re} \{r_{10}^{1} \}  & \simeq \frac{1}{2} \mathrm{Re}\{-t_{01}(t_{11}-t_{1-1})^* \}/N   \label{a6} \,, \displaybreak[2]\\
              r_{1-1}^1      & \simeq \frac {1}{2} \{|t_{11}|^2+|t_{1-1}|^2-|u_{11}|^2 \}/N \label{a7} \,, \displaybreak[2] \\
\mathrm{Im} \{r_{10}^2 \}    & \simeq \frac{1}{2} \mathrm{Re} \{t_{01}(t_{11}+t_{1-1})^* \}/N  \label{a8} \,, \displaybreak[2]\\
\mathrm{Im} \{r_{1-1}^2 \}   & \simeq \frac {1}{2} \{ -|t_{11}|^2 +|t_{1-1}|^2 +|u_{11}|^2 \}/N \label{a9} \,, \displaybreak[2]  \\
              r_{11}^{5}     & \simeq \frac{1}{\sqrt{2}}  \mathrm{Re} \{ t_{10} (t_{11}-t_{1-1})^*\}/N \label{a10} \,,\displaybreak[2] \\
r_{00}^5                     & \simeq \sqrt{2}  \mathrm{Re} \{ t_{01}\}/N   \label{a11} \,, \displaybreak[2]\\
\mathrm{Re} \{r_{10}^{5}\}   & \simeq \frac{1}{\sqrt{8}} \mathrm{Re} \{2t_{10} t_{01}^*+(t_{11}-t_{1-1})\} /N  \label{a12} \,,  \displaybreak[2] \\
              r_{1-1}^{5}    & \simeq \frac{1}{\sqrt{2}} \mathrm{Re} \{ -t_{10} (t_{11}-t_{1-1})^*\}/N \label{a13} \,, \displaybreak[2]\\
\mathrm{Im} \{r_{10}^{6}\}   & \simeq -\frac{1}{\sqrt{8}} \mathrm{Re} \{t_{11}+t_{1-1}\}/N   \label{a14} \,, \displaybreak[2]\\
\mathrm{Im} \{r^6_{1-1}\}    & \simeq \frac{1}{\sqrt{2}} \mathrm{Re} \{ t_{10} (t_{11}+t_{1-1})^* \}/N \label{a15} \,,\displaybreak[2] \\
\mathrm{Im} \{r_{10}^{3}\}   & \simeq -\frac{1}{2} \mathrm{Im} \{t_{01}(t_{11}+t_{1-1})^* \}/N \label{a16}   \,, \displaybreak[2]\\
\mathrm{Im} \{r_{1-1}^{3}\}  & \simeq -\mathrm{Im} \{t_{1-1}t_{11}^* \} /N \;,\label{a17} \displaybreak[2] \\
\mathrm{Im} \{r_{10}^{7}\}   & \simeq \frac{1}{\sqrt{8}} \mathrm{Im} \{t_{11}+t_{1-1}\}/N  \label{a118} \,,  \displaybreak[2] \\
\mathrm{Im} \{r_{1-1}^{7}\}  & \simeq \frac{1}{\sqrt{2}} \mathrm{Im} \{t_{10}(t_{11}+t_{1-1})^*\}/N \label{a19},\displaybreak[2] \\
              r^8_{11}       &  \simeq   -\frac{1}{\sqrt{2}} \mathrm{Im} \{t_{10}(t_{11}-t_{1-1})^*\}/N \label{a20} \,, \displaybreak[2]\\
              r_{00}^{8}     &  \simeq   \sqrt{2}  \mathrm{Im} \{ t_{01}  \}/N \label{a21} \,, \displaybreak[2]\\
\mathrm{Re} \{r_{10}^{8}\}   &  \simeq   \frac{1}{\sqrt{8}} \mathrm{Im} \{-2t_{10}t_{01}^*+t_{11}-t_{1-1}\}/N  \label{a22} \,, \displaybreak[2]\\
              r_{1-1}^{8}    &  \simeq  \frac{1}{\sqrt{2}}\mathrm{Im} \{t_{10}(t_{11}-t_{1-1})^*\}/N  \label{a23}
\end{align}
where the normalization factor $N=\mathcal{N}/|T_{00}|^2$ is defined in accordance with Eqs.
(\ref{ntotal})-(\ref{sigmalong}) by
\begin{eqnarray}
\nonumber
N &=&   N_T + \epsilon N_L,\\ \nonumber
 N_{T} &  \simeq  &
  |t_{11}|^2+|t_{01}|^2+|t_{1-1}|^2 +|u_{11}|^2, \\
N_L  &  \simeq  &  1+2|t_{10}|^2.
 \label{a24}
\end{eqnarray}
Here $t_{\lambda_V \lambda_{\gamma}} \equiv T_{\lambda_V \lambda_{\gamma}}/T_{00}$ and $|u_{11}|^2 \equiv |U_{11}|^2/|T_{00}|^2$  with 
$|U_{11}|^2$ defined in  
 Eq.~(\ref{U-short}). There are nine independent real functions in Eqs.~(\ref{a1}-\ref{a24}), namely: 
 $\mathrm{Re}(t_{11})$, $\mathrm{Im}(t_{11})$,  $\mathrm{Re}(t_{01})$, $\mathrm{Im}(t_{01})$,  
 $\mathrm{Re}(t_{10})$, $\mathrm{Im}(t_{10})$, $ \mathrm{Re}(t_{1-1})$, $\mathrm{Im}(t_{1-1})$, and $|u_{11}|$. 
 Equations~(\ref{a1}-\ref{a24}) are the basis for the extraction of the helicity amplitude ratios from the 
 measured angular distributions and also for the calculation of SDMEs from amplitude ratios (see 
 Sec.~\ref{sec:comp-sdme-am}).
%%%%%%%%%%%%%%%%%%%%%%%%%%%%%%%%%%%%%%%%%%%%%%%%%%%%%%%%%%%%%%%%%%%%%%%%%%%%%%%%%%%
\section{The HERMES Experiment }
A detailed description of the HERMES experiment at DESY can be found in Ref.~\cite{herspec}. The data analyzed in this paper were collected between the years 1996 and 2005. A longitudinally polarized positron or electron beam of $27.6$~GeV was scattered from pure hydrogen or deuterium gas targets internal to the HERA lepton storage ring. The lepton beam was transversely polarized due to an asymmetry in the emission of synchrotron-radiation~\cite{ST} in the field of the dipole magnets. The transverse beam polarization was transformed locally into longitudinal polarization at the interaction point by spin rotators located upstream and downstream of the HERMES apparatus. The helicity of the beam was typically reversed every two months. For both  positive and negative beam helicities, the beam polarization was continuously measured by two Compton polarimeters~\cite{BAR,BECK}. The average beam polarization for the hydrogen (deuterium) data set was 0.45 (0.47) after requiring $0.15 < P_{B} < 0.8$  in the analysis process, and the fractional  uncertainty of the beam polarization was 3.4\% (2.0\%)~\cite{BAR,BECK}. Some of the data  were collected with longitudinally or  transversely polarized targets. However, the time-averaged polarization of the polarized targets was negligible, while the rapid (60-180s) reversal of the polarization direction minimized polarization bias due to detector effects.

HERMES was a forward spectrometer~\cite{herspec} in which  both scattered lepton and produced hadrons were detected within an angular acceptance of $\pm$170~mrad horizontally, and $\pm(40 - 140)$~mrad vertically. The tracking system had a momentum resolution of about $1.5\%$ and an angular resolution of about $1$~mrad. Lepton identification was accomplished using a transition-radiation detector, a  preshower scintillator counter, and the electromagnetic calorimeter. The particle identification system included a gas threshold $\breve {\mathrm C}$e\-ren\-kov counter, which was replaced in 1998 by a dual-radiator ring-imaging  $\breve {\mathrm C}$erenkov  detector~\cite{RICH}. Combining the responses of these detectors in a likelihood method led to  an average lepton identification efficiency of 98\% with a hadron contamination of less than 1\%. The event sample used in this analysis is exactly the same as that used in Ref.~\cite{DC-24}.
%%%%%%%%%%%%%%%%%%%%%%%%%%%%%%%%%%%%%%%%%%%%%%%%%%%%%%%%%%%%%%%
\section{Extraction of Amplitude Ratios}
\label{sec:amp-extrc}
The measurement of the angular distribution of the scattered electrons/positrons and the  pions from the decay 
$\rho^0 \to \pi^+ \pi^-$ is the basis for the extraction of
 spin  density matrix elements $r^{\eta}_{\lambda_V \mu_V}$ in the SDME method and of helicity amplitude ratios
 $t_{\lambda_V \lambda_{\gamma}} $ and  $|u_{11}|$  in the amplitude method. For a polarized lepton beam and an 
unpolarized target,  the angular distribution of the pions is given by  relations~(\ref{eqang1}-\ref{eqang3}) in 
Sec.~\ref{app:app1} (see also Refs.~\cite{SW,DC-24}), where the SDMEs are expressed in terms 
 of helicity amplitude ratios according to Eqs.~(\ref{a1}-\ref{a24}).
%%%%%%%%%%%%%%%%%%%%%%%%%%%%%%%
\subsection{Binned Maximum Likelihood Method}
Amplitude ratios are extracted with the same binned maximum likelihood method  as in the previous SDME 
analysis (see Sec.~6 of Ref.~\cite{DC-24} ). In brief, the amplitude ratios in each of the kinematic 
bins are obtained by minimizing the difference between the 3-dimensional $(\cos\theta,\phi,\Phi)$ angular 
distribution of the experimental events and that  of a sample  of fully reconstructed Monte Carlo events. The 
 latter  are generated isotropically in $(\cos\theta,\phi,\Phi)$ using the rhoMC 
 generator~\cite{tytgat,ami-phd} for exclusive $\rho^0$ simulated in the instrumental context of the
 spectrometer, and passed through the same reconstruction chain as the experimental data. The generated Monte
 Carlo events are iteratively re\-weigh\-ted with  the angular distribution given by 
 Eqs.~(\ref{eqang1}-\ref{eqang3}), with the amplitude ratios treated as free parameters.

The minimization itself and the uncertainty calculation are performed using the MINUIT package~\cite{CERN-CN}. More details can be found in Ref.~\cite{DC-24}.
%%%%%%%%%%%%%%%%%%%%%%%%%%%%%%%%%%%%%%%%%%%%%%%%%%%%%%%%%%%
\subsection{Results on Amplitude Ratios in Bins of $Q^2$ and $-t^\prime$ }
%\begin{landscape}
\begin{table*}[hbtc!]
 \renewcommand{\arraystretch}{1.2}
\begin{center}
{\footnotesize
\begin{tabular}{|c|c|c|c|c|c|}
\hline
bin& $\langle Q^2 \rangle$, GeV$^2$&$ \langle -t^\prime \rangle$, GeV$^2$&
bin& $\langle Q^2 \rangle$, GeV$^2$ &$\langle -t^\prime \rangle$, GeV$^2$\\
\hline
$q_1$$t_1$&0.817&0.019&$q_3$$t_1$&1.658&0.019\\
\hline
$q_1$$t_2$&0.823&0.068&$q_3$$t_2$&1.660&0.068\\
\hline
$q_1$$t_3$&0.821&0.146&$q_3$$t_3$&1.663&0.146\\
\hline
$q_1$$t_4$&0.815&0.280&$q_3$$t_4$&1.663&0.282\\
\hline
$q_2$$t_1$&1.184&0.019&$q_4$$t_1$&2.996&0.019\\
\hline
$q_2$$t_2$&1.188&0.068&$q_4$$t_2$&3.056&0.068\\
\hline
$q_2$$t_3$&1.189&0.145&$q_4$$t_3$&3.076&0.146\\
\hline
$q_2$$t_4$&1.188&0.282&$q_4$$t_4$&3.134&0.284\\
\hline
\end{tabular} 
\\[2pt]
}
\caption{ \label{meanbin} \footnotesize{
Mean values of kinematic variables for 16 bins. The limits of the $q_1$, $q_2$, $q_3$, and
$q_4$ bins  for $Q^2$  are 0.5; 1.0; 1.4; 2.0; 7.0~GeV$^2$ while those of the $t_1$, $t_2$, $t_3$, and $t_4$
bins for $-t'$  are the following: 0.0; 0.04; 0.10; 0.20; 0.40~GeV$^2$.}}
\end{center}
\end{table*}
%\end{landscape}

Ratios of helicity amplitudes are extracted in a two-dimen\-sio\-nal ($Q^2$, $-t^\prime$) binning from the same HERMES proton and deuteron
data sets as were used for the SDME analysis of Ref.~\cite{DC-24}. The four $Q^2$ bins are denoted as  $q_1$, $q_2$, $q_3$, and $q_4$  defined by the limits   $0.5$, $1.0$, $1.4$, $2.0$, $7.0$~GeV$^2$.  The four $-t^\prime$ bins  are denoted as $t_1$, $t_2$, $t_3$, and $t_4$   defined by the limits $0.0$, $0.04$, $0.10$, $0.20$, $0.40$~GeV$^2$. The mean values of $Q^2$ and $-t^\prime$ in  each bin are presented in Tab.~\ref{meanbin}. The results of the extraction of the amplitude ratios in 16 bins are presented in Tab.~\ref{ampl-hydr} for the proton data and in Tab.~\ref{ampl-deutr} for the deuteron data. In every ($Q^2$, $-t^\prime$) bin, the nine free parameters are obtained from a fit to the 3-dimensional angular distribution without subtracting the background from semi-inclusive deep-inelastic scattering (SIDIS). The SIDIS background under the exclusive peak, i.e., in the region $-1$ GeV $< \Delta E <0.6$ GeV, is estimated using the PYTHIA Monte Carlo generator~\cite{pyth} (see Ref.~\cite{DC-24} for more details) and its effect is assigned as a systematic uncertainty.

\begin{table*}[hbtc!]
 \renewcommand{\arraystretch}{1.2}
\begin{center}
{\footnotesize
\begin{tabular}{|l|r|r|r|r|}
\hline
Ratio & $q_1$$t_1$ \hspace{1.2cm} & $q_1$$t_2$ \hspace{1.2cm} & $q_1$$t_3$
\hspace{1.2cm} & $q_1$$t_4$
\hspace{1.2cm} \\
\hline
$\mathrm{Re}(t_{11}) $&$0.975\pm0.121\pm0.297 $&$0.961\pm0.101\pm0.213 $&$1.363\pm0.140\pm0.309 $&$1.037\pm0.176 \pm
0.230 $\\
\hline
$\mathrm{Im}(t_{11})$&$0.542\pm0.187\pm0.052 $&$0.285\pm0.145\pm0.174 $&$0.082\pm0.285\pm0.104 $&$0.783\pm0.147\pm0.213 $\\
\hline
$\mathrm{Re}( t_{01} )$&$0.025\pm0.044\pm0.056 $&$0.112\pm0.036\pm0.047 $&$0.214\pm0.064\pm0.050 $&$0.182\pm0.048\pm0.116 $\\
\hline
$\mathrm{Im}( t_{01} )$&$-0.098\pm0170\pm0.211 $&$0.111\pm0.111\pm0.136 $&$0.315\pm0.120\pm0.073 $&$0.107\pm0.108\pm0.107 $\\
\hline
$\mathrm{Re}( t_{10} )$&$0.037\pm0.038\pm0.050 $&$-0.049\pm0.039\pm0.024 $&$0.024\pm0.039\pm0.026 $&$-0.010\pm0.056\pm0.043
$\\
\hline
$\mathrm{Im}( t_{10} )$&$-0.067\pm0.064\pm0.156 $&$-0.066\pm0.068\pm0.009 $&$0.005\pm0.069\pm0.023
$&$0.050\pm0.067\pm0.021$\\
\hline
$\mathrm{Re}( t_{1-1} )$&$-0.110\pm0.042\pm0.045 $&$-0.021\pm0.037\pm0.079 $&$-0.037\pm0.043\pm0.043 $&$0.020\pm0.056\pm0.040
$\\
\hline
$\mathrm{Im}( t_{1-1} )$&$0.178\pm0.087\pm0.349 $&$-0.147\pm0.092\pm0.055 $&$-0.124\pm0.100\pm0.079 $&$-0.172\pm0.079\pm0.104
$\\
\hline
$|u_{11}|$&$0.329\pm0.070\pm0.021 $&$0.424\pm0.049\pm0.032 $&$0.391\pm0.063\pm0.124 $&$0.357\pm0.077 \pm 0.070 $\\
\hline
\hline
Ratio  & $q_2$$t_1$ \hspace{1.2cm} & $q_2$$t_2$ \hspace{1.2cm} & $q_2$$t_3$
\hspace{1.2cm} & $q_2$$t_4$ \hspace{1.2cm} \\
\hline
%%%% 2-line
$\mathrm{Re}(t_{11})$&$1.138\pm0.143\pm0.021 $&$0.996\pm0.106\pm0.084 $&$1.079\pm0.094\pm0.135 $&$0.925\pm0.097\pm0.139 $\\
\hline
$\mathrm{Im}(t_{11})$&$0.282\pm0.268\pm0.212 $&$0.386\pm0.125\pm0.039 $&$0.342\pm0.162\pm0.159 $&$0.289\pm0.156\pm0.141 $\\
\hline
$\mathrm{Re}( t_{01} )$&$0.044\pm0.051\pm0.009 $&$0.116\pm0.037\pm0.061 $&$0.113\pm0.035\pm0.030 $&$0.250\pm0.043\pm0.075 $\\
\hline
$\mathrm{Im}( t_{01} )$&$0.073\pm0.146\pm0.137 $&$0.327\pm0.099\pm0.083 $&$-0.009\pm0.141\pm0.069 $&$ 0.265\pm0.107\pm0.055
$\\
\hline
$\mathrm{Re}( t_{10} )$&$0.001\pm0.054\pm0.011 $&$-0.054\pm0.034\pm0.033 $&$0.032\pm0.030\pm0.015 $&$-0.065\pm0.032\pm0.020
$\\
\hline
$\mathrm{Im}( t_{10} )$&$-0.081\pm0.073\pm0.154 $&$0.064\pm0.073\pm0.058 $&$0.006\pm0.070\pm0.027 $&$0.063\pm0.073\pm0.057
$\\
\hline
$\mathrm{Re}( t_{1-1} )$&$0.001\pm0.045\pm0.015 $&$0.023\pm0.040\pm0.018 $&$-0.017\pm0.036\pm0.021 $&$-0.073\pm0.035\pm0.019
$\\
\hline
$\mathrm{Im}( t_{1-1} )$&$-0.041\pm0.137\pm0.271 $&$0.002\pm0.078\pm0.019 $&$-0.084\pm0.107\pm0.024
$&$-0.099\pm0.069\pm0.034
$\\
\hline
$|u_{11}|$&$0.499\pm0.040\pm0.039 $&$0.429\pm0.042\pm0.081 $&$0.418\pm0.046\pm0.119 $&$0.359\pm0.055\pm0.055 $\\
\hline
%%% 3-line
\hline
Ratio  & $q_3$$t_1$ \hspace{1.2cm}  & $q_3$$t_2$ \hspace{1.2cm} & $q_3$$t_3$
\hspace{1.2cm} & $q_3$$t_4$ \hspace{1.2cm} \\
\hline
$\mathrm{Re}(t_{11})$&$1.029\pm0.104\pm0.039 $&$1.088\pm0.098\pm0.083 $&$0.878\pm0.079\pm0.202 $&$0.975\pm0.096 \pm 0.228 $\\
\hline
$\mathrm{Im}(t_{11})$&$0.257\pm0.131\pm0.151 $&$0.479\pm0.111\pm0.050 $&$0.513\pm0.102\pm0.385 $&$0.280\pm0.176\pm0.365 $\\
\hline
$\mathrm{Re}( t_{01})$&$-0.052\pm0.039\pm0.052 $&$0.008\pm0.032\pm0.060 $&$0.113\pm0.035\pm0.081 $&$0.194\pm0.044\pm0.064
$\\
\hline
$\mathrm{Im}( t_{01} )$&$0.357\pm0.127\pm0.480 $&$0.100\pm0.100\pm0.004 $&$0.123\pm0.092\pm0.275 $&$-0.100\pm0.108\pm0.406
$\\
\hline
$\mathrm{Re}( t_{10} 0$&$0.074\pm0.028\pm0.021 $&$0.043\pm0.028\pm0.011 $&$-0.007\pm0.034\pm0.032 $&$-0.032\pm0.038\pm0.036
$\\
\hline
$\mathrm{Im}( t_{10} )$&$-0.051\pm0.067\pm0.042 $&$-0.129\pm0.058\pm0.031 $&$0.075\pm0.053\pm0.173 $&$0.025\pm0.113\pm0.120
$\\
\hline
$\mathrm{Re}( t_{1-1} )$&$-0.013\pm0.038\pm0.021 $&$-0.019\pm0.035\pm0.037 $&$ 0.034\pm0.037\pm0.101
$&$-0.044\pm0.036\pm0.065 $\\
\hline
$\mathrm{Im}( t_{1-1} )$&$0.108\pm0.090\pm0.236 $&$0.100\pm0.071\pm0.031 $&$-0.186\pm0.071\pm0.071 $&$0.037\pm0.098\pm0.185
$\\
\hline
$|u_{11}|$&$0.423\pm0.055\pm0.048 $&$0.323\pm0.068\pm0.084 $&$0.346\pm0.056\pm0.085 $&$0.445\pm0.050\pm0.191 $\\
\hline
%%% 4-line
\hline
Ratio & $q_4$$t_1$ \hspace{1.2cm} & $q_4$$t_2$ \hspace{1.2cm} & $q_4$$t_3$
\hspace{1.2cm} & $q_4$$t_4$ \hspace{1.2cm} \\
\hline
$\mathrm{Re}(t_{11})$&$ 0.723\pm0.071\pm0.053 $&$0.706\pm0.068\pm0.039 $&$0.582\pm0.074\pm0.156 $&$0.650\pm0.063\pm0.117 $\\
\hline
$\mathrm{Im}(t_{11})$&$0.570\pm0.071\pm0.025 $&$0.657\pm0.061\pm0.066 $&$0.583\pm0.110\pm0.120 $&$0.488\pm0.067\pm0.119 $\\
\hline
$\mathrm{Re}( t_{01} )$&$0.062\pm0.031\pm0.007 $&$0.121\pm0.031\pm0.006 $&$0.190\pm0.054\pm0.021 $&$0.282\pm0.036\pm0.060
$\\
\hline
$\mathrm{Im}( t_{01} )$&$0.011\pm0.069\pm0.019 $&$0.084\pm0.079\pm0.019 $&$-0.129\pm0.0.57\pm0.154 $&$-0.099\pm0.081\pm0.044
$\\
\hline
$\mathrm{Re}( t_{10} )$&$-0.013\pm0.030\pm0.011 $&$-0.046\pm0.035\pm0.008 $&$0.095\pm0.025\pm0.118 $&$0.007\pm0.031\pm0.012
$\\
\hline
$\mathrm{Im}( t_{10} )$&$0.010\pm0.052\pm0.010 $&$0.050\pm0.045\pm0.015 $&$-0.181\pm0.060\pm0.258 $&$0.003\pm0.044\pm0.058
$\\
\hline
$\mathrm{Re}( t_{1-1} )$&$0.000\pm0.034\pm0.007 $&$0.065\pm0.034\pm0.007 $&$0.027\pm0.040\pm0.033 $&$0.024\pm0.031\pm0.011
$\\
\hline
$\mathrm{Im}( t_{1-1} )$&$0.029\pm0.056\pm0.005 $&$-0.068\pm0.048\pm0.015 $&$-0.038\pm0.040\pm0.142 $&$-0.036\pm0.051\pm0.066
$\\
\hline
$|u_{11}|$&$0.451\pm0.049\pm0.024 $&$0.306\pm0.061\pm0.032 $&$0.383\pm0.067\pm0.097 $&$0.380\pm0.044\pm0.091 $\\
\hline
\end{tabular} \\[2pt]
}
\end{center}
\caption{ \label{ampl-hydr} \small{
Ratios of helicity amplitudes in $Q^2$ and $-t'$ bins for proton data. The mean values of $Q^2$ and $-t'$ for all
16 bins are given in Tab.~\ref{meanbin}. The notations $t_{kj}$ and $|u_{11}|$ are used for the amplitude ratios
$T_{kj}/T_{00}$ and $|U_{11}/T_{00}|$, respectively. The first uncertainties are statistical, the  second  are systematic.
%\vspace*{10.0cm}
}}
\end{table*}

\begin{table*}[hbtc!]
 \renewcommand{\arraystretch}{1.2}
\begin{center}
{\footnotesize
\begin{tabular}{|l|r|r|r|r|}
\hline
Ratio& $q_1$$t_1$\hspace{1.2cm} & $q_1$$t_2$\hspace{1.2cm} & $q_1$$t_3$\hspace{1.2cm} &
$q_1$$t_4$\hspace{1.2cm} \\
\hline
$\mathrm{Re}(t_{11})$&$ 0.860 \pm 0.078 \pm 0.163$&$1.256 \pm 0.095 \pm 0.055$&$1.220 \pm 0.113 \pm 0.145$&$1.197 \pm 0.124
\pm 0.218$\\
\hline
$\mathrm{Im}(t_{11})$&$0.509 \pm 0.118 \pm 0.203$&$0.304 \pm0.159 \pm 0.049$&$0.192 \pm 0.228 \pm 0.114 $&$0.222 \pm 0.320
\pm 0.208$\\ 
\hline
$\mathrm{Re}( t_{01} )$&$-0.011\pm0.029\pm0.027$&$0.020\pm0.032\pm0.069$&$0.174\pm0.044\pm 0.080$&$0.151\pm 0.052\pm 0.042$\\
\hline
$\mathrm{Im}( t_{01} )$&$0.023\pm0.076\pm0.070$&$0.088\pm0.076\pm0.035$&$0.239\pm0.112\pm0.086$&$0.222\pm0.143\pm0.164 $\\
\hline
$\mathrm{Re}( t_{10} )$&$-0.011\pm0.029\pm0.020$&$0.002\pm0.027\pm0.014$&$0.025\pm0.032\pm0.048$&$0.017\pm0.033\pm0.019 $\\
\hline
$\mathrm{Im}( t_{10} )$&$0.020\pm0.049\pm0.044$&$-0.017\pm0.063\pm0.047$&$-0.048\pm0.083\pm0.135$&$0.014\pm0.076\pm0.095 $\\
\hline
$\mathrm{Re}( t_{1-1} )$&$-0.017\pm0.029\pm0.018 $&$0.056\pm0.030\pm0.051 $&$-0.052\pm0.032\pm0.039$&$0.027\pm0.050\pm0.078 
$\\
\hline
$\mathrm{Im}( t_{1-1} )$&$-0.062\pm0.064\pm0.020 $&$-0.098\pm0.078\pm0.039 $&$-0.093\pm0.093\pm0.153 
$&$-0.228\pm0.105\pm0.054 $\\
\hline
$|u_{11}|$&$0.364\pm0.037\pm0.090 $&$0.343\pm0.046\pm0.037 $&$0.419\pm0.048\pm0.020 $&$0.388\pm0.060\pm0.105 $\\
\hline
\hline
Ratio& $q_2$$t_1$\hspace{1.2cm} & $q_2$$t_2$\hspace{1.2cm} & $q_2$$t_3$\hspace{1.2cm} & 
$q_2$$t_4$\hspace{1.2cm}\\
\hline
$\mathrm{Re}(t_{11})$&$0.984\pm0.087\pm0.048 $&$0.888\pm0.080\pm0.126 $&$1.119\pm0.074\pm0.129 $&$0.990\pm0.063\pm0.140 $\\
\hline
$\mathrm{Im}(t_{11})$&$0.341\pm0.106\pm0.054 $&$0.778\pm0.078\pm0.153 $&$0.252\pm0.105\pm0.163 $&$0.039\pm0.142\pm0.482 $\\
\hline
$\mathrm{Re}( t_{01} )$&$0.042\pm0.023\pm0.022 $&$0.071\pm0.028\pm0.028 $&$0.109\pm0.025\pm0.037 $&$0.262\pm0.031\pm0.043 $\\
\hline
$\mathrm{Im}( t_{01} )$&$0.289\pm0.130\pm0.084 $&$-0.062\pm0.069\pm0.158 $&$0.136\pm0.074\pm0.035 $&$0.070\pm0.086\pm0.064 
$\\
\hline
$\mathrm{Re}( t_{10} )$&$-0.006\pm0.027\pm0.020 $&$0.039\pm0.025\pm0.038 $&$0.039\pm0.019\pm0.012 $&$0.022\pm0.029\pm0.048 
$\\
\hline
$\mathrm{Im}( t_{10} )$&$0.029\pm0.101\pm0.052 $&$-0.009\pm0.034\pm0.013 $&$-0.087\pm0.044\pm0.007 $&$-0.201\pm0.065\pm0.171 
$\\
\hline
$\mathrm{Re}( t_{1-1} )$&$0.013\pm0.046\pm0.017 $&$0.021\pm0.028\pm0.060 $&$-0.077\pm0.007\pm0.012 $&$-0.041\pm0.028\pm0.039 
$\\
\hline
$\mathrm{Im}9 t_{1-1} )$&$0.063\pm0.100\pm0.034 $&$-0.042\pm0.046\pm0.042 $&$0.030\pm0.070\pm0.012 $&$-0.071\pm0.072\pm0.106 
$\\
\hline
$|u_{11}|$&$0.395\pm0.035\pm0.066 $&$0.262\pm0.050\pm0.048 $&$0.401\pm0.036\pm0.134 $&$0.373\pm0.048\pm0.085 $\\
\hline
\hline
Ratio& $q_3$$t_1$\hspace{1.2cm} & $q_3$$t_2$\hspace{1.2cm} & $q_3$$t_3$\hspace{1.2cm} & 
$q_3$$t_4$\hspace{1.2cm} \\
\hline
$\mathrm{Re}(t_{11})$&$0.787\pm0.059\pm0.052 $&$0.819\pm0.065\pm0.067 $&$0.839\pm0.076\pm0.076 $&$0.835\pm0.060 \pm  0.128 
$\\
\hline
$\mathrm{Im}(t_{11})$&$0.487\pm0.072\pm0.081 $&$0.518\pm0.085\pm0.079 $&$0.553\pm0.088\pm0.111 $&$0.203\pm0.097\pm0.129 $\\
\hline
$\mathrm{Re}( t_{01} )$&$0.028\pm0.024\pm0.030 $&$0.094\pm0.025\pm0.053 $&$0.129\pm0.028\pm0.017 $&$0.222\pm0.028\pm0.034 $\\
\hline
$\mathrm{Im}( t_{01} )$&$0.102\pm0.080\pm0.021 $&$0.001\pm0.091\pm0.042 $&$-0.017\pm0.078\pm0.072 $&$0.080\pm0.072\pm0.040 
$\\
\hline
$\mathrm{Re}( t_{10} )$&$-0.006\pm0.024\pm0.003 $&$-0.007\pm0.023\pm0.020 $&$0.027\pm0.028\pm0.010 $&$0.053\pm0.026\pm0.038 
$\\
\hline
$\mathrm{Im}( t_{10} )$&$0.008\pm0.048\pm0.018 $&$-0.025\pm0.045\pm0.042 $&$-0.092\pm0.046\pm0.014 $&$-0.228\pm0.058\pm0.035 
$\\
\hline
$\mathrm{Re}( t_{1-1} )$&$0.031\pm0.026\pm0.026 $&$-0.045\pm0.027\pm0.012 $&$-0.018\pm0.032\pm0.052 $&$-0.008\pm0.027\pm0.025 
$\\
\hline
$\mathrm{Im}( t_{1-1} )$&$0.000\pm0.052\pm0.037 $&$0.102\pm0.061\pm0.015 $&$-0.038\pm0.065\pm0.048 $&$-0.068\pm0.067\pm0.047 
$\\
\hline
$|u_{11}|$&$0.386\pm0.034\pm0.020 $&$0.402\pm0.039\pm0.076 $&$0.355\pm0.045\pm0.023 $&$0.353\pm0.050\pm0.072 $\\
\hline
\hline
Ratio& $q_4$$t_1$\hspace{1.2cm} & $q_4$$t_2$\hspace{1.2cm} & $q_4$$t_3$\hspace{1.2cm} & 
$q_4$$t_4$\hspace{1.2cm} \\
\hline
$\mathrm{Re}(t_{11})$&$0.655\pm0.054\pm0.044 $&$0.842\pm0.068\pm0.043 $&$0.768\pm0.060\pm0.079$&$0.820\pm0.075\pm0.315 $\\
\hline
$\mathrm{Im}(t_{11})$&$0.629\pm0.049\pm0.011 $&$0.614\pm0.066\pm0.052 $&$0.592\pm0.059\pm0.089 $&$0.614\pm0.070\pm0.273 $\\
\hline
$\mathrm{Re}( t_{01} )$&$0.060\pm0.025\pm0.003 $&$0.147\pm0.030\pm0.010 $&$0.144\pm0.029\pm0.018 $&$0.243\pm0.031\pm0.105 $\\
\hline
$\mathrm{Im}( t_{01} )$&$-0.091\pm0.075\pm0.011 $&$-0.065\pm0.067\pm0.018 $&$-0.173\pm0.076\pm0.028 $&$0.042\pm0.078\pm0.382 
$\\
\hline
$\mathrm{Re}( t_{10} )$&$0.025\pm0.027\pm0.009 $&$0.042\pm0.030\pm0.012 $&$0.073\pm0.031\pm0.016 $&$-0.014\pm0.034\pm0.063 
$\\
\hline
$\mathrm{Im}( t_{10} )$&$-0.088\pm0.038\pm0.014 $&$-0.148\pm0.047\pm0.010 $&$-0.131\pm0.041\pm0.017 $&$-0.025\pm0.049\pm0.043 
$\\
\hline
$\mathrm{Re}( t_{1-1} )$&$-0.009\pm0.030\pm0.001 $&$-0.004\pm0.032\pm0.017 $&$-0.010\pm0.030\pm0.008 
$&$-0.009\pm0.030\pm0.046 $\\
\hline
$\mathrm{Im}( t_{1-1} )$&$0.016\pm0.038\pm0.006 $&$-0.024\pm0.051\pm0.026 $&$-0.050\pm0.042\pm0.075 $&$-0.055\pm0.048\pm0.109 
$\\
\hline
$|u_{11}|$&$0.436\pm0.038\pm0.042 $&$0.422\pm0.044\pm0.045 $&$0.389\pm0.042\pm0.070 $&$0.420\pm0.045\pm0.076 $\\
\hline
\end{tabular} \\[2pt]
}
\end{center}
\caption{ \label{ampl-deutr} \small{
Ratios of helicity amplitudes in $Q^2$ and $-t'$ bins for deuteron data. The mean values of $Q^2$ and $-t'$ for all
16 bins are given in Tab.~\ref{meanbin}. The notations $t_{kj}$ and $|u_{11}|$ are used for the amplitude ratios
$T_{kj}/T_{00}$ and $|U_{11}/T_{00}|$, respectively. The first uncertainties are statistical, the second are systematic.
}
}
\end{table*}

\begin{figure*}
  \begin{center}
\vspace{0.6cm}\includegraphics[width=\textwidth]{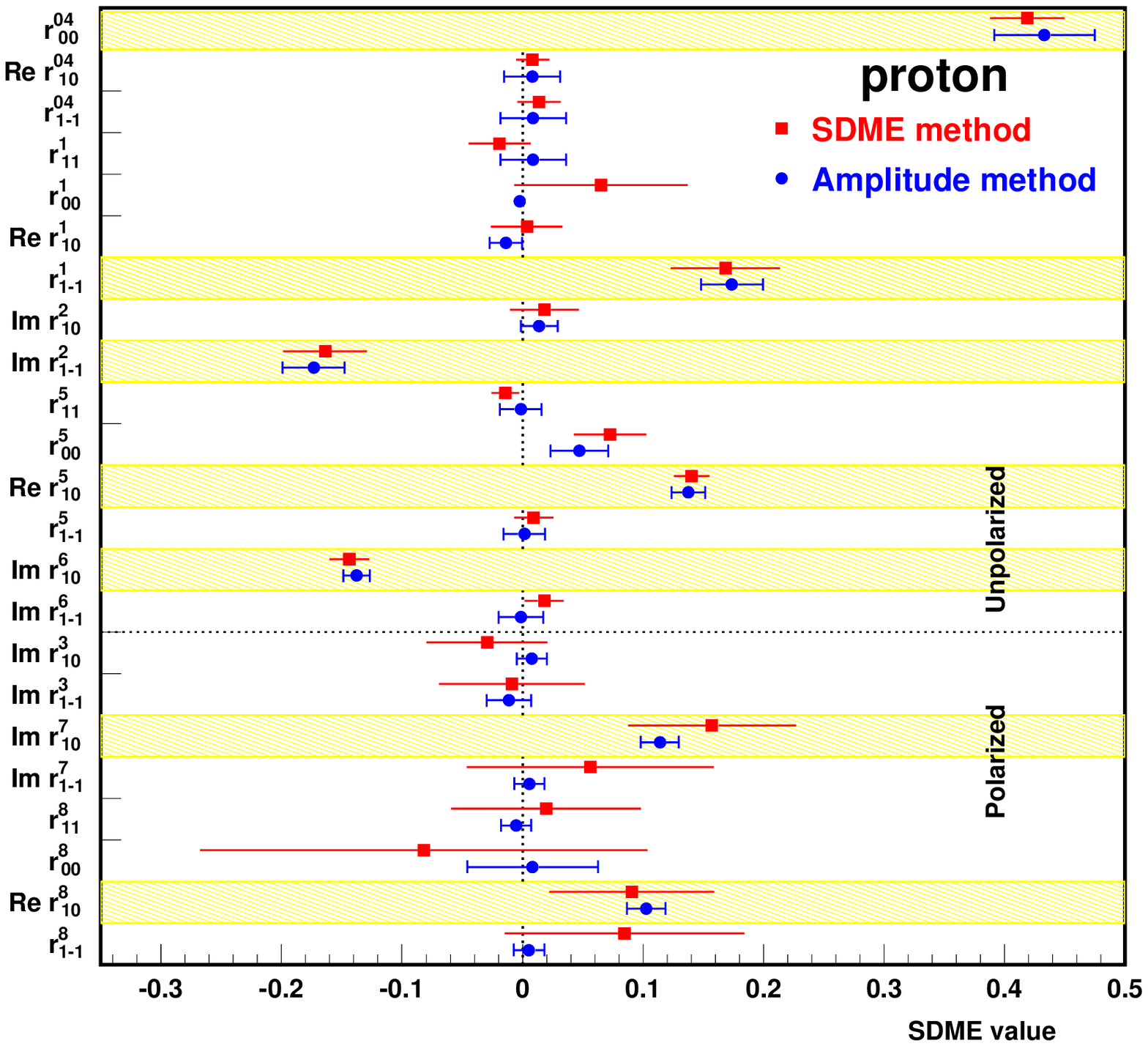}
\caption{Comparison of SDME and amplitude methods. Red squares show the results of the SDME analysis~\cite{DC-24}.  Blue circles (amplitude method)
 are obtained in  the present work from amplitude ratios fitted directly to the three-dimensional angular distribution in every $Q^2$ and $-t^\prime$ bin. The proton data at $\langle Q^2 \rangle = 3$ GeV$^2$, $\langle -t^\prime \rangle = 0.019$ GeV$^2$ are presented. Yellow bands mark those SDMEs that are non-zero in the SCHC approximation. Total uncertainties are depicted.
}
\label{two-meth-q4t1}
  \end{center}
\end{figure*}
\begin{figure*}
  \begin{center}
\vspace{0.6cm} \includegraphics[width=\textwidth]{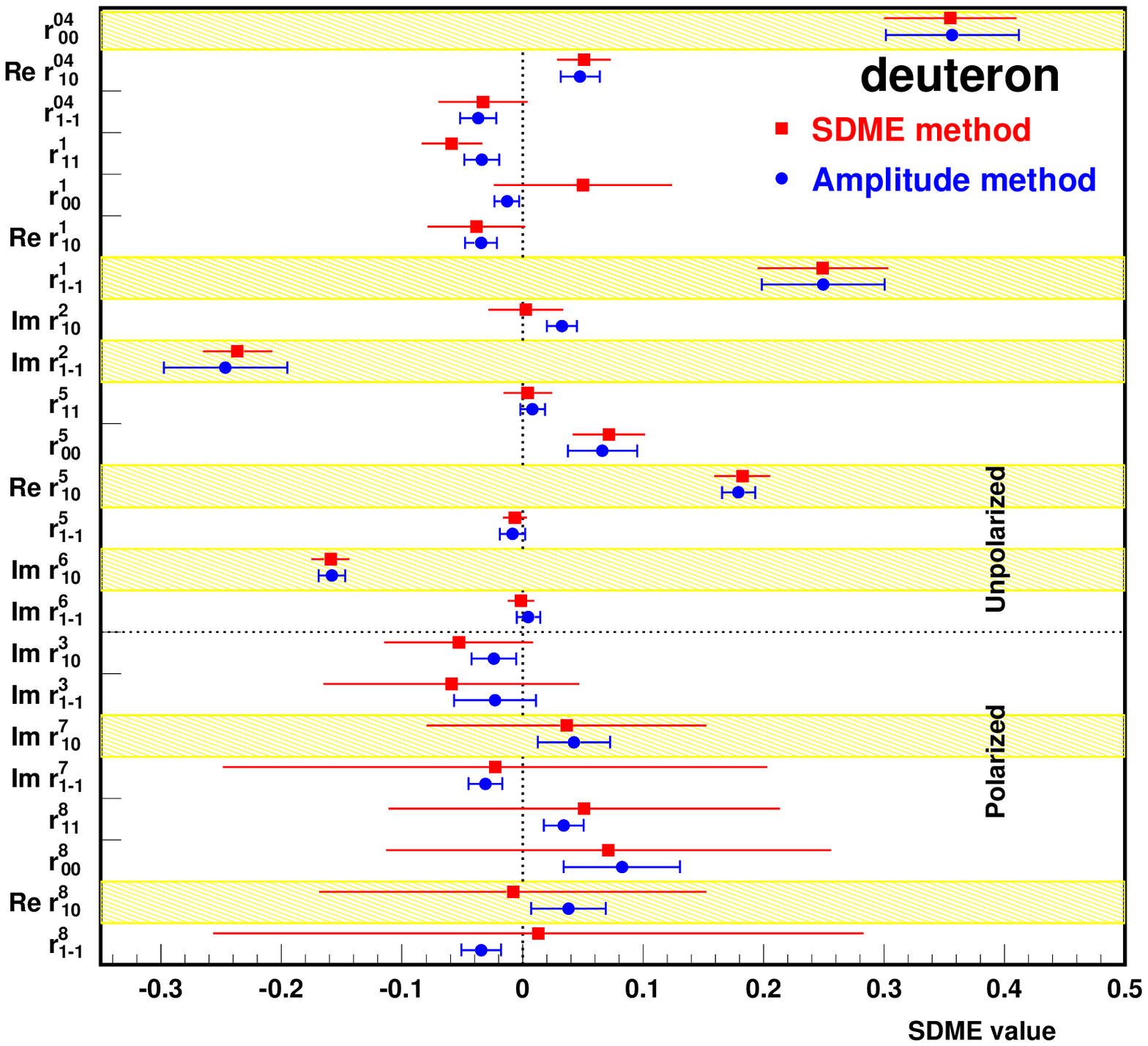}
\caption{
Comparison of SDME and amplitude methods. Red squares show the results of the SDME analysis~\cite{DC-24}.  Blue circles  (amplitude method) are obtained in the present work from amplitude ratios fitted directly to the three-dimensional angular distribution in every $Q^2$ and $-t^\prime$ bin.  The deuteron data at $\langle Q^2 \rangle = 1.19$ GeV$^2$, $\langle -t^\prime \rangle = 0.145$ GeV$^2$  are presented. Yellow bands mark those SDMEs that are non-zero in the SCHC approximation. Total uncertainties are depicted.}
\label{two-meth-q2t3}
  \end{center}
\end{figure*} 
%%%%%%%%%%%%%%%%%%
\subsection{Testing the Extraction Method } 
The self-consistency of the extraction of  amplitude ratios is tested with rhoMC Monte Carlo data. The fit results obtained for the nine free parameters in several ($Q^2$, $-t^\prime$) bins spanning the experimental kinematic range are used to calculate the 23 SDMEs according to Eqs.~(\ref{a1}-\ref{a24}). These SDMEs are then used as input for a Monte Carlo simulation of the angular distribution described in Eqs.~(\ref{eqang1}-\ref{eqang3}). This angular distribution is used to extract the nine free parameters again. The resulting amplitude ratios are found to be consistent with the input amplitude ratios within statistical uncertainties.
%%%%%%%%%%%%%%%%%%%
\subsection{Systematic Uncertainties}
In the extraction of SDMEs in Ref.~\cite{DC-24}, two main contributions to the  total systematic uncertainty of the results were  discussed. One was the uncertainty in the background contribution  to the signal in the region of the exclusive peak in the $\Delta E$ distribution. The background was simulated with the PYTHIA Monte Carlo generator. In the present analysis, the   amplitude ratios are determined with and without background subtraction. The systematic uncertainty due to the background is chosen to be the difference between the amplitude ratios of these two sets. The other main uncertainty arises from the reliance on Monte Carlo simulations to perform the extraction of amplitude ratios from data. There are uncertainties in the parameters of the  description of $\rho^0$ production in the Monte Carlo that propagate through to the amplitude-ratio values presented in this  paper. The parameterization of the total electroproduction cross section in rhoMC is chosen  in the context of a  vector-meson-dominance model~\cite{Bauer}  that incorporates a propagator-type $Q^2$ dependence, and also contains a $Q^2$  dependence of the ratio of the longitudinally-polarized virtual photon and the transversely-polarized virtual photon cross-sections. As the HERMES spectrometer  acceptance depends on $Q^2$,  different input parameters result in slightly different reconstructed isotropic angular distributions. The corresponding systematic uncertainty of the resulting helicity amplitude ratios is obtained by varying these parameters within one standard deviation in the total uncertainty of the parameters given in Refs.~\cite{rho-xsec,ACK,tytgat,ami-phd}. The effect of a possible kinematic dependence of the $t^\prime$ slope of the cross section in the rhoMC generator is found to be negligible in the kinematic region of our experiment.

A third main contribution to the systematic uncertainty arises in the amplitude method as a result of the neglect of the small  NPE nucleon helicity-flip amplitudes and all the UPE amplitudes except $U_{1\frac{1}{2}1\pm\frac{1}{2}}$. As shown in Appendix~\ref{app:app2},  the real and imaginary parts of the deviation $\delta t_{\lambda_V  \lambda_{\gamma}}$ and also $\delta |u_{11}|$  of the obtained amplitude ratios from the true values $t_{\lambda_V
\lambda_{\gamma}}$ and $|u_{11}|$, respectively, can be estimated using the relations
\begin{eqnarray}
\label{syst-amp-meth01}
|\delta \mathrm{Re}(t_{\lambda_V \lambda_{\gamma}})| = \frac{v_T^2}{2M^2} |t_{\lambda_V \lambda_{\gamma}}|,\\
\label{syst-amp-meth02}
|\delta \mathrm{Im}(t_{\lambda_V \lambda_{\gamma}})| = \frac{v_T^2}{2M^2} |t_{\lambda_V \lambda_{\gamma}}|,\\
\delta |u_{11}|=\frac{v_T^2}{8M^2}|u_{11}|.
\label{syst-amp-meth03}
\end{eqnarray}
These equations are used to calculate the systematic uncertainty for the amplitude method described in this paper. 

The three main systematic uncertainties are added in quadrature to give the total systematic uncertainty  presented in Tabs.~\ref{ampl-hydr} and \ref{ampl-deutr}.

The effect of radiative corrections onto the values of $\rho^0$ SDMEs extracted was shown to be negligible~\cite{DC-24}. This is mainly due to the exclusivity cut $\Delta E < 0.6$~GeV that excludes photons radiated with more than 0.6~GeV. The amplitude ratios in this work are extracted using exactly the same data sets and cuts, by a very similar fit comparing the shapes of the 3-dimensional experimental and simulated angular distributions. Hence we conclude that any possible systematic uncertainty of amplitude ratios due to radiative corrections can be safely neglected.

%%%%%%%%%%%%%%%%%%%%%%%%%%%%%%%%%%%%%%%%%%%%%%
\subsection{Calculation of SDMEs from Extracted Helicity Amplitude Ratios}
\label{sec:comp-sdme-am}
In the following, the SDME results obtained from SDME and amplitude methods are compared. The SDMEs are calculated for every ($Q^2$, $-t^\prime$) bin by applying formulas~(\ref{a1}-\ref{a24}) and using the values of the real and imaginary parts of the amplitude ratios given in Tabs.~\ref{ampl-hydr} and
\ref{ampl-deutr}. The results of both methods are found to agree over the full kinematic range of the experiment. The comparison is shown for two representative bins, for the proton data using the bin  $q_4t_1$ ($\langle Q^2  \rangle = 3$ GeV$^2$, $\langle -t^\prime \rangle = 0.019$ GeV$^2$) in Fig.~\ref{two-meth-q4t1} and for deuteron data using the bin $q_2t_3$ ($\langle Q^2 \rangle = 1.19$ GeV$^2$, $\langle -t^\prime \rangle = 0.145$
GeV$^2$) in Fig.~\ref{two-meth-q2t3}.  The bands in the figures mark those SDMEs that may be non-zero in the SCHC approximation,  while the line at zero is drawn for those SDMEs that vanish at $t^\prime=0$.

For the unpolarized SDMEs, the total uncertainties obtained with the amplitude method are in most cases comparable to those obtained with the SDME method. An exception seen in Fig.~\ref{two-meth-q4t1} is $r^1_{00}$ which is proportional to $|t_{01}|^2$ (see Eq.~(\ref{a5})) and hence extremely small at small $|t^\prime|$. For polarized SDMEs, the uncertainties obtained using the SDME method are generally much larger than those obtained using the amplitude method. This difference reflects a major difference in the methods themselves. In the SDME method,  polarized and unpolarized SDMEs are fitted  as independent free parameters.  In this case, the error bars of the polarized SDMEs are inflated due to the factors $P_B=0.47$ and
$\sqrt{1-\epsilon}\approx 0.45$ (see Eqs.~(\ref{eqang1}-\ref{eqang3})). However, Eqs.~(\ref{a1}-\ref{a24}) show that the polarized SDMEs depend on amplitude ratios constrained by both polarized and unpolarized data. Hence the polarized SDMEs calculated from the amplitude ratios have uncertainties comparable to those of the unpolarized SDMEs. Nevertheless, the polarized data serves the important function of constraining the sign of the imaginary parts of amplitude ratios.

%%%%%%%%%%%%%%%%%%%%%%%%%%%%%%%%%%%%%%%%%%%%%%%%%%%%%%%%%%%%%%%%%%%%%%%%%%%%%%%%%%%%%%%%%%
\section{Kinematic Dependences of  Amplitude Ratios}
%%%%%%%%%%%%%%%%%%%%%%%%%%%%%%%%
\subsection{Predictions for Asymptotic Behavior}
At small $|t^{\prime}|$, the behavior of NPE amplitudes $T_{\lambda_{V} \lambda_{\gamma}}$ can be described~\cite{Diehl} by
\begin{eqnarray}  \label{t-asympt}
T_{\lambda_{V} \lambda_{\gamma}} \propto (-t^\prime)^{|\lambda_{V}- \lambda_{\gamma}|/2},
\end{eqnarray}
which reflects angular momentum conservation in the process in Eq.~(\ref{react01}). Equation (\ref{t-asympt}) shows that  $T_{00}$ behaves as a constant at small $|t^{\prime}|$. Hence each ratio $t_{\lambda_{V} \lambda_{\gamma}}=T_{\lambda_{V} \lambda_{\gamma}}/T_{00}$ has the same asymptotic  behavior at $t^{\prime} \rightarrow 0$ as the amplitude $T_{\lambda_{V} \lambda_{\gamma}}$.

The asymptotic behaviour of the amplitudes of vec\-tor-meson electroproduction at large $Q^2$ and small $x_B$ was considered in Refs.~\cite{IK,KNZ} in the framework of pQCD, using the approximation of leading logarithms in which the logarithms $\ln(1/x_B)$ and $\ln(Q^2)$ are large, and the amplitudes are considered to be nearly imaginary. It was predicted~\cite{IK,KNZ,RK} that the amplitude ratios should have the following behavior at large $Q^2$ and small
$x_B$:
\begin{eqnarray}
\label{Q-asympt11}
t_{11} & \equiv & T_{11}/T_{00} \propto \frac{M_V}{Q}, \\
\label{Q-asympt01}
t_{01} &\equiv& T_{01}/T_{00} \propto \frac{\sqrt{-t^{\prime}}}{Q}, \\
\label{Q-asympt10}
t_{10} &\equiv& T_{10}/T_{00} \propto \frac{M_V \sqrt{-t^{\prime}}}{Q^2+M^2_V},
\end{eqnarray}
\begin{eqnarray}
\label{Q-asympt1-1}
t_{1-1} &\equiv& T_{1-1}/T_{00} \propto \frac{ -t^{\prime}M_V }{Q}
\Bigg[ \frac{C_1}{Q^2+M^2_V}+\frac{C_2}{\mu^2}\Bigg].\;\;
\end{eqnarray}
 The  functions $C_1$ and $C_2$ in Eq.~(\ref{Q-asympt1-1}) are dimensionless and the parameter $\mu$ is  between $0.7$ GeV and  $1.0$ GeV. It  was noted in Refs.~\cite{IK,KNZ,RK} that  $C_1$ is essentially  a constant while $C_2$ is a ratio  whose numerator contains the gluon transversity distribution~\cite{Kivel} $G_{T}(x_B,\mu^2)$ at the scale $\mu$  and the denominator contains the unpolarized gluon density $G(x_B,Q_V^2)$ measured at the conjectured~\cite{KNZ} scale $Q_V$ for vector-meson electroproduction in the non-asymptotic region, with
\begin{equation}
\label{eq:q_v}
Q_V^2=(Q^2+M_V^2)/4 \;.
\end{equation}

As explained in Sec.~\ref{sec:herm-twist}, the twist-decomposition parameter $\beta=M_V/Q$ is not really small in the kinematic region of the HERMES experiment. Therefore, HERMES results on amplitude ratios are expected to be not always in agreement with the pQCD predictions given by Eqs.~(\ref{Q-asympt11}-\ref{Q-asympt1-1}). In the case of a disagreement, we use a parameterization that does not contradict general principles and describes the amplitude ratios with reasonable $\chi^2$ per degree of freedom. Any NPE amplitude ratio has to obey Eq.~(\ref{t-asympt}). On very general principles, at finite $Q$ the amplitudes $T_{11}$, $T_{01}$, and $T_{1-1}$ are even functions in $Q$. In contrast, the amplitudes $T_{00}$ and $T_{10}$ are odd functions in $Q$ due to the extra factor $1/Q$ in Eq.~(\ref{longphot}). Therefore, the ratios $t_{11}$, $t_{01}$, and $t_{1-1}$ ($t_{10}$) are  odd (even) functions in $Q$. The fit functions will be chosen in agreement with this property of amplitude ratios. Whenever the chosen fit function does not agree with a pQCD prediction, the resulting curve is shown not by a solid but a dash-dotted line.

%%%%%%%%%%%%%%%%%%%%%%%%%%%%%%%%%%
\subsection{ Kinematic Dependence of $T_{11}/T_{00}$}
The amplitudes $T_{00}$ and $T_{11}$ describe the diagonal helicity transitions  $\gamma_L^* \rightarrow \rho^0_L$ and $\gamma_T^* \rightarrow \rho^0_T$ respectively, and are  the largest amplitudes of $\rho^0$ meson  production in the HERMES kinematic region. The $Q^2$ dependence of the extracted  amplitude ratio $t_{11}$ in four $-t^{\prime}$ bins is presented in Fig.~\ref{fig-t11-16} for the proton (left) and the deuteron (right). The points correspond to the amplitude  ratios extracted  from the data in 16 bins and given in Tabs.~\ref{ampl-hydr} and ~\ref{ampl-deutr}. In order to test the predictions of  Refs.~\cite{IK,KNZ},  the $Q^2$ dependence of the ratio $\mathrm{Re}(t_{11})$ was fitted in every $-t^\prime$ bin  with the function
\begin{eqnarray} \label{fit-Q-Ret11}
 \mathrm{Re}(t_{11}) = \frac{a}{Q}.
\end{eqnarray}
For the four $-t^{\prime}$ bins, the values of the parameter $a$ are found to be consistent with each other within experimental accuracy. Therefore, data in all $-t^\prime$ bins are fitted simultaneously to improve the statistical accuracy in determining the value of $a$. The results of  the fit for proton and deuteron data are presented in Tab.~\ref{t11} and shown in  Fig.~\ref{fig-t11-16} for  proton (left) and  deuteron (right).  The  figure shows that the $Q^2$ dependence of $\mathrm{Re}(t_{11})$ is well described by the parameterization given by Eq.~(\ref{fit-Q-Ret11}).

\begin{figure*}[hbtc!] 
\vspace{-0.7cm}
\hspace{-0.5cm}
\includegraphics[width=0.52\textwidth]{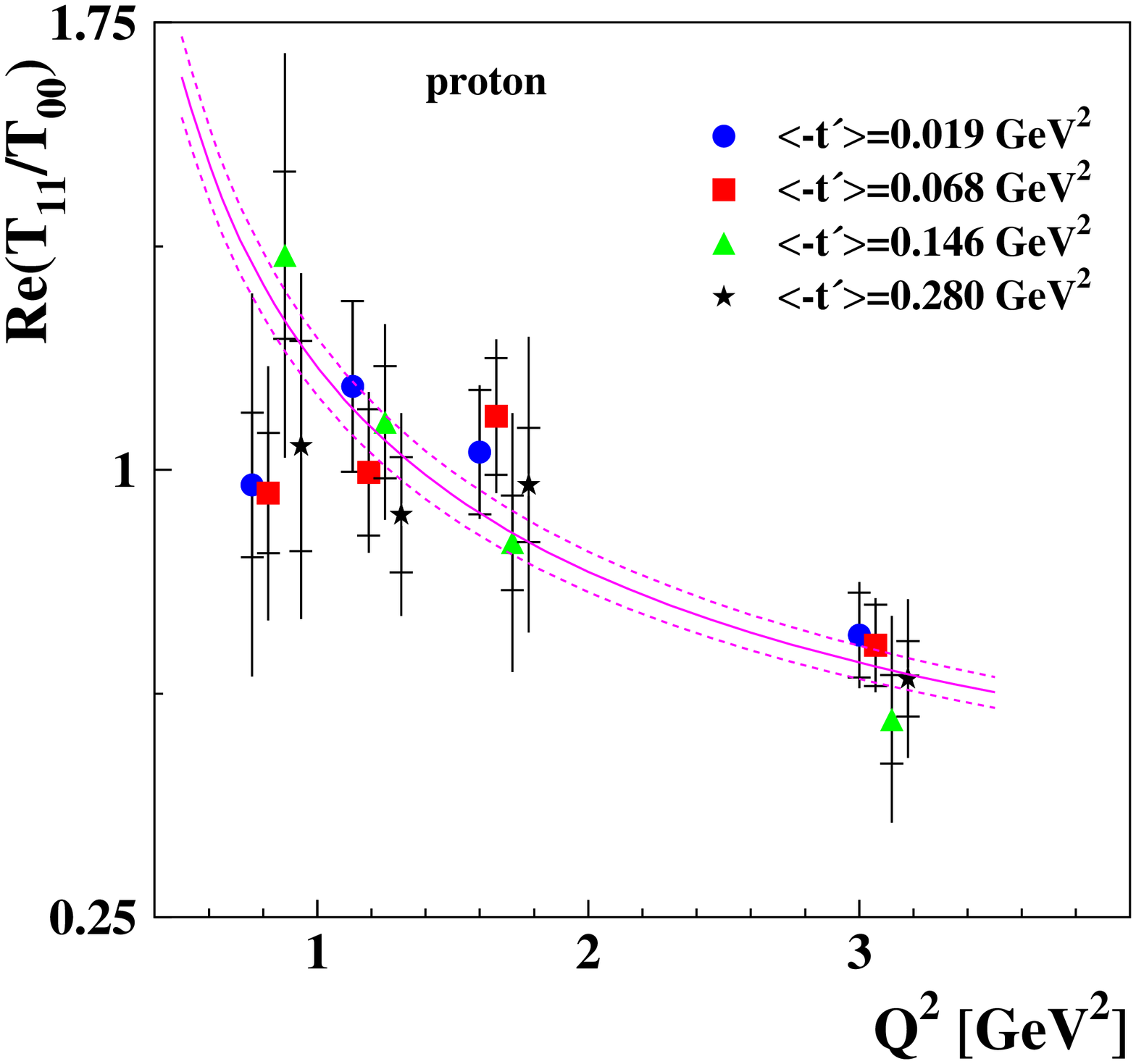}
\includegraphics[width=0.52\textwidth]{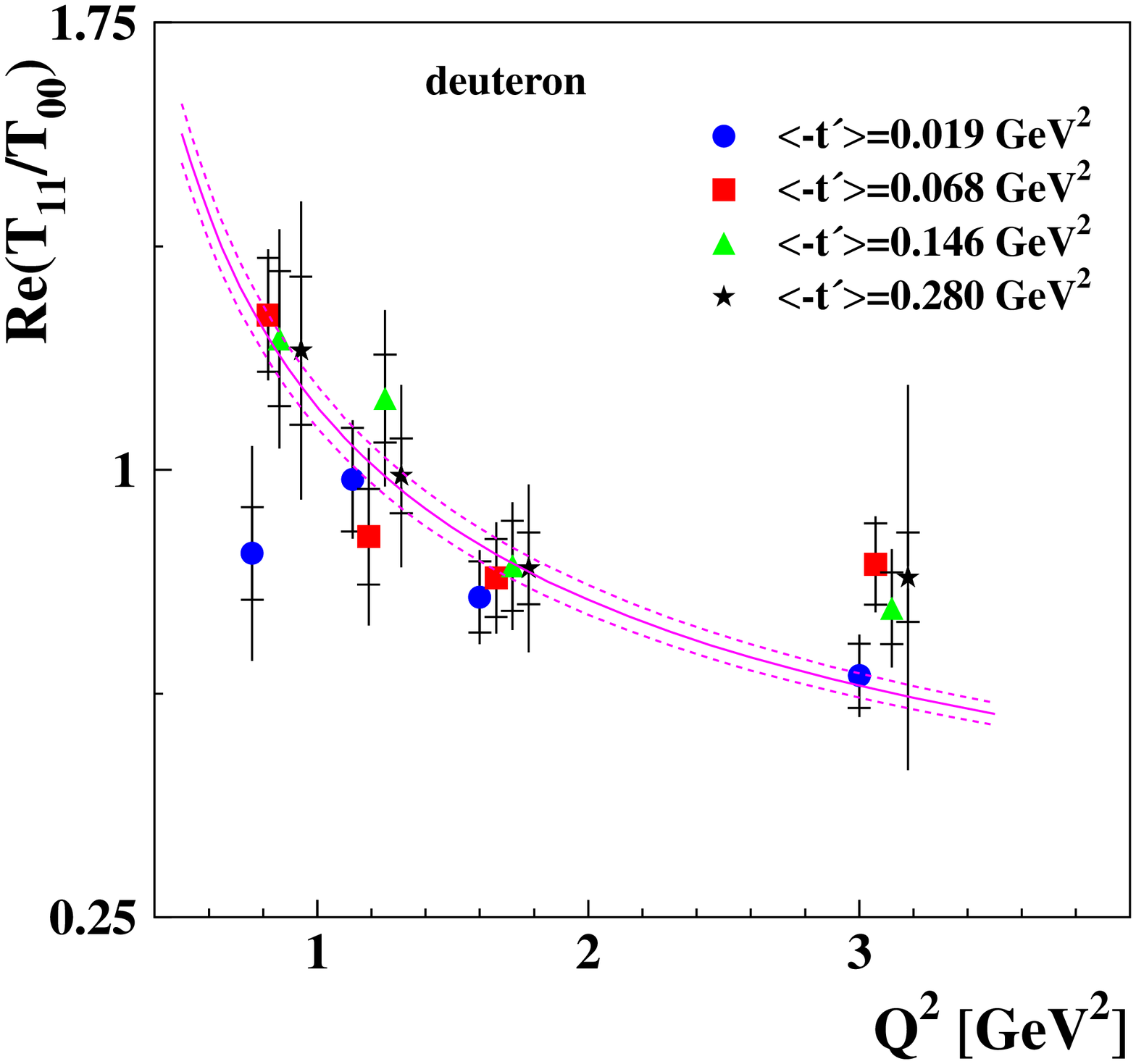}
  \caption{ 
The $Q^2$ dependence of $\mathrm{Re}( T_{11}/T_{00} )$  for proton (left panel)  and deuteron (right panel) data, showing the result for the 16 ($Q^2$, $-t^\prime$) bins. Inner error bars show the statistical uncertainty and the total error bars represent statistical and systematic uncertainties added in quadrature. The parameterization of the curves is given by Eq.~(\ref{fit-Q-Ret11}) and their parameters  are given in Tab.~\ref{t11}. Central lines are calculated with the fitted  values of parameters, while the dashed lines correspond to one standard deviation of the curve parameter. Except for the second $-t^\prime$ bin, the data points are shifted for better visibility.
}
\label{fig-t11-16}
\end{figure*}

\begin{figure*}[hbtc!]
\hspace{-0.3cm}
\includegraphics[width=0.52\textwidth]{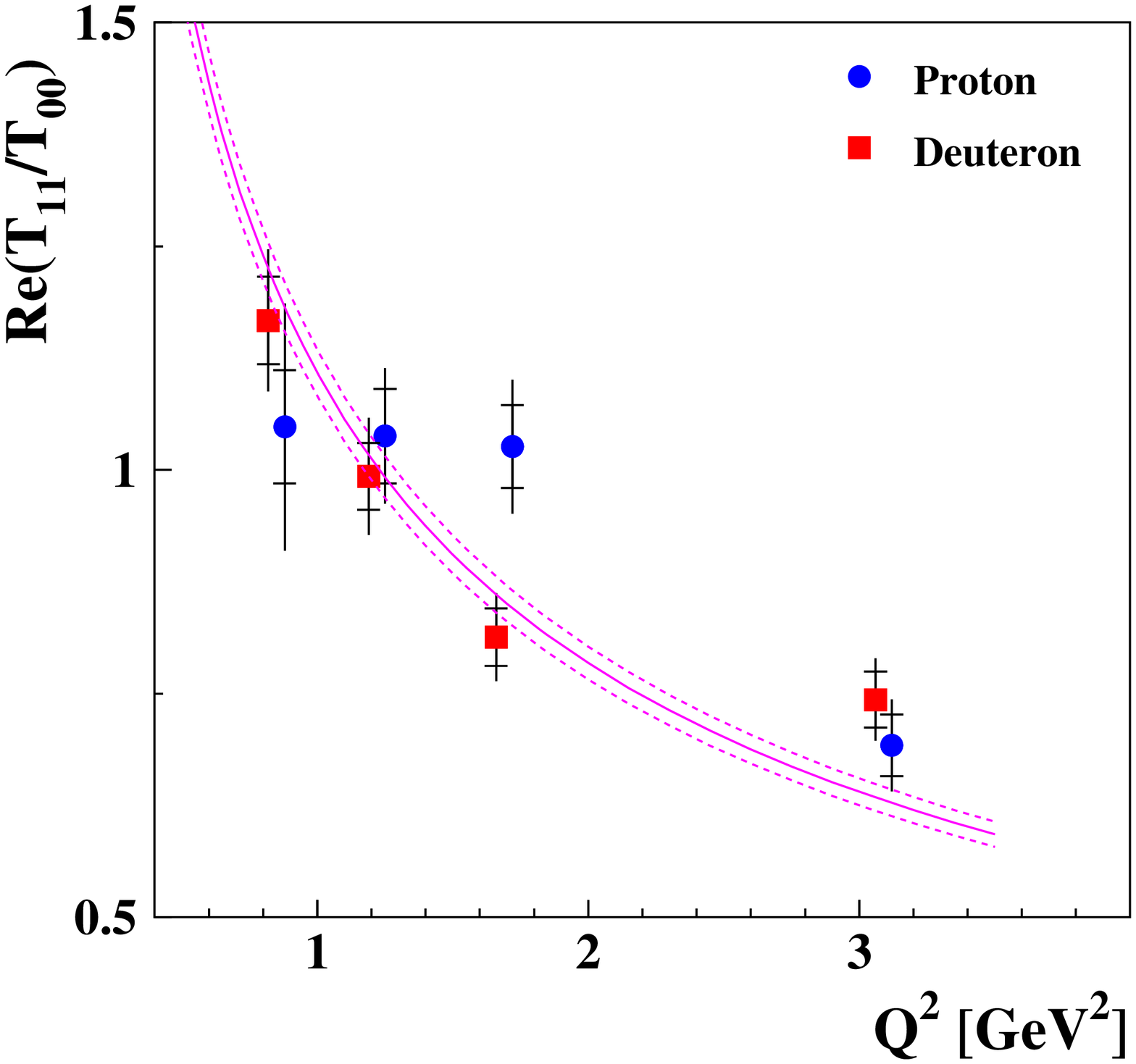}
\includegraphics[width=0.52\textwidth]{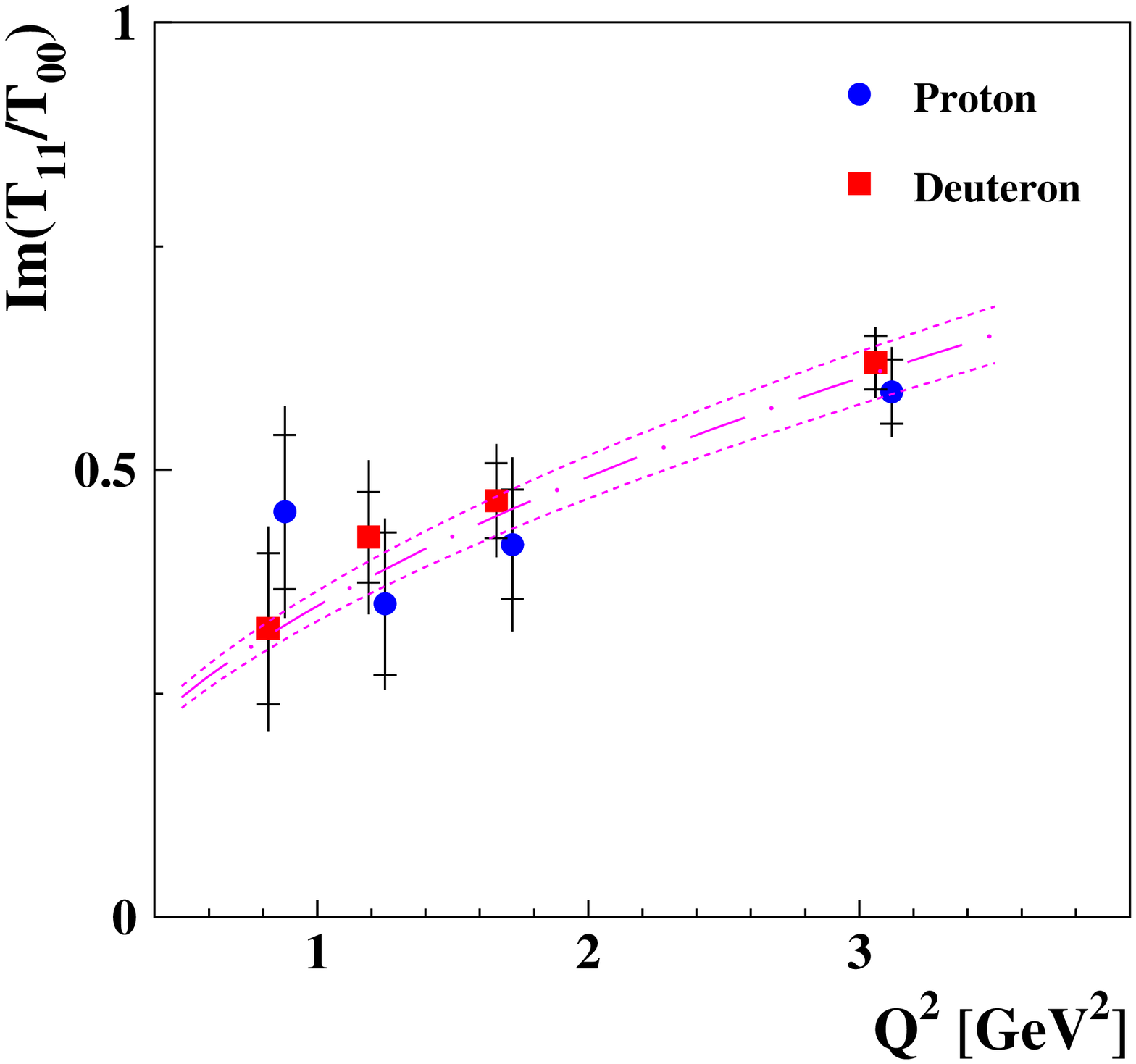}
  \caption{
The $Q^2$ dependence of $\mathrm{Re} (T_{11}/T_{00} )$ (left panel) and $ \mathrm{Im}( T_{11}/T_{00})$ (right panel) for proton and deuteron data. Points show the amplitude ratios from Tabs.~\ref{ampl-hydr} and \ref{ampl-deutr} after averaging over four $-t^\prime$ bins using Eqs.~(\ref{avermean}) and (\ref{avererror}). The parameterization of the curves obtained from combined proton and deuteron data is given by Eqs.~(\ref{fit-Q-Ret11}) and (\ref{fit-Q-Imt11}) while the parameters are given in Tab.~\ref{t11}. The meaning of the error bars and the explanation of the curves are the same as for Fig.~\ref{fig-t11-16}. Here and hereafter proton data points are slightly displaced to the right for better visibility. As explained in the text, curves are shown as solid (dash-dotted) lines if the parameterization is based (not based) on pQCD predictions.
}
\label{fig-t11}
\end{figure*}

\begin{table*}[hbtc!]
 \renewcommand{\arraystretch}{1.2}
 \vspace{-0.3cm}
\begin{center}
\begin{tabular}{|c|c|c|c|c|}
    \hline
target&ratio&$a$, GeV&$\delta a$, GeV &$\chi^2/N_{df}$\\
    \hline
proton &$\mathrm {Re}(T_{11}/T_{00})$&1.173& $\pm 0.048$ &0.63\\
    \hline
deuteron &$\mathrm {Re}(T_{11}/T_{00})$&1.106& $\pm 0.035$ &1.04\\
    \hline
proton+deuteron &$\mathrm {Re}(T_{11}/T_{00})$&1.109& $\pm 0.026$ &0.78\\
    \hline
    \hline
target & ratio & $b$, GeV$^{-1}$& $\delta b$, GeV$^{-1}$ &$\chi^2/N_{df}$\\
    \hline	

proton &$\mathrm {Im}(T_{11}/T_{00})$&0.340&$\pm 0.025$ &0.54\\
    \hline
deuteron &$\mathrm {Im}(T_{11}/T_{00})$&0.359&$\pm 0.020$ &0.71\\
    \hline
proton+deuteron &$\mathrm {Im}(T_{11}/T_{00})$&0.348&$\pm 0.017$ &0.75\\
    \hline
\end{tabular}\\
\end{center}
\caption{ \label{t11}
\small{
The $Q^2$ dependence of $\mathrm {Re}(T_{11}/T_{00})$ and $\mathrm {Im}(T_{11}/T_{00})$
for proton, deuteron and combined data sets parameterized as given by Eqs.~(\ref{fit-Q-Ret11}) and (\ref{fit-Q-Imt11}).
The values of parameters with their total uncertainties are presented. 
The last column shows the  value of  $\chi^2$  per degree of freedom.}}
\end{table*}

The dependences of the other amplitude ratios, which are described in the following, are examined in a similar manner as was done for $\mathrm{Re}(t_{11})$. In every case, it was checked that the respective four datasets containing 4 points each can be combined into one dataset containing 16 points. When proton and deuteron results are consistent, the data are combined and 32 points are used in  the fit. All the fits to the $Q^2$ or $t^\prime$ dependences use 16 data points for the proton or deuteron separately and 32 data points for the combined data sets.  Only for the sake of a more clear  representation of the kinematic dependence of amplitude ratios in  one variable $Q^2$ (or $t^\prime$), we average the values  of the amplitude ratios $\mathcal{T}=t_{\lambda_V\lambda_{\gamma}}$ or $|u_{11}|$ over four $-t^\prime$ (or $Q^2$) bins at the same value  of $Q^2$ (or $t^\prime$)  using the standard relations for the mean value $\langle \mathcal{T} \rangle $ and squared total uncertainty $(\delta \mathcal{T} )^2$: 
\begin{eqnarray}
\label{avermean}
\langle \mathcal{T} \rangle &=& \frac{\sum_j  \mathcal{T}_j/(\sigma_j)^2}{\sum_m 1/(\sigma_m)^2}\;, \\
(\delta \mathcal{T} )^2 &=& \frac{1}{\sum_m 1/(\sigma_m)^2}\;.
\label{avererror}
\end{eqnarray}
Here, $\sigma_j$ denotes the total uncertainty of $\mathcal{T}_j$ in the $j^{th}$ bin and $\delta \mathcal{T}$ is the total uncertainty of the averaged ratio $\mathcal{T}$. The same formula~(\ref{avererror}) is applied for the calculation of the statistical uncertainty $\delta \mathcal{T}_{stat}$ where $\sigma_m$ is the statistical uncertainty of $\mathcal{T}_m$.

In the left  panel of Fig.~\ref{fig-t11}, the average values of the ratio $\mathrm{Re}(t_{11})$ in each of the $Q^2$ bins for both the proton and deuteron targets are shown. As the results for the two targets are  found to be compatible, a fit to the combined proton and  deuteron data is performed and the result extracted for $a$ in Eq.~(\ref{fit-Q-Ret11}) is also presented  in Tab.~\ref{t11}. Reasonable values of  $\chi^2$ per degree of freedom,  $\chi^2/N_{df} \approx 1$, are obtained. A comparison of  the average  values of ratio $\mathrm{Re}(t_{11})$ across $-t^\prime$ bins and the curve calculated with  Eq.~(\ref{fit-Q-Ret11}) using the value of the parameter $a$ obtained from the combined proton and deuteron data is shown in the left panel of  Fig.~\ref{fig-t11}. The $Q^2$ dependence of $\mathrm{Re}(t_{11})$ is found to be in a good agreement with the asymptotic  behavior  expected from pQCD.

As shown in the right panel of Fig.~\ref{fig-t11}, the amplitude ratio $\mathrm{Im}(t_{11})$ rises with $Q^2$. This dependence of $\mathrm{Im}(t_{11})$ disagrees with the prediction~\cite{IK,KNZ} given by Eq.~(\ref{Q-asympt11}), whereas the dependence of the real part of the same ratio $t_{11}$ agrees with predictions based on the same formula. A fit of the function in Eq.~(\ref{fit-Q-Ret11}) to the experimental data does not provide reasonable values of $\chi^2$ per degree of freedom: $\chi^2/N_{df} = 2.46$  for the proton data and $\chi^2/N_{df} = 4.25$  for the deuteron data. Instead, a fit function of the type
\begin{eqnarray}  
\mathrm{Im}(t_{11}) = bQ\;
\label{fit-Q-Imt11}
\end{eqnarray}
gives reasonable results as can be seen from Tab.~\ref{t11}, where fit results for the proton, deuteron and combined datasets are presented.  The right panel of Fig.~\ref{fig-t11} shows the curve calculated with the parameter $b$ obtained for the combined proton and deuteron data   compared with the  $-t^\prime$ bin  averaged ratio $\mathrm{Im}(t_{11})$ for both  proton and deuteron data. The disagreement with the prediction  of Eq.~(\ref{Q-asympt11}) may be due to the  fact  that the $Q^2$ range of the HERMES experiment is not  in the asymptotic region.   This is in agreement with the conjecture of Ref.~\cite{KNZ} that the hard scale for vector-meson electroproduction is $Q_V$ given by Eq.~(\ref{eq:q_v}) which is smaller than $Q$. Another possible explanation can be found in  the discussion  later in this  paper of potential final-state interactions between the struck parton and the target remnant.

The phase difference $\delta_{11}$ between the helicity amplitudes $T_{11}$ and $T_{00}$ is given by the relation
\begin{eqnarray}
\tan \delta_{11}= \mathrm{Im}(t_{11})/\mathrm{Re}(t_{11}) \,.
\label{delta-11-dep}
\end{eqnarray}
It increases with $Q^2$ as shown in the left panel of Fig.~\ref{fig-delta11}. Also shown is a fit to the $Q^2$ dependence of $\delta_{11}$ calculated with the functional form tan $\delta_{11} = bQ^2/a$ deduced from Eqs.~(\ref{fit-Q-Ret11}) and (\ref{fit-Q-Imt11}). The result of the fit is $\delta_{11} = (31.5 \pm 1.4) $ degrees at $\langle Q^2 \rangle = 1.95$ GeV$^2$, obtained with the parameters $a$ and $b$ given in Tab.~\ref{t11} for the combined proton and deuteron data.  It is consistent within one standard deviation  with the published  result~\cite{DC-24} obtained with the SDME method using the same proton and deuteron data as in the present analysis.  A large phase difference $\delta_{11} \approx 20$ degrees was measured by the H1 collaboration for exclusive $\rho^0$ and $\phi$ meson electroproduction at 2.5~GeV$^2 <Q^2 < 60$~GeV$^2$~\cite{H1-amp}. A large value of  $\delta_{11} =(33.0 \pm 7.4)$ degrees was measured also in $\phi$ meson production at HERMES~\cite{phi-sdmes}.  As two-gluon exchange dominates in $\phi$ meson production at HERMES kinematics, the measured large value of $\delta_{11}$ for $\rho^0$ cannot be attributed solely to  the quark-antiquark exchange essential for $\rho^0$ meson production.  This value of $\delta_{11}$  is in  clear disagreement with a calculation ~\cite{GK} performed using a GPD-based approach in pQCD that predicts a very small value for $\delta_{11}$. At present, there exists no model capable of explaining the value and $Q^2$ dependence of $\delta_{11}$.

In  existing  pQCD calculations for exclusive vector-meson electroproduction, only two-gluon and/or quark-anti\-quark exchange have been taken  into account. It can be argued~\cite{Brodsky} that even in inclusive deep-inelastic lepton-nucleon scattering, the final-state interaction  of the  struck quark alters the cross section due to  multi-gluon exchanges with the target remnant.
 This is at variance with the traditional understanding obtained from inclusive DIS. It is shown~\cite{Brodsky} that a summation of multi-gluon-exchange amplitudes results in an ei\-ko\-nal-like correction. For vector-meson production, if rescattering on the nucleon occured also for the quark-antiquark pair that transforms to the final vector meson, the eikonal-like correction might be responsible for the measured large  phases observed in the present work.

\begin{figure*}[hbtc!]
\vspace{-0.7cm}
\hspace{-0.5cm}
\includegraphics[width=0.53\textwidth]{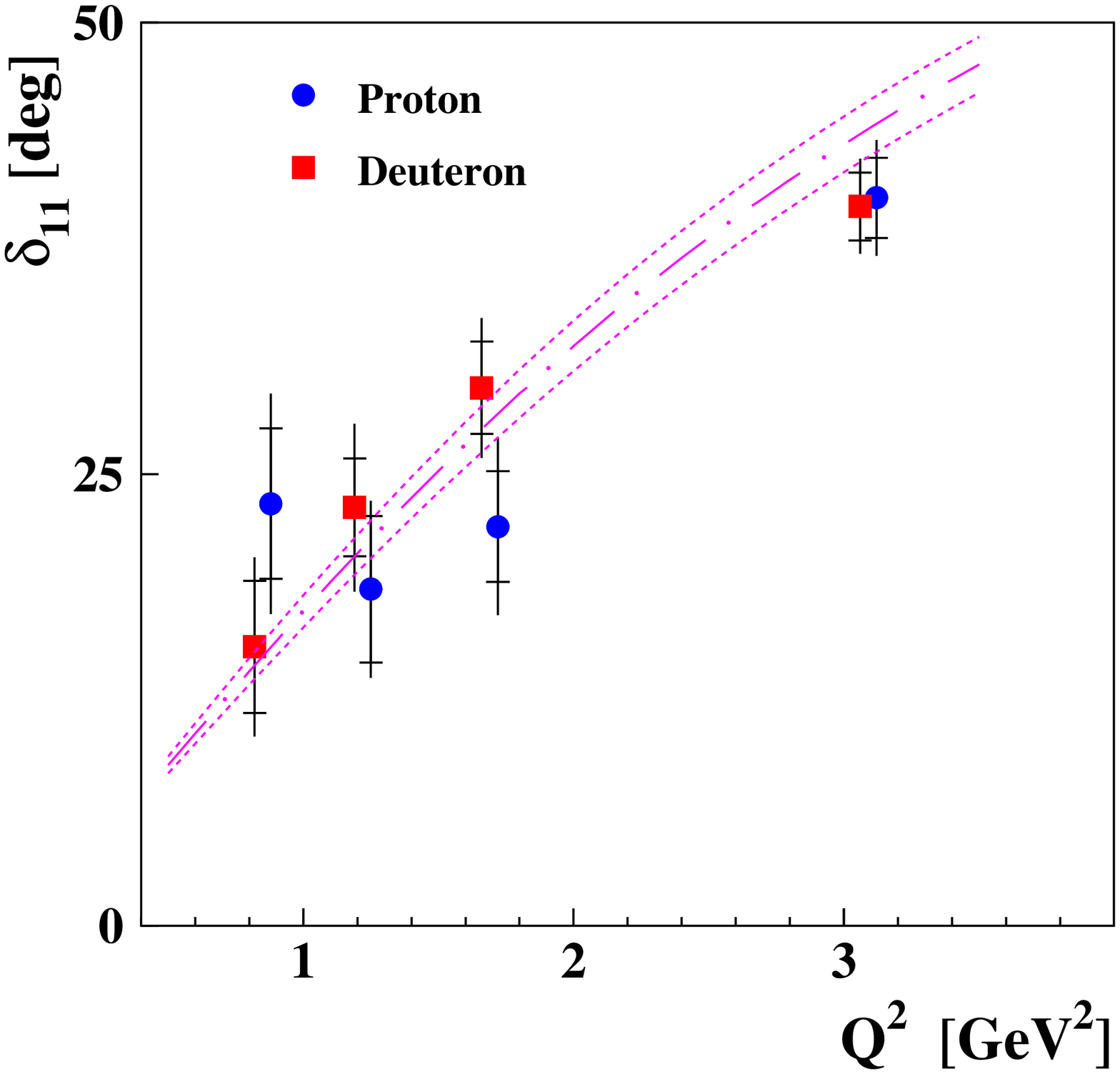}
\includegraphics[width=0.53\textwidth]{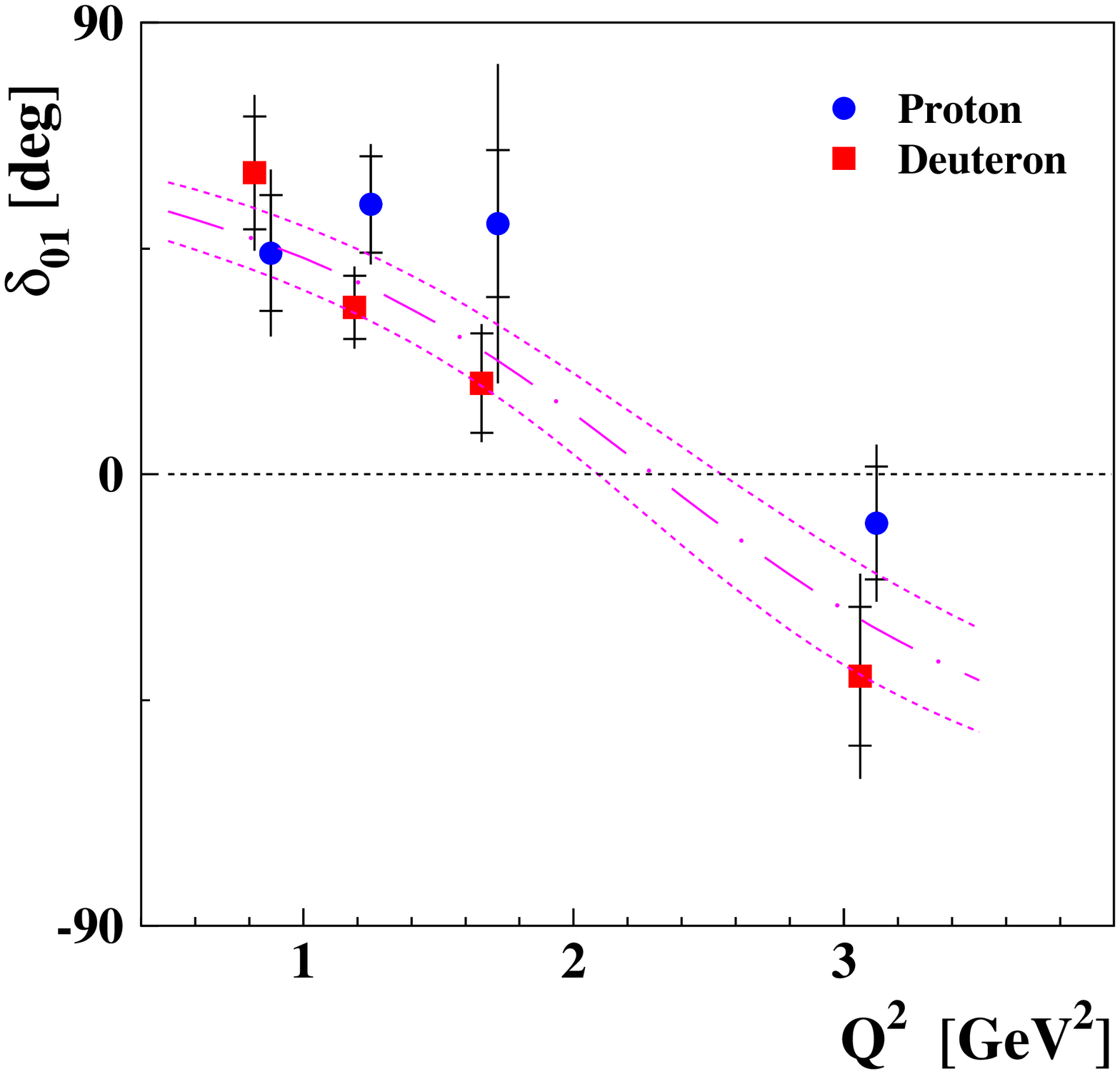}
\caption{
The $Q^2$ dependence of  the  phase difference $\delta_{11}$ (left panel)  and $\delta_{01}$ (right panel, see Sec.~\ref{sec:kin-dep-t01}) between the amplitudes $T_{11}$ and $T_{01}$, respectively, and $T_{00}$  obtained for proton and deuteron data. Points show the phase differences $\delta_{11}$ and $\delta_{01}$ calculated from ratios of amplitudes  given in  Tabs.~\ref{ampl-hydr}  and \ref{ampl-deutr} after averaging over $-t^\prime$ bins. Inner error bars show the statistical uncertainty and the outer ones show the statistical and systematic uncertainties added in quadrature. The fitted parameterization is given by Eqs.~(\ref{delta-11-dep}) and (\ref{phase-Imt01}) respectively for $\delta_{11}$ and $\delta_{01}$. The parameters of the curves are given in Tabs.~\ref{t11} and \ref{t01} for combined proton and deuteron data.  The central lines are calculated with the fitted  values of the parameters, while the dashed lines correspond to one standard deviation  in the uncertainty of the curve parameter.
}
\label{fig-delta11}
\end{figure*}

\begin{figure*}[hbtc!]
\hspace{-0.3cm}
\includegraphics[width=0.53\textwidth]{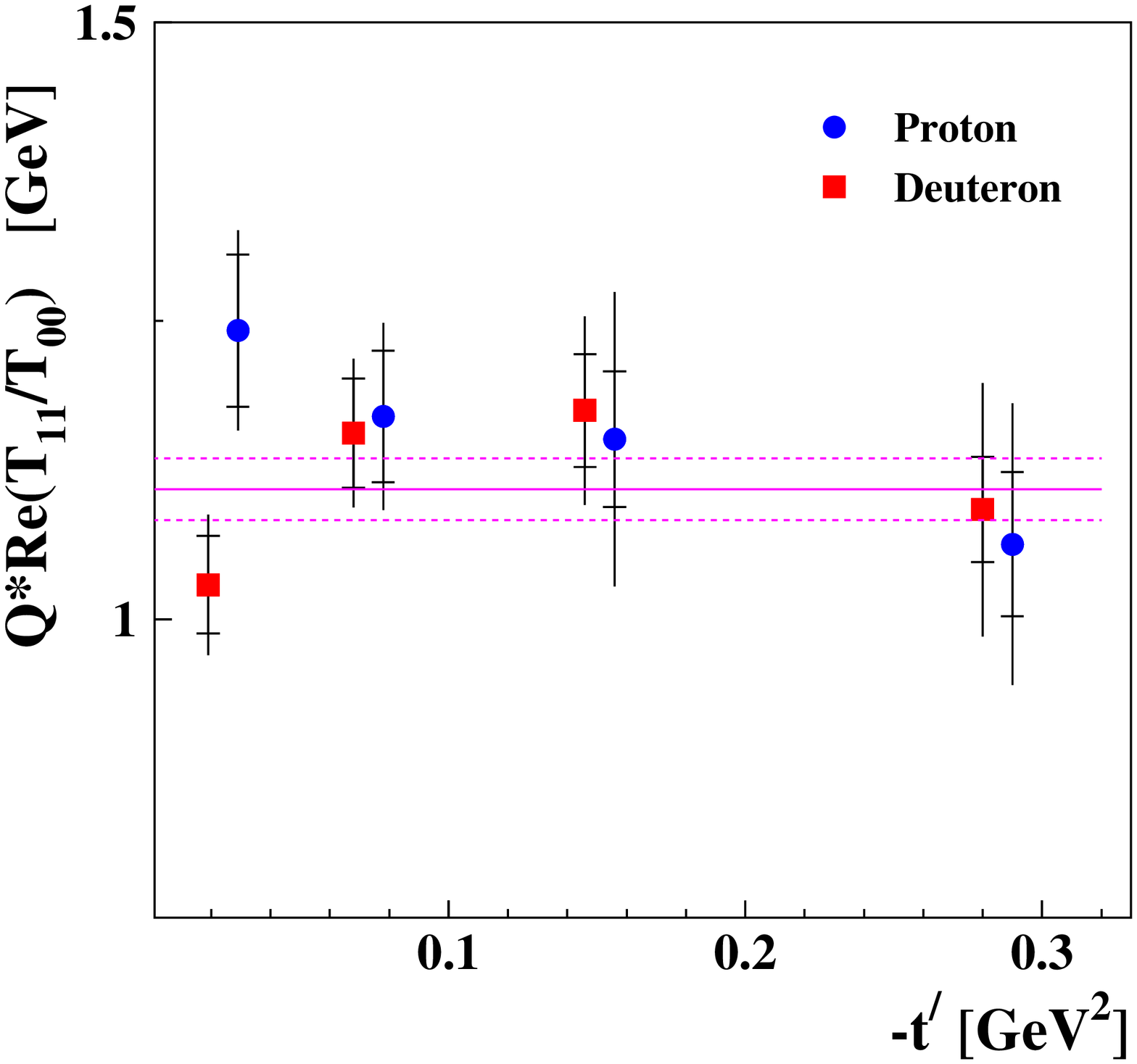}
\includegraphics[width=0.53\textwidth]{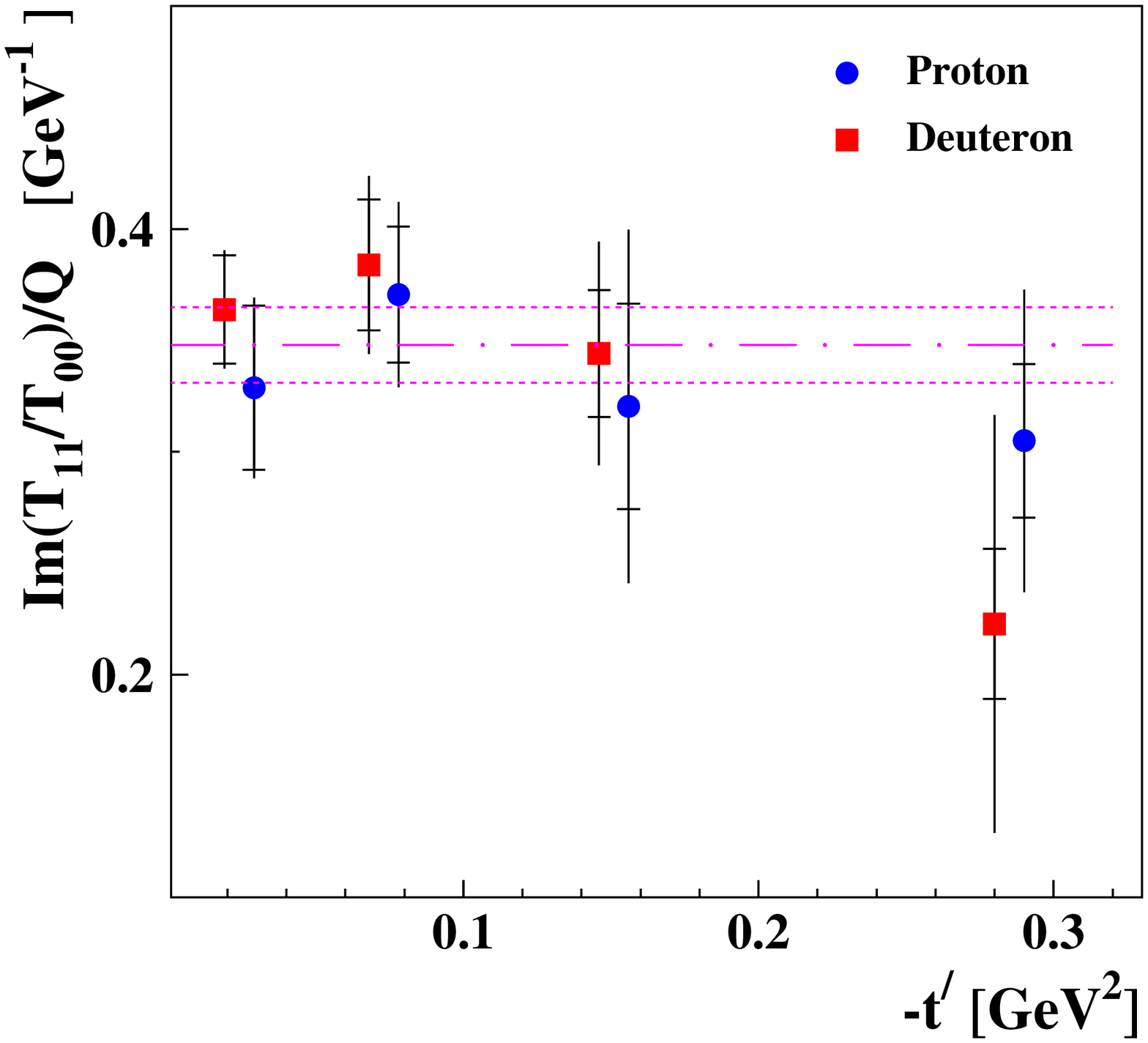}
 \caption{
The $t^\prime$ dependence of $Q\cdot \mathrm{Re}(T_{11}/T_{00})$ (left panel)  and $\mathrm{Im}(T_{11}/T_{00})/Q $ (right panel) for proton and deuteron data. Points show the amplitude ratios from Tabs.~\ref{ampl-hydr} and \ref{ampl-deutr} after averaging over four $Q^2$ bins using Eqs.~(\ref{avermean}) and (\ref{avererror}). The straight lines in the left and right panel show the value of $a$ and $b$, respectively, from Eqs.~(\ref{fit-Q-Ret11}) and (\ref{fit-Q-Imt11}) while the parameters $a$  and $b$ are given in   Tab.~\ref{t11}. The meaning of the error bars and the explanation of the curves are the same as for Fig.~\ref{fig-t11-16}.}
\label{fig-t11-tpr}
\end{figure*}

\begin{figure*}[hbtc!]
\vspace{-0.7cm}
\hspace{-0.5cm}
\includegraphics[width=0.53\textwidth]{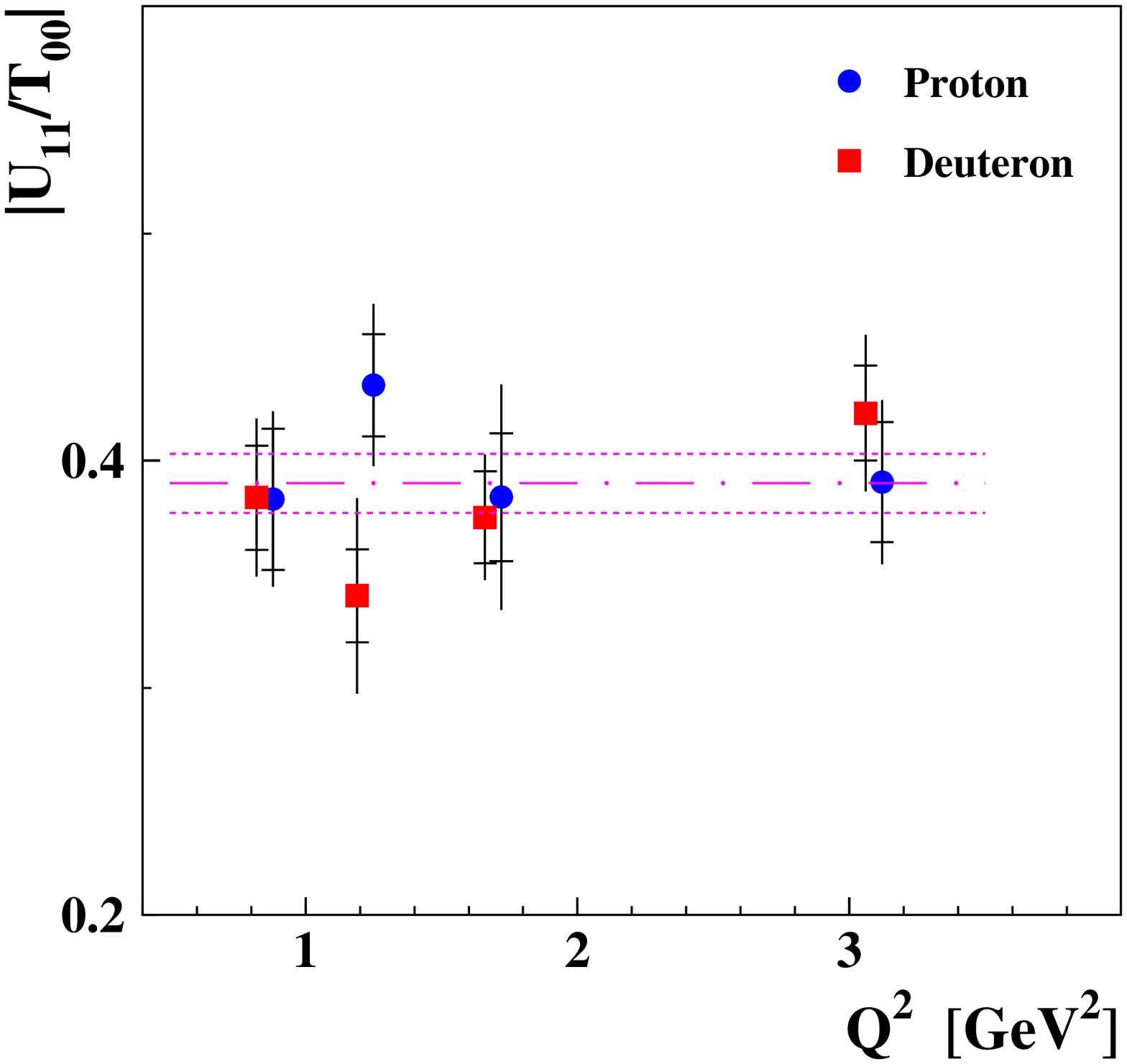}
\includegraphics[width=0.53\textwidth]{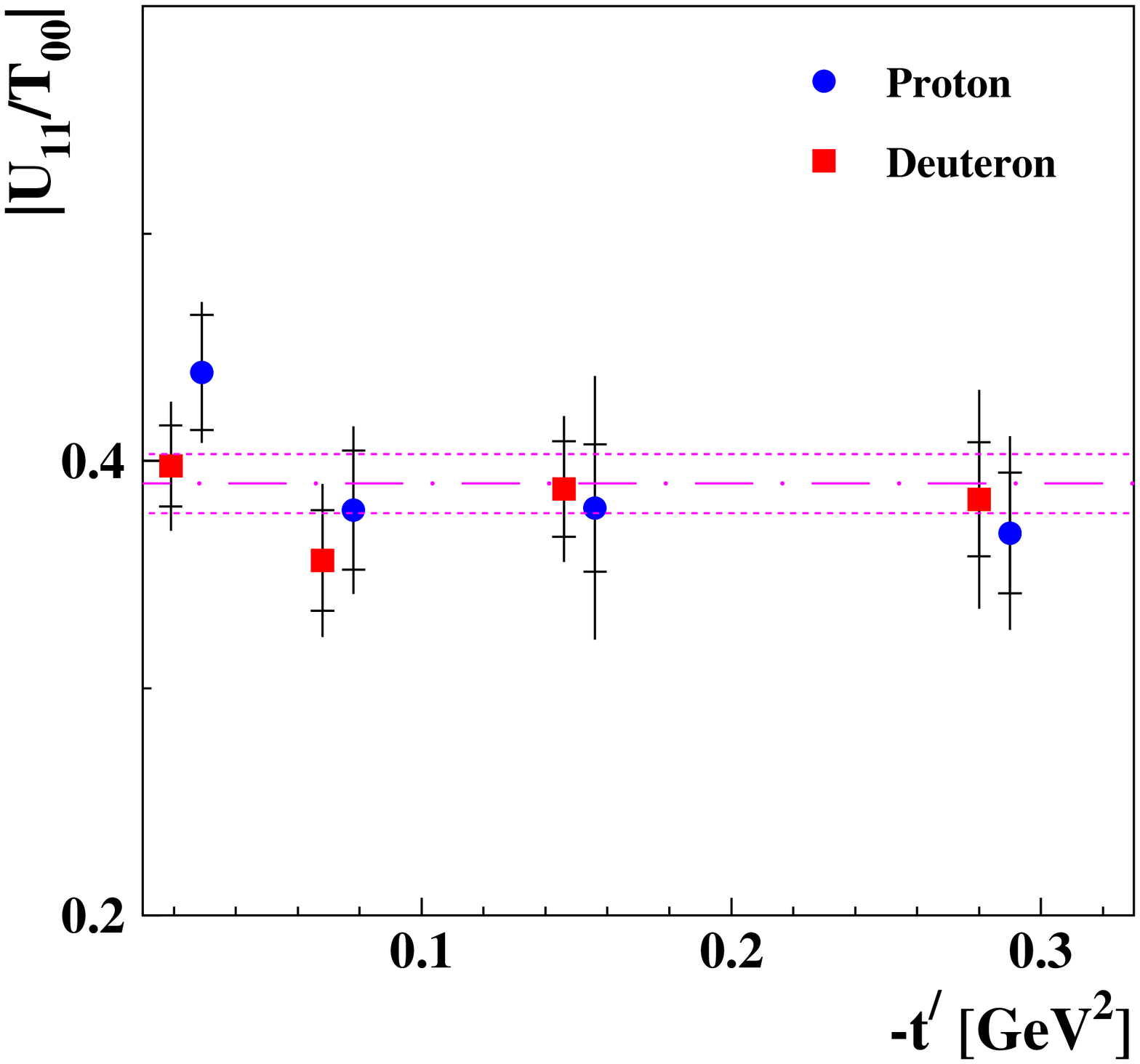}
  \caption{
 The dependences of $|U_{11}/T_{00}|$ on $Q^2$ and $t^\prime$ for proton and deuteron data. The points show the amplitude ratios given in Tabs.~\ref{ampl-hydr} and \ref{ampl-deutr} after averaging over $-t^\prime$ ($Q^2$) bins in the left (right) panel. The inner error bars  show the statistical uncertainty and the outer ones show the statistical and systematic uncertainties added in quadrature. The results  fitting the combined data set with a constant (central line),  $|U_{11}/T_{00}|=g$, are given in Tab.~\ref{u11}. The dashed lines correspond to one standard deviation in the total uncertainty.}
\label{fig-u11}
\end{figure*}

\begin{table*}[hbtc!]
 \renewcommand{\arraystretch}{1.2}
\begin{center}
\begin{tabular}{|c|c|c|c|c|}
    \hline
target&ratio&$g$&$\delta g$ &$\chi^2/N_{df}$\\
    \hline
proton &$|U_{11}/T_{00}|$&0.400& $\pm 0.020$ &0.60\\
    \hline
deuteron &$|U_{11}/T_{00}|$&0.383& $\pm 0.017$ &0.40\\
    \hline
proton+deuteron &$|U_{11}/T_{00}|$ &0.390& $\pm 0.013$ &0.49\\
    \hline
\end{tabular}\\
\end{center}
\caption{
\small{
Results of fitting the ratio  $|U_{11}/T_{00}|$ to a constant for proton, deuteron and combined data sets.
The values of parameters with their total uncertainties are presented.
The last column shows the  value of  $\chi^2$  per degree of freedom.}}
\label{u11}
\end{table*}

Figure~\ref{fig-t11-tpr} shows the $t^\prime$ dependence of the real and imaginary parts of the ratio $t_{11}$. Since $\mathrm{Re}(t_{11})$ and $\mathrm{Im}(t_{11})$ depend on $Q^2$ according to Eqs.~(\ref{fit-Q-Ret11}) and (\ref{fit-Q-Imt11}), they are shown in Fig.~\ref{fig-t11-tpr} multiplied or divided by $Q$, respectively. No noticeable $t^\prime$ dependence is observed for $\mathrm{Re}(t_{11})$ and $\mathrm{Im}(t_{11})$. Since the differential cross section of the process in Eq.~(\ref{react01}) for high energies and small $|t^\prime|$ is usually described by an exponential factor $\exp\{\beta t^\prime\}$, the helicity amplitudes should have exponential factors $T_{00} \propto \exp\{\beta_L t^\prime/2 \}$ and
$T_{11} \propto \exp\{\beta_T t^\prime/2 \}$. The absence of a $t^\prime$ dependence of the ratio $T_{11}/T_{00}$ means that the slope parameters $\beta_L$ and $\beta_T$ for the amplitudes $T_{00}$ and $T_{11}$ are close to each other. For very small $|t^\prime|$, it is reasonable to use the linear approximation
\begin{align}
\mathrm{Re}(T_{11}/T_{00})&     =     & \frac{a}{Q}\exp\{-\frac{1}{2} \Delta \beta_1  t^\prime\} \nonumber \\
                                     &\approx & \frac{a}{Q}(1+\frac{1}{2}\Delta \beta_1 |t^\prime|)
\label{bLbT-slope-Ret11},\displaybreak[2] \\
\mathrm{Im}(T_{11}/T_{00})&     =     & b Q\exp\{-\frac{1}{2}\Delta \beta_2  t^\prime\}  \nonumber \displaybreak[2] \\
                                     &\approx & bQ(1+\frac{1}{2}\Delta \beta_2 |t^\prime|).
\label{bLbT-slope-ImT11}
\end{align}
The proton results are $\Delta\beta_1=(-1.02 \pm 0.85)$~GeV$^{-2}$ and $\Delta\beta_2=(-0.91 \pm 2.00)$~GeV$^{-2}$, while the fit  for the deute\-ron data gives $\Delta\beta_1=(0.58 \pm 0.80)$~GeV$^{-2}$ and $\Delta\beta_2=(-1.96 \pm 1.58)$~GeV$^{-2}$. The results of the fits show that all four numbers are consistent with one another. We now assume that, within experimental accuracy, the slope parameters for the real and imaginary parts of the ratio coincide across both target types. In this case we have $\Delta\beta_1 \approx \Delta\beta_2 \approx \beta_L-\beta_T$. Combining these four numbers making use  of Eqs.~(\ref{avermean})  and (\ref{avererror}) we get an estimate for $\beta_L-\beta_T =(-0.4 \pm 0.5)$ GeV$^{-2}$.  This result on $\beta_L-\beta_T $  is in agreement with the prediction published in Ref.~\cite{bzk},  which ranges  from $-0.7$ GeV$^{-2}$ at $Q^2=0.8$ GeV$^{2}$ to $-0.4$ GeV$^{-2}$ at $Q^2=5$ GeV$^{2}$.

\begin{table*}[hbtc!]
 \renewcommand{\arraystretch}{1.2}
\begin{center}
\begin{tabular}{|c|c|c|c|c|c|c|c|}
    \hline
target & ratio & $c$, GeV$^{-1}$&$\delta c$, GeV$^{-1}$ &&&& $\chi^2/N_{df}$\\
    \hline
proton &$\mathrm{Re}(T_{01}/T_{00})$&0.405& $\pm 0.046$ &&&&0.56\\
    \hline
deuteron &$\mathrm{Re}(T_{01}/T_{00})$&0.378& $\pm 0.033$ &&&&0.69\\
    \hline
proton+deuteron &$\mathrm{Re}(T_{01}/T_{00})$&0.394& $\pm 0.024$ &&&&0.63\\
    \hline
    \hline
target&ratio&$f$& $\delta f$ &&&&$\chi^2/N_{df}$\\
    \hline
proton &$\mathrm{Im}(T_{01}/T_{00})$&0.317&$\pm 0.115$ &&&&0.97\\
    \hline
deuteron &$\mathrm{Im}(T_{01}/T_{00})$&0.161&$\pm 0.090$ &&&&1.20\\
    \hline
proton+deuteron &$\mathrm{Im}(T_{01}/T_{00})$&0.221&$\pm 0.069$ &&&&1.10\\
    \hline
target&ratio&$f_1$ GeV$^{-1}$& $\delta f_1$ GeV$^{-1}$ &$f_2$ GeV$^{-3}$
&$\delta f_2$ GeV$^{-3}$&$\rho_c$&$\chi^2/N_{df}$\\
    \hline
proton+deuteron &$\mathrm{Im}(T_{01}/T_{00})$&0.653&$\pm 0.132$ &-0.285&$\pm0.065$&-0.903&0.66\\
    \hline

\end{tabular}\\
\end{center}
\caption{  \small{
The kinematic dependence of $\mathrm{Re}(T_{01}/T_{00})$ and $\mathrm{Im}(T_{01}/T_{00})$ for
the proton, deuteron and combined data sets parameterized as given by Eqs.~$(\ref{fit-t/Q-Imt01}- \ref{fit-t-Ret01})$.
The values of parameters with their total uncertainties are presented. 
The last column shows the value of  $\chi^2$ per degree of freedom and $\rho_c$ is the correlation parameter.
}}
\label{t01}
\end{table*}

%%%%%%%%%%%%%%%%%%
\subsection{Kinematic Dependence of $|U_{11}/T_{00}|$}
 The unnatural-parity-exchange amplitude, $U_{11}$, des\-crib\-es the transition from a transversely polarized photon to a transversely polarized  $\rho^0$ meson ($\gamma_T^* \rightarrow \rho^0_T$). At large $W$ and $Q^2$, this transition should be suppressed by a factor of $M_V/Q$ compared to the dominant amplitude $T_{00}$  as explained in Sec.~\ref{sec:twistdec}.  The UPE contributions to the amplitude may be sizable at intermediate energies~\cite{IW,KS} typical for HERMES.

The  ratio $|u_{11}|$ versus $Q^2$ and $t^\prime$ is presented in  Fig.~\ref{fig-u11}.  The value of  $|U_{11}|$ is found to be smaller than   $|T_{00}|$  by a factor of approximately $2.5$. No kinematic dependences of the ratio $|u_{11}|$ are seen and therefore it is fitted to a  constant
\begin{eqnarray} 
\label{fit-Q-u11}  
 |u_{11}| = g.
\end{eqnarray}
 The results of the fit to proton and deuteron combined data are given in Tab.~\ref{u11}, and the constant $g$ is  shown in Fig.~\ref{fig-u11}  by  straight lines.  Fits with the same function $a/Q$ as for $\mathrm{Re}(t_{11})$  (see Eq.~(\ref{fit-Q-Ret11})), corresponding to the behavior expected  in pQCD, give values for $\chi^2$ per degree of freedom $2.03$, $2.67$, and $2.25$ for the proton, deuteron, and combined data, respectively.  This disagreement may reflect the fact that the HERMES $Q^2$ region is far from the asymptotic one.

From a study of soft hadron scattering, it is known~\cite{IW,KS,Kaidalov} that the most important contribution to UPE amplitudes at intermediate energies comes from pion ex\-cha\-nge. The amplitude of one-pion-exchange in the $t$-chan\-nel contains the propagator $1/(t-m_{\pi}^2)$, which become large for small values of $-t$ that approach the pole at the unphysical value of $-t = -m_{\pi}^2 \approx -0.018$~GeV$^2$. The mean value  of $-t^\prime$ for the first $-t^\prime$ bin is 0.019~GeV$^2$, and hence might appear to be small enough to approach the vicinity of the pole. However, as shown  in the right panel of Fig.~\ref{fig-u11}, no evidence for a pion-pole-like dependence can be seen in the data within the statistical precision of the measurement. Such a  dependence could be weakened by various effects:
\newline
i) even in the first bin in $-t^\prime$, the value of $\langle -t \rangle$ equal to $0.06$~GeV$^2$  is substantial, and hence not close enough to the pion pole at unphysical positive $t$,
\newline
ii) the amplitude $T_{00}$ has a strong exponential dependence $\propto \exp \{ \beta_L t^\prime/2 \}$  with $\beta_L \approx 7$~GeV$^{-2}$~\cite{DC-24,tytgat} and decreases with $|t^\prime|$, hence the ratio $U_{11}/T_{00}$ is flatter than $U_{11}$ itself,
\newline
 iii) in addition to one-pion exchange, other exchange processes can contribute to the amplitude $U_{11}$.

Using the amplitude method, the signal of unnatu\-ral-parity exchange has a significance of more than 20 standard deviations in the total experimental uncertainty separately for each of the proton and deuteron data sets (see  Tab.~\ref{u11} and  Fig.~\ref{fig-u11}). In contrast,  the existence of UPE was established~\cite{DC-24} with a  significance of 3 standard deviations for the combined proton and deuteron data in the analysis using the SDME method. 

%%%%%%%%%%%%%%%%
\subsection{Kinematic dependence of $T_{01}/T_{00}$}
\label{sec:kin-dep-t01}
The amplitude $T_{01} \equiv T_{0\frac{1}{2}1\frac{1}{2}}$ describing the transition $\gamma_T^* \rightarrow \rho^0_L$ is expected to be the largest SCHC-violating amplitude. In accordance with the  asymptotic formula (\ref{Q-asympt01}), the parameterization
\begin{eqnarray}
\mathrm{Im}(t_{01})=f\frac{\sqrt{-t^\prime}}{Q}
\label{fit-t/Q-Imt01}
\end{eqnarray}
is used. A fit to both proton and deuteron data using this parameterization gives acceptable $\chi^2$ values, as seen in Tab.~\ref{t01}. Proton and deuteron results are compatible within one standard deviation in the total uncertainty, although neither measurement is particularly precise as
shown in Tab.~\ref{t01} and the right panel of Fig.~\ref{fig-t01} in which the data are multiplied by $Q$ in order to demonstrate the $\sqrt{-t^\prime}$ dependence. The fit of the combined proton and deuteron data set yields a value of the parameter $f$ that is three standard deviations from zero with $\chi^2 /N_{df} \approx 0.66$  (see Tab.~\ref{t01}). We note that these results do not necessarily confirm the validity of Eq.~(\ref{Q-asympt01}) although the data do not contradict the pQCD prediction.

\begin{figure*}[hbtc!] 
\vspace{-0.7cm}
\hspace{-0.5cm}
\includegraphics[width=0.53\textwidth]{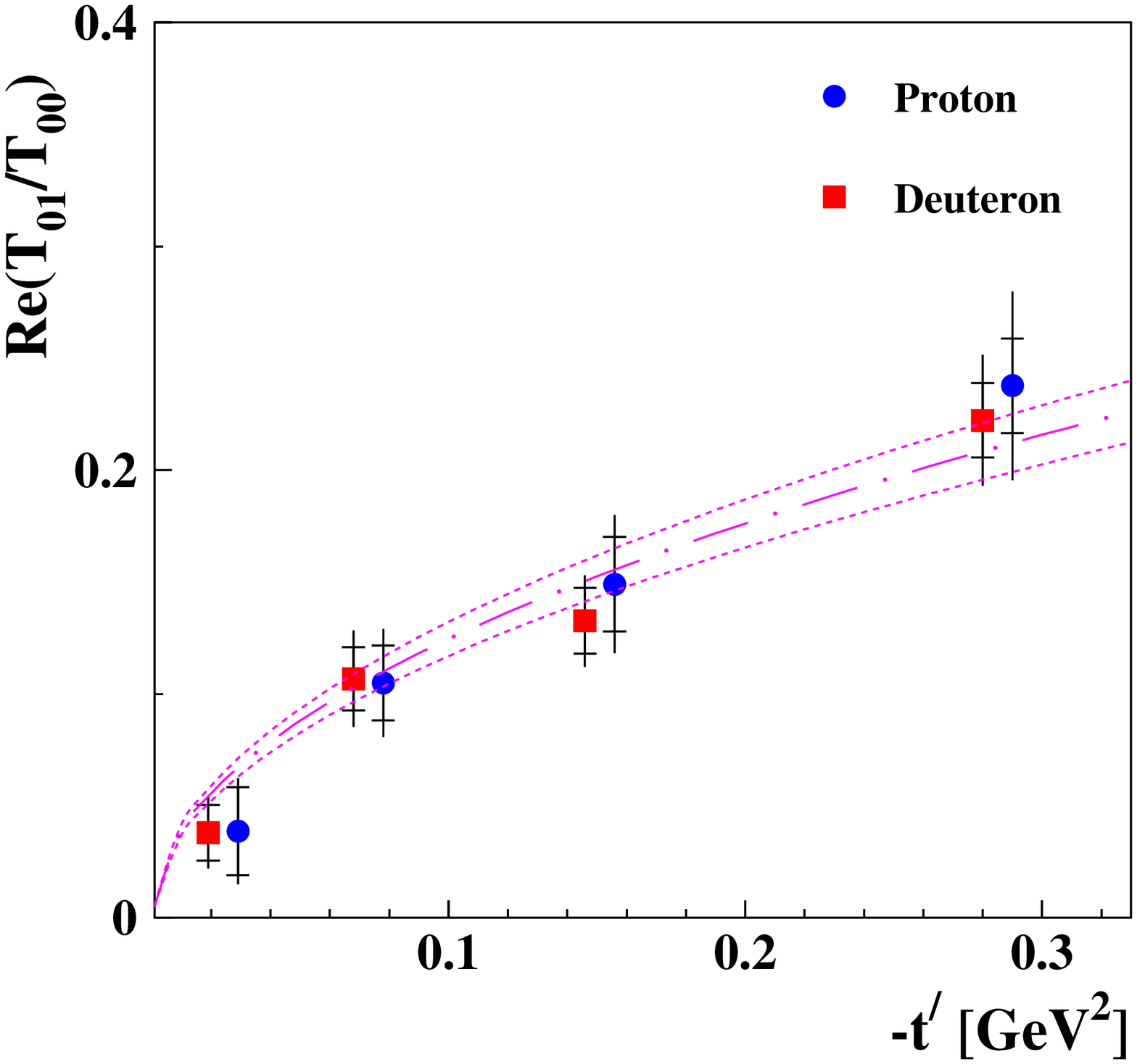}
\includegraphics[width=0.53\textwidth]{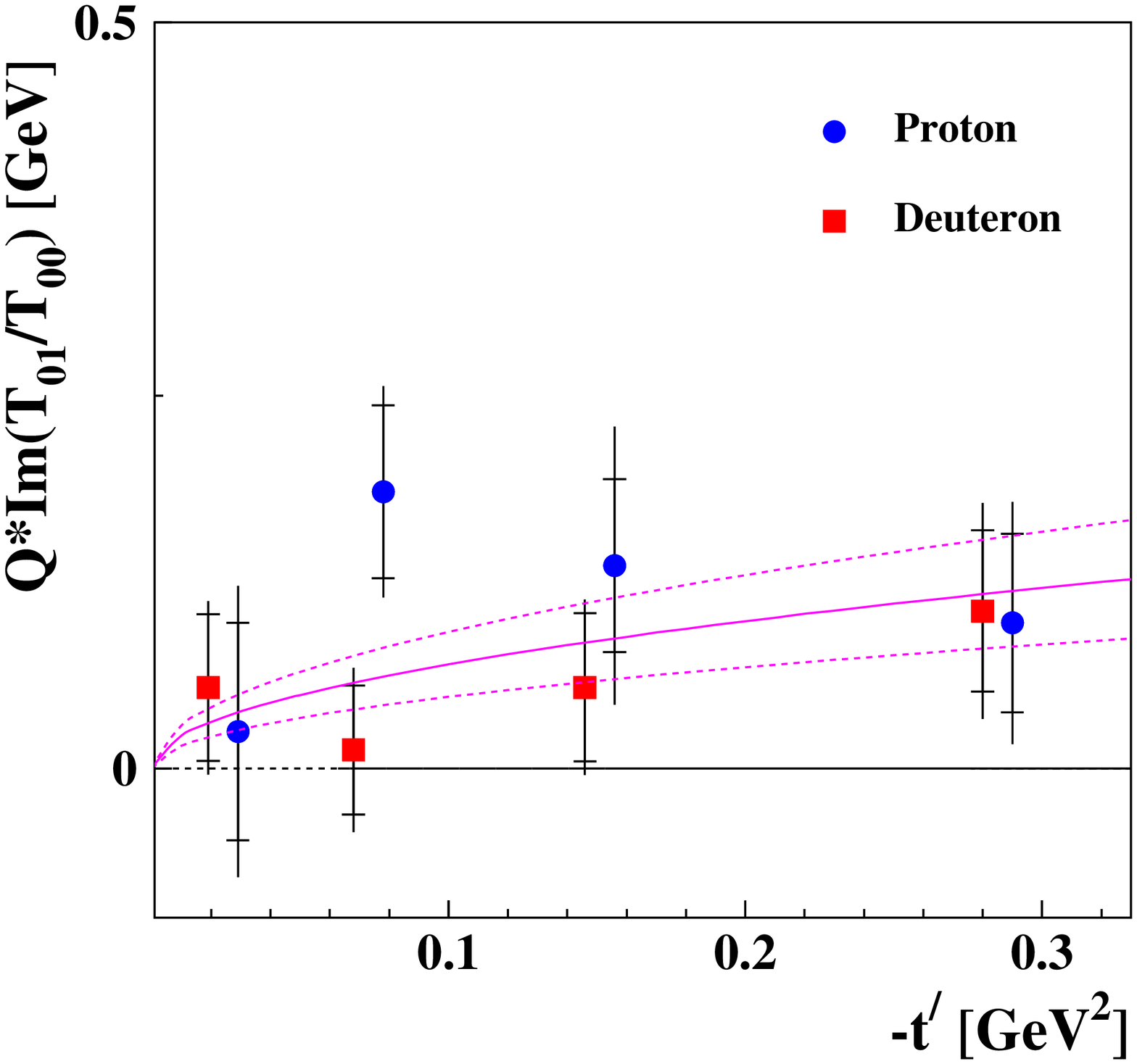}
\caption{
The $t^\prime$ dependence of $\mathrm{Re}(T_{01}/T_{00})$ (left panel) and $Q \cdot \mathrm{Im} (T_{01}/T_{00} )$ (right panel) for proton and deuteron data. Points show amplitude ratios from Tabs.~\ref{ampl-hydr} and \ref{ampl-deutr} after averaging over $Q^2$ bins.  Inner error bars show the statistical uncertainty and the outer bars show statistical and systematic uncertainties added in quadrature.  The parameterization is given by Eqs.~(\ref{fit-t/Q-Imt01}) and (\ref{fit-t-Ret01}). The parameters of the curves are given in Tab.~\ref{t01} for combined proton and deuteron data. Central lines are calculated with fitted  values of parameters, while the dashed lines correspond to one standard deviation of the curve parameter.
}
\label{fig-t01}
\end{figure*}

\begin{figure*}[hbtc!]
\hspace{-0.3cm}
\includegraphics[width=0.53\textwidth]{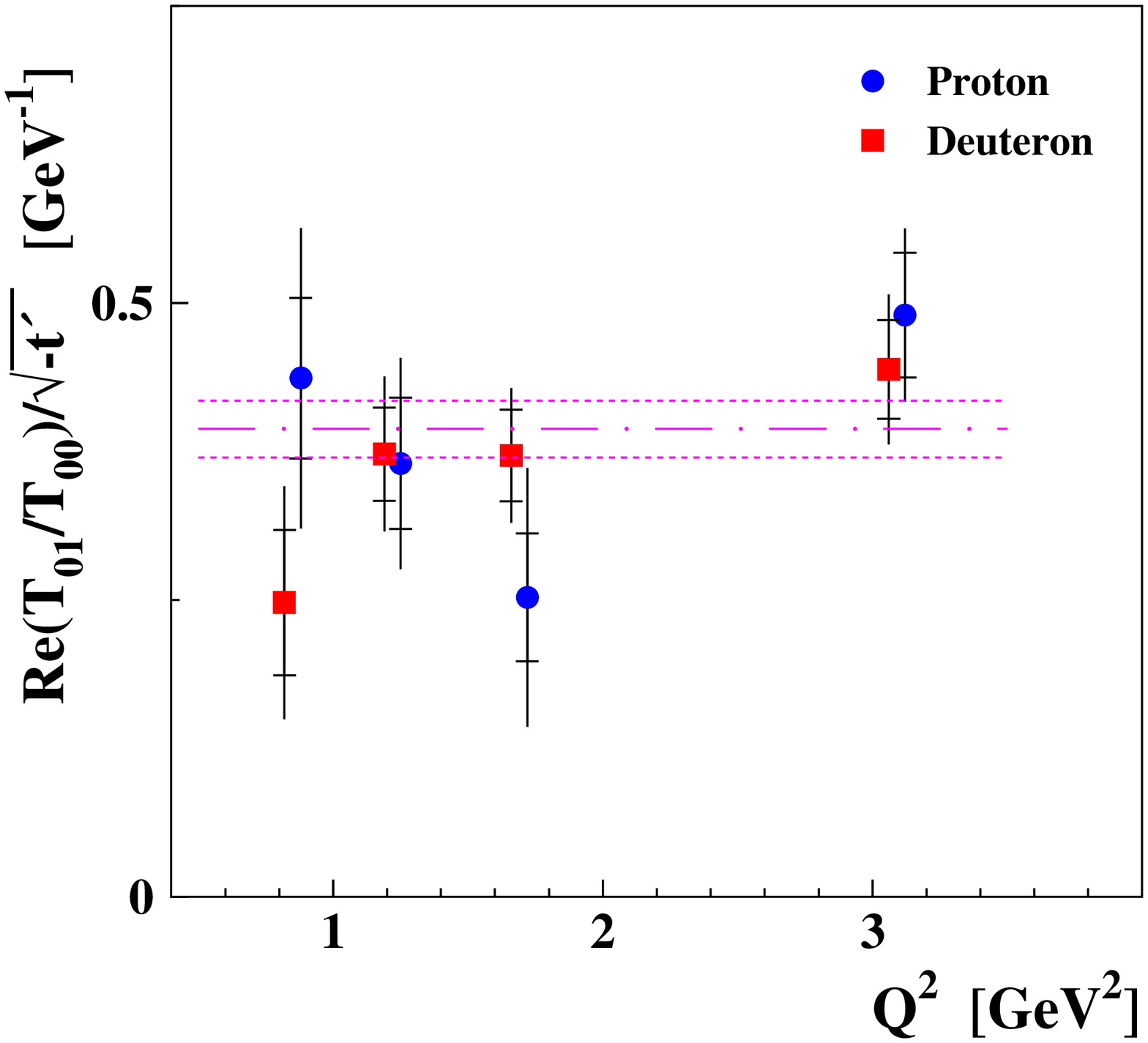}
\includegraphics[width=0.53\textwidth]{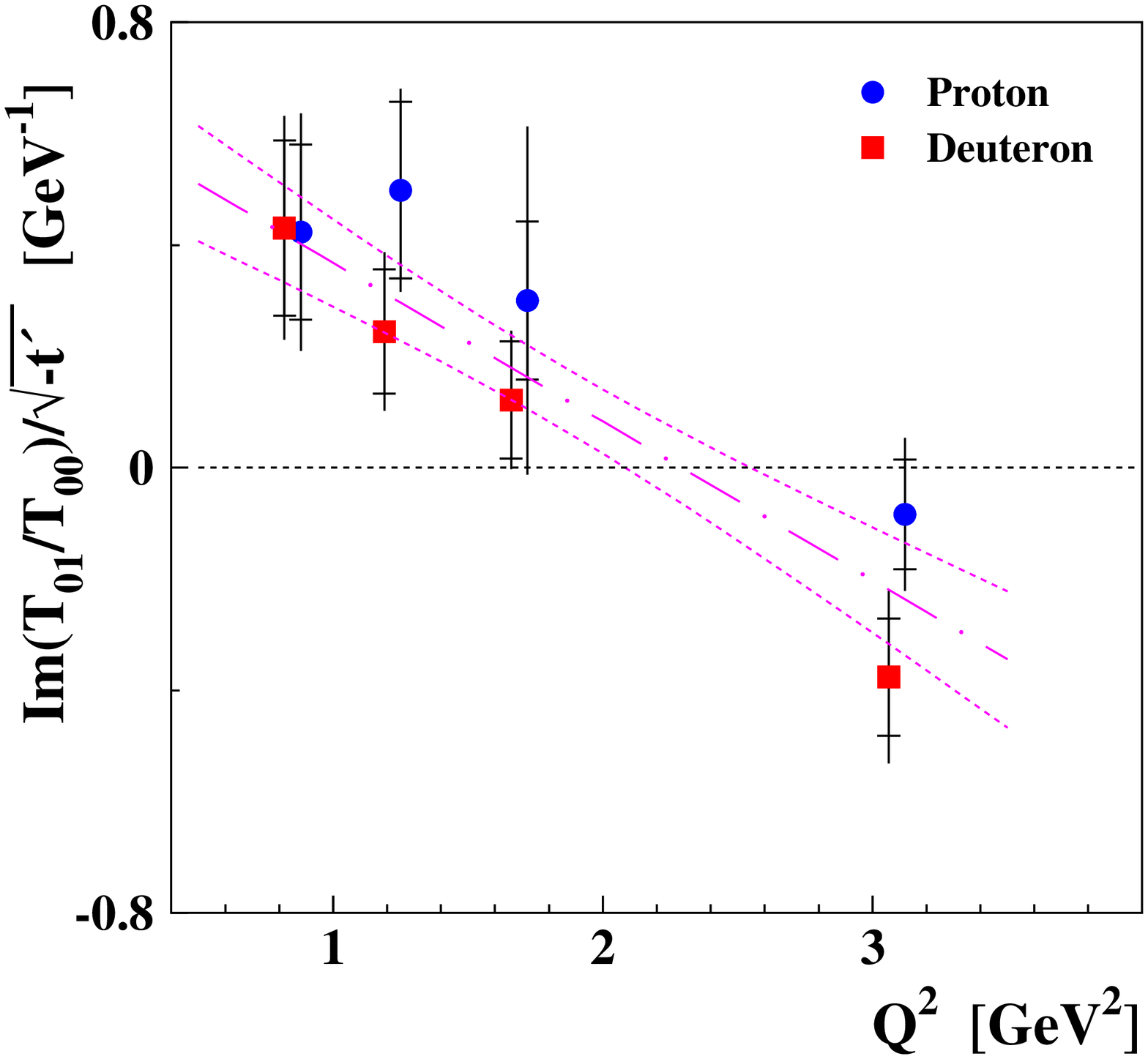}
\caption{
The $Q^2$ dependence of $\mathrm{Re}(T_{01}/T_{00})/\sqrt{-t^\prime}$ (left panel) and $\mathrm{Im}(T_{01}/T_{00})/\sqrt{-t^\prime}$ (right panel) for proton and deuteron data. Points show amplitude ratios  from Tabs.~\ref{ampl-hydr} and \ref{ampl-deutr} after averaging over $-t^\prime$ bins. Inner error bars show the statistical uncertainty and the outer bars show statistical and systematic uncertainties added in quadrature. The parameterization is given by Eqs.~(\ref{fit-f1f2-Imt01}) and (\ref{fit-t-Ret01}). The parameters of the curves are given in Tab.~\ref{t01} for combined proton and deuteron data. Central lines are calculated with fitted values of parameters, while the dashed lines correspond to one standard deviation of the curve parameter.
}
\label{fig-t01-Q2}
\end{figure*}

The $Q^2$ dependence of the amplitude ratio $\mathrm{Im}(t_{01})$ for the proton and deuteron data is shown in the right panel of Fig.~\ref{fig-t01-Q2}. As  shown in  the figure, the point for the deuteron in the largest $Q^2$ bin is slightly negative which favours a fit using the equation
\begin{eqnarray}
\mathrm{Im}(t_{01})=\sqrt{-t^\prime}(f_1+f_2Q^2)
\label{fit-f1f2-Imt01}
\end{eqnarray}
over that using parameterization (\ref{fit-t/Q-Imt01}). The result of the fit with function (\ref{fit-f1f2-Imt01}) is shown in the right panel of Fig.~\ref{fig-t01-Q2}. The parameters $f_1$ and $f_2$ are strongly correlated.  The correlation parameter $\rho_c$ is presented in Tab.~\ref{t01} and is taken into account in the calculation of the uncertainty of $\mathrm{Im}(t_{01})$ using Eq.~(\ref{fit-f1f2-Imt01}).

\begin{table*}[hbtc!]
 \renewcommand{\arraystretch}{1.2}
\begin{center}
\begin{tabular}{|c|c|c|c|c|}
    \hline
target&ratio&$r$, GeV$^{-1}$&$\delta r$, GeV$^{-1}$ &$\chi^2/N_{df}$\\
    \hline
proton &$\mathrm {Re}(T_{10}/T_{00})$&-0.012& $\pm 0.030$ &1.07\\
    \hline
proton &$\mathrm {Im}(T_{10}/T_{00})$&0.019& $\pm 0.061$ &0.58\\
    \hline
    \hline
target & ratio & $s$, GeV$^{-3}$& $\delta s$, GeV$^{-3}$ &$\chi^2/N_{df}$\\
    \hline

deuteron &$\mathrm {Re}(T_{10}/T_{00})$&0.045&$\pm 0.015$ &0.32\\
    \hline
deuteron &$\mathrm{Im}(T_{10}/T_{00})$&-0.109&$\pm 0.021$ &1.02\\
    \hline
\end{tabular}\\
\end{center}
\caption{ \label{t10}
\small{
The kinematic dependences of $\mathrm{Re}(T_{10}/T_{00})$ and $\mathrm{Im}(T_{10}/T_{00})$
for proton and  deuteron parameterized as given by Eqs.~(\ref{fit-t-re10}) and (\ref{fit-Q-re10}).
The values of parameters with their total uncertainties are presented. 
The last column shows the  value of  $\chi^2$  per degree of freedom.}
}
\end{table*}

\begin{table*}[hbtc!]
 \renewcommand{\arraystretch}{1.2}
\begin{center}
\begin{tabular}{|c|c|c|c|c|}
    \hline
target&ratio&$h$, GeV$^{-1}$&$\delta h$, GeV$^{-1}$ &$\chi^2/N_{df}$\\
    \hline
proton+deuteron &$\mathrm {Re}(T_{1-1}/T_{00})$&-0.156& $\pm 0.059$ &0.79\\
    \hline
proton+deuteron&$\mathrm {Im}(T_{1-1}/T_{00})$&-0.418& $\pm 0.126$ &0.54 \\

    \hline

\end{tabular}\\
\end{center}
\caption{ \label{t1m1}
\small{
The kinematic dependence of $\mathrm{Re}(T_{1-1}/T_{00})$ and $\mathrm{Im}(T_{1-1}/T_{00})$
for combined proton and deuteron data parameterized as given by Eq.~(\ref{fit-Q-t1-1}).
The values of parameters with their total uncertainties are presented. 
The last column shows the  value of  $\chi^2$  per degree of freedom.}
}
\end{table*}

The $Q^2$ dependence of the ratio $\mathrm{Re}(t_{01})/\sqrt{-t^\prime}$ is presented in the left panel of Fig.~\ref{fig-t01-Q2}. The quantity  $\mathrm{Re}(t_{01})$\\$/\sqrt{-t^\prime}$ does not decrease with $Q^2$ and can be well described by a constant which is also shown in the figure.
Using for $\mathrm{Re}(t_{01})$ the functional form of Eq.~(\ref{fit-t/Q-Imt01}) yields $\chi^2/N_{df}$ values of $1.08$, $1.55$, and $1.31$   for separate fits to the proton, deuteron  and combined data sets. Using instead the simpler parameterization
\begin{eqnarray}
\mathrm{Re}(t_{01})=c\sqrt{-t^\prime}
\label{fit-t-Ret01}
\end{eqnarray}
decreases the value of  $\chi^2$ by a factor of approximately  two, indicating a better description of the data. The results of this fit are 
shown in Tab.~\ref{t01} and the left panels of Figs.~\ref{fig-t01} and \ref{fig-t01-Q2}. The small values of $\chi^2/N_{df}$ as shown in Tab.~\ref{t01} indicate that our systematic uncertainty might be overestimated.

The phase difference $\delta _{01}$ between the amplitudes $T_{01}$ and $T_{00}$ can be calculated using the formula
\begin{eqnarray}
\tan \delta _{01}=\mathrm{Im}(t_{10})/\mathrm{Re}(t_{10}).
\label{phase-t01}
\end{eqnarray}
 The results for the $Q^2$ dependence of $\delta _{01}$ after averaging over $-t^\prime$ bins are shown in the
 right panel of Fig.~\ref{fig-delta11}. The phase difference $\delta _{01}$ is non-zero and decreases with $Q^2$. 
 It is slightly negative for the deuteron data at the largest value of $Q^2$. Using instead the results from the 
fits to 
formulas~(\ref{fit-f1f2-Imt01}, \ref{fit-t-Ret01}) given in Tab.~\ref{t01}, the phase difference is obtained as
\begin{eqnarray}
\tan \delta _{01}=(f_1+f_2Q^2)/c \; ,
\label{phase-Imt01}
\end{eqnarray}
which is shown as a curve in the right panel of Fig.~\ref{fig-delta11}. As can be seen, the fit function~(\ref{phase-Imt01}) is able to describes the $Q^2$ dependence of the phase difference $\delta _{01}$.

%%%%%%%%%%%%%%%%%%%%%%%%%%%%%%%%%%%%%%%%%%%%%%%%%%%%
\subsection{Small Amplitude Ratios $T_{10}/T_{00}$ and $T_{1-1}/T_{00}$ }
According to  the hierarchy given in Eq.~(\ref{HERMES-hier}), $T_{10}$ and $T_{1-1}$ are the smallest 
amplitudes  that give linear contributions to the numerators of SDMEs 
in the case of an unpolarized target. The ratio $t_{10}$ is expected
to be proportional to $\sqrt{-t^\prime}$ in accordance with 
Eq.~(\ref{t-asympt}), while $t_{1-1}$ is expected to be proportional to $t^\prime$. 
These ratios are supposed to depend on $Q^2$ in the asymptotic region according to 
Eqs.~(\ref{Q-asympt10}) and (\ref{Q-asympt1-1}). Figure~\ref{fig-t10} shows $\mathrm{Re}(t_{10})/\sqrt{-t^\prime}$
and $\mathrm{Im}(t_{10})/\sqrt{-t^\prime}$ versus $Q^2$ for
both proton and deuteron data. As shown in Fig.~\ref{fig-t10},
the values of $\mathrm{Re}(t_{10})/\sqrt{-t^\prime}$ and
$\mathrm{Im}(t_{10})/\sqrt{-t^\prime}$ on the proton are compatible with zero. 
The results of the fit of $\mathrm{Re}(t_{10})$ with the function
\begin{eqnarray}
\mathrm{Re}(t_{10}) =r\sqrt{-t^\prime}
\label{fit-t-re10}
\end{eqnarray}
and $\mathrm{Im}(t_{10})$ with the same function are presented in Tab.~\ref{t10}.
The ratio $\mathrm{Re}(t_{10})/\sqrt{-t^\prime}$ for the deuteron 
is slightly positive (except the first point at the smallest value of $Q^2$) and increases with $Q^2$. The
fit of the deuteron data with the function
\begin{eqnarray}
\mathrm{Re}(t_{10})=sQ^2\sqrt{-t^\prime}
\label{fit-Q-re10}
\end{eqnarray}
provides the positive result which is presented in Tab.~\ref{t10}.
As shown in Fig.~\ref{fig-t10}, the quantity $\mathrm{Im}(t_{10})/\sqrt{-t^\prime}$ is negative
for the three points at the largest values of $Q^2$.
The fit of  $\mathrm{Im}(t_{10})/\sqrt{-t^\prime}$ using the same function (\ref{fit-Q-re10})
gives the negative result presented in Tab.~\ref{t10}.
This behavior contradicts the pQCD prediction given by Eq.~(\ref{Q-asympt10}), and indicates that in the case of the amplitude  $T_{10}$ the HERMES kinematic range may be far from the large-$Q$ asymptotic region.

We recall that the amplitudes $T_{00}$ and $T_{10}$ vanish in the small $Q$-limit ($Q^2 \rightarrow 0$) because the real photon
has no longitudinal polarization. If the behavior at   fixed $t^\prime$ and small $Q^2$ were described by
\begin{eqnarray}
T_{00} &\propto &  Q \;, \label{T00-smal-Q} \\
T_{10} &\propto & Q^3, \label{T10-smal-Q}
\end{eqnarray}
the behavior of the ratio $t_{10} \equiv T_{10}/T_{00}$ would be just as assumed by Eq.~(\ref{fit-Q-re10}) and presented in Fig.~\ref{fig-t10}.

\begin{figure*}[hbtc!]
\vspace{-0.8cm}
\hspace{-0.5cm}
\includegraphics[width=0.53\textwidth]{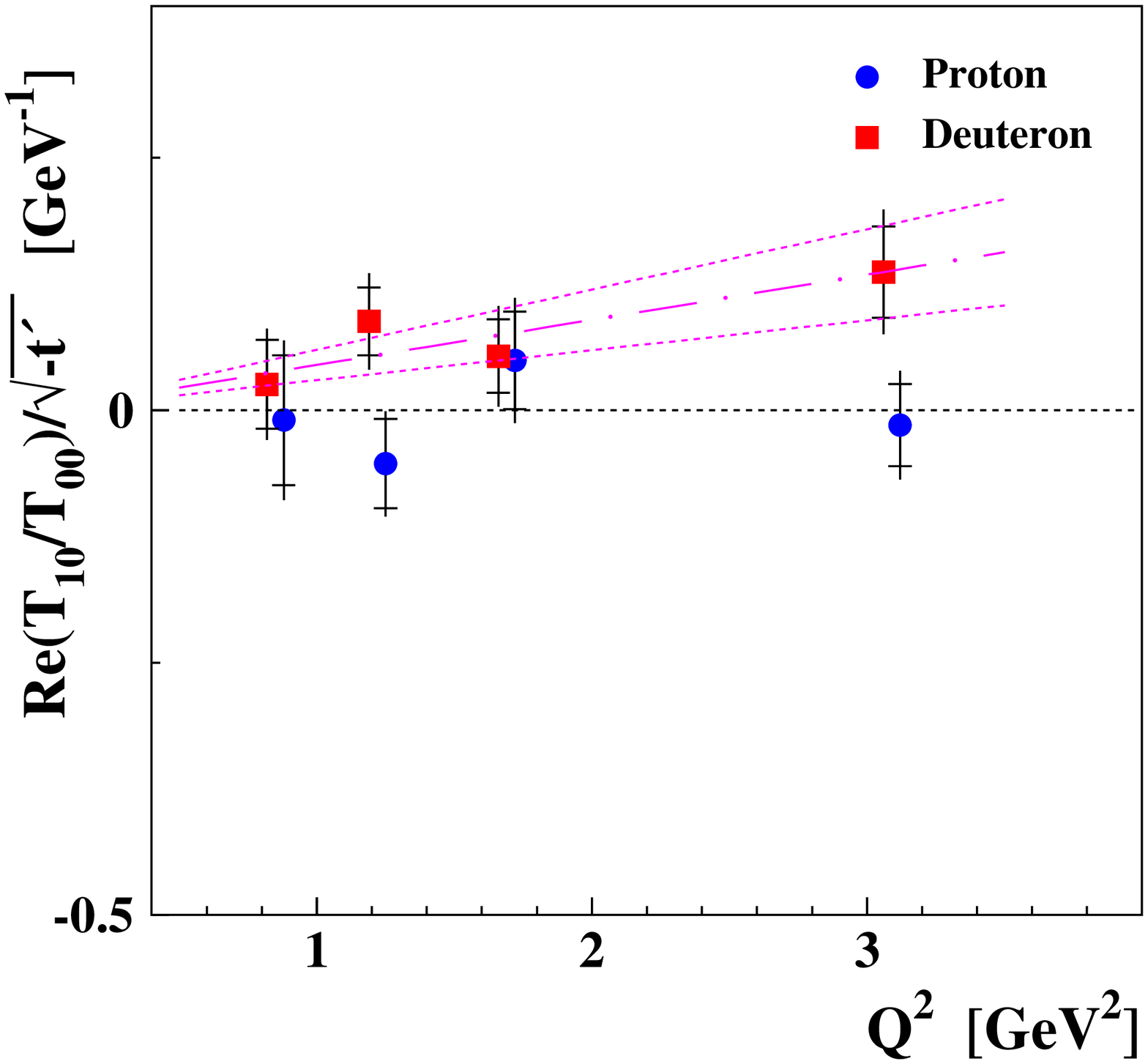}
\includegraphics[width=0.53\textwidth]{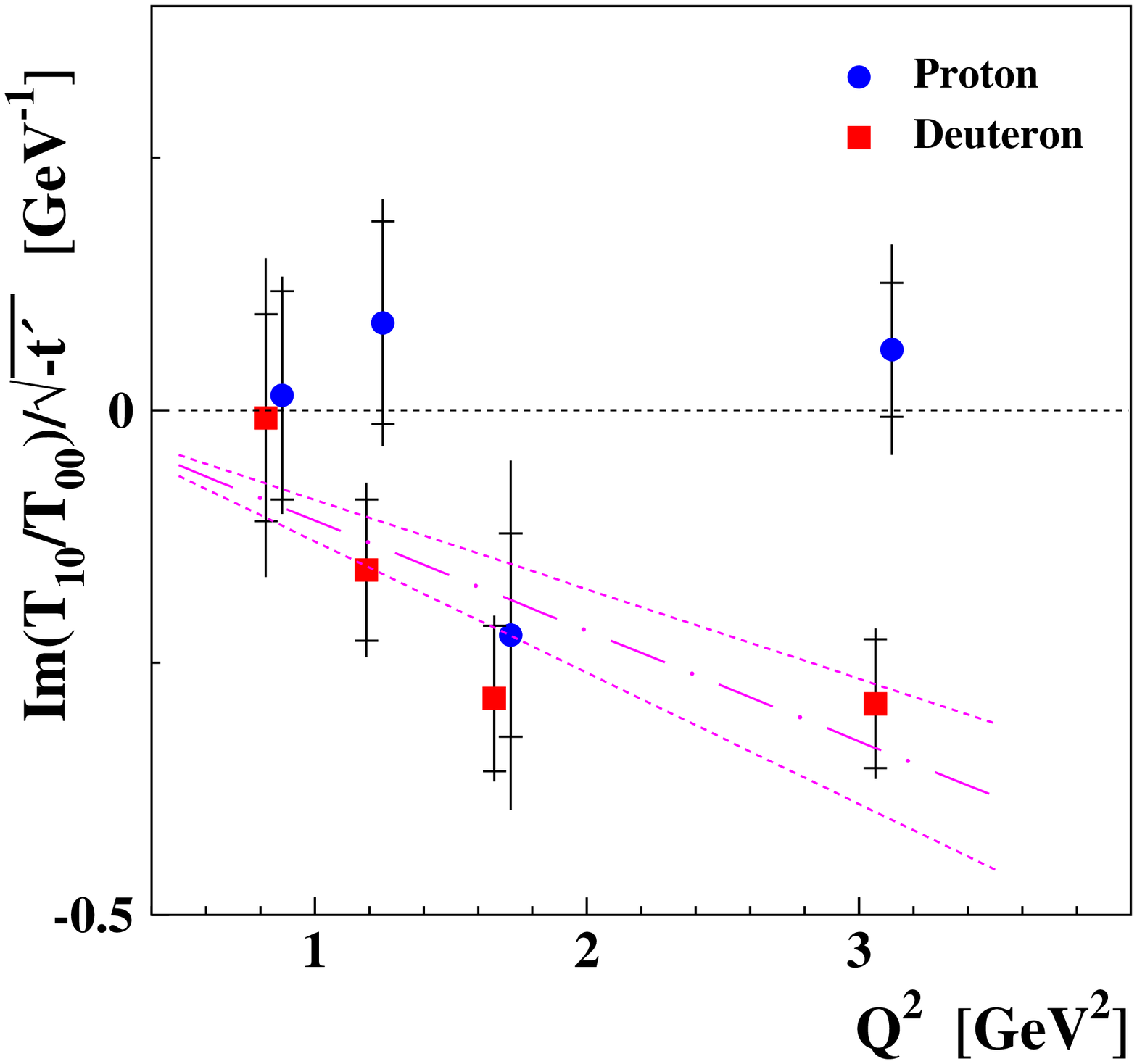} 
  \caption{
 The $Q^2$ dependence of $\mathrm{Re}(T_{10}/T_{00})/\sqrt{-t^\prime}$ (left panel) and 
 $\mathrm{Im}(T_{10}/T_{00})/\sqrt{-t^\prime}$ (right panel) for proton and deuteron data. Points show amplitude 
 ratios from Tabs.~\ref{ampl-hydr} and \ref{ampl-deutr} after averaging over $-t^\prime$ bins. Inner error bars show
 the statistical uncertainty and the outer ones indicate statistical and systematic uncertainties added in 
 quadrature. The parameterization of the curves describing the deuteron data both on
 $\mathrm{Re}(T_{10}/T_{00})$ and $\mathrm{Im}(T_{10}/T_{00})$ is given by Eq.~(\ref{fit-Q-re10}) and their 
 parameters  are given in Tab.~\ref{t10}. Central lines are calculated with the  fitted  values of  parameters,
 while the dashed lines correspond to one standard deviation of the curve parameter.
}
\label{fig-t10}
\end{figure*}

\begin{figure*}[hbtc!]
\vspace{-0.3cm}
\hspace{-0.3cm}
\includegraphics[width=0.53\textwidth]{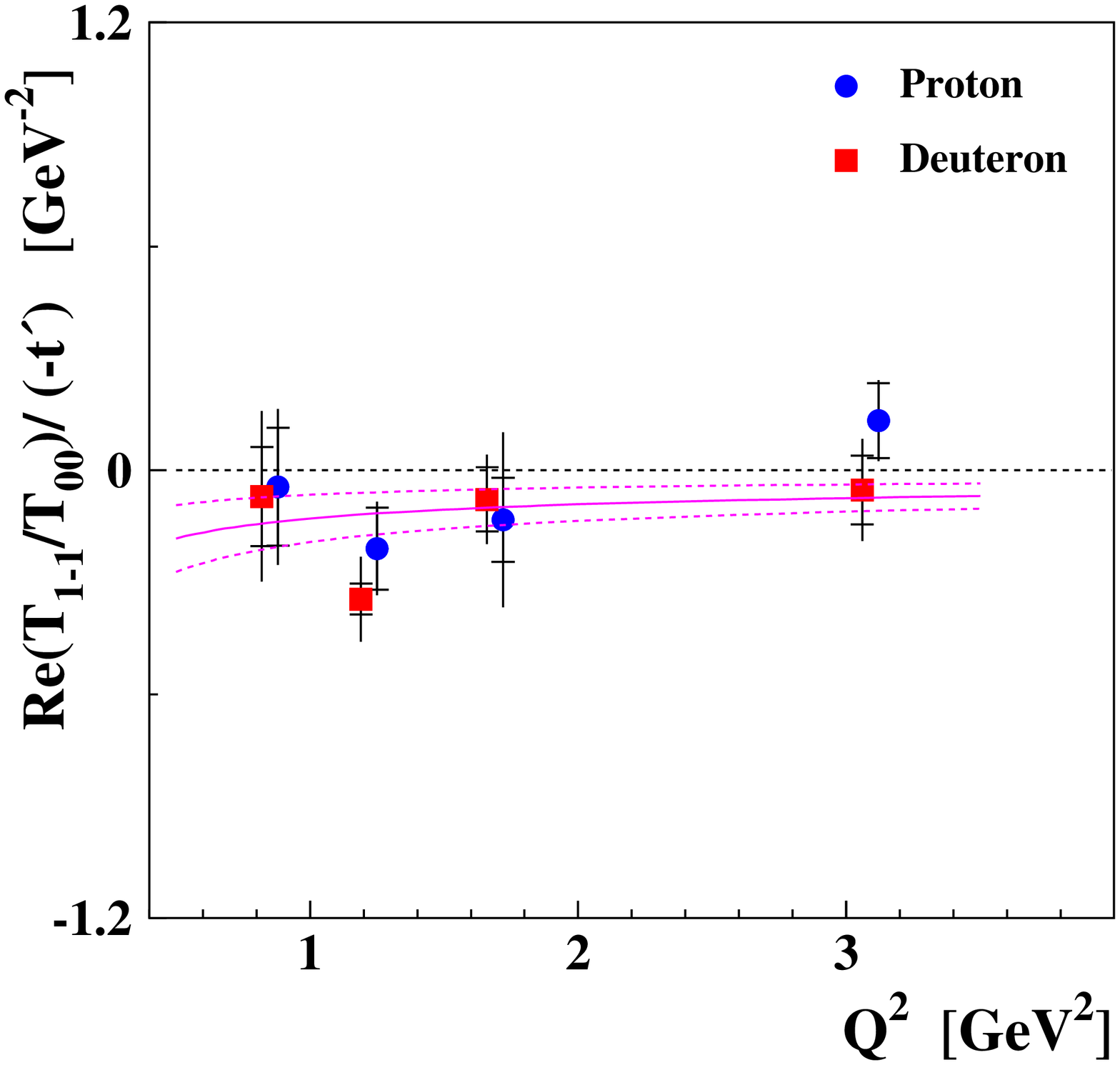}
\includegraphics[width=0.53\textwidth]{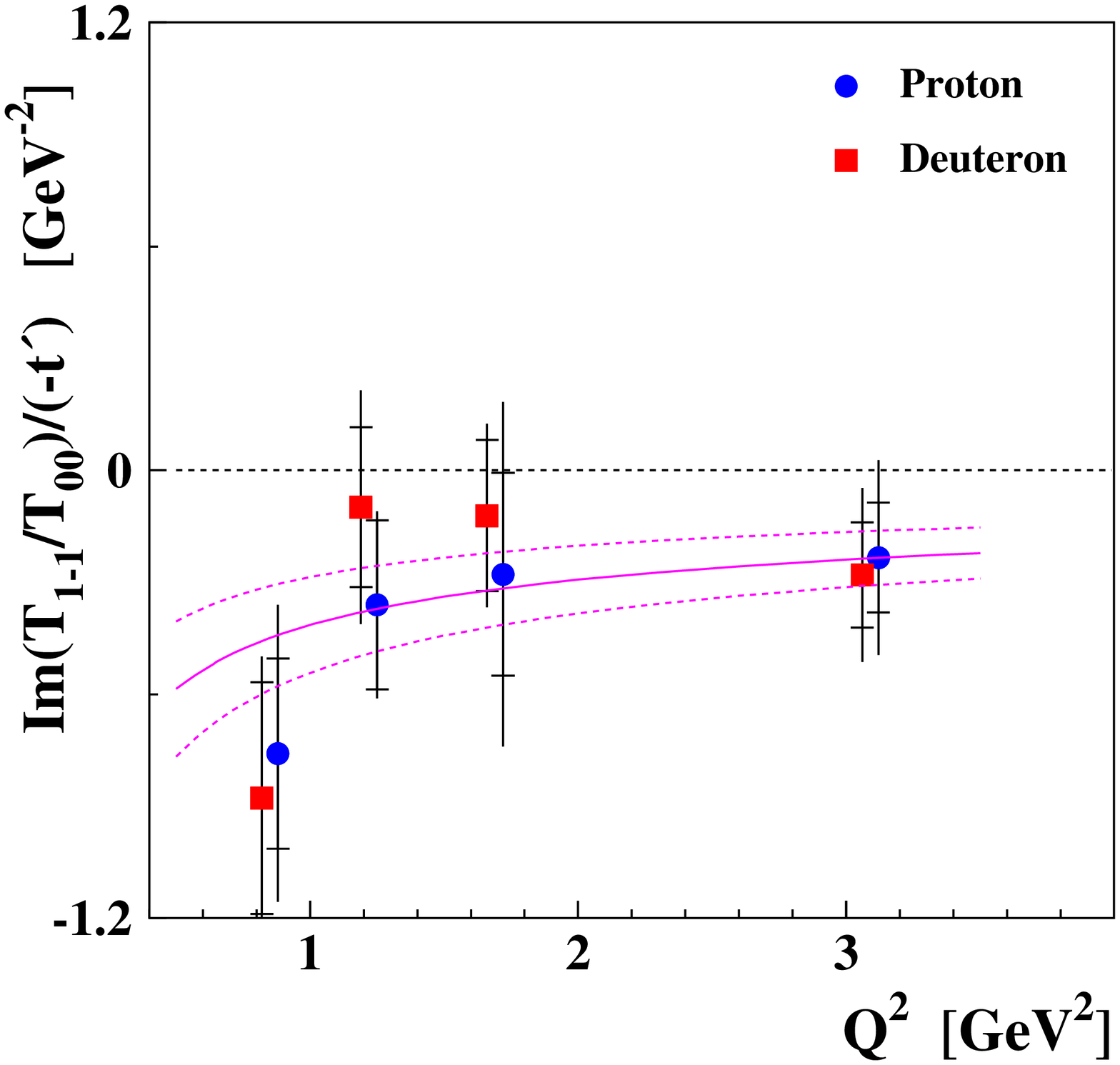}
  \caption{
 The $Q^2$ dependence of $\mathrm{Re}(T_{1-1}/T_{00})/(-t^\prime)$ (left panel) and 
 $\mathrm{Im}(T_{1-1}/T_{00})/(-t^\prime)$ (right panel) for
 proton and deuteron data. Points show amplitude ratios from Tabs.~\ref{ampl-hydr} and \ref{ampl-deutr} after 
 averaging over $-t^\prime$ bins. Inner error bars show the statistical uncertainty and the outer ones indicate
 statistical and systematic uncertainties added in quadrature. The parameterization of the curves describing the
 combined proton and deuteron data both on $\mathrm{Re}(T_{1-1}/T_{00})$ and $\mathrm{Im}(T_{1-1}/T_{00})$ is 
 given by Eq.~(\ref{fit-Q-t1-1}) and their parameters  are given in Tab.~\ref{t1m1}. Central lines are calculated
 with the fitted  values of parameters, while the dashed lines correspond to one standard deviation of the curve 
 parameter.
}
\label{fig-t1m1}
\end{figure*}

%%%%%%%%%%%%%%%%%%%%%%%%%%
\begin{figure*}[hbtc!]
\vspace{-0.7cm}
\hspace{-0.5cm}
\includegraphics[width=0.53\textwidth]{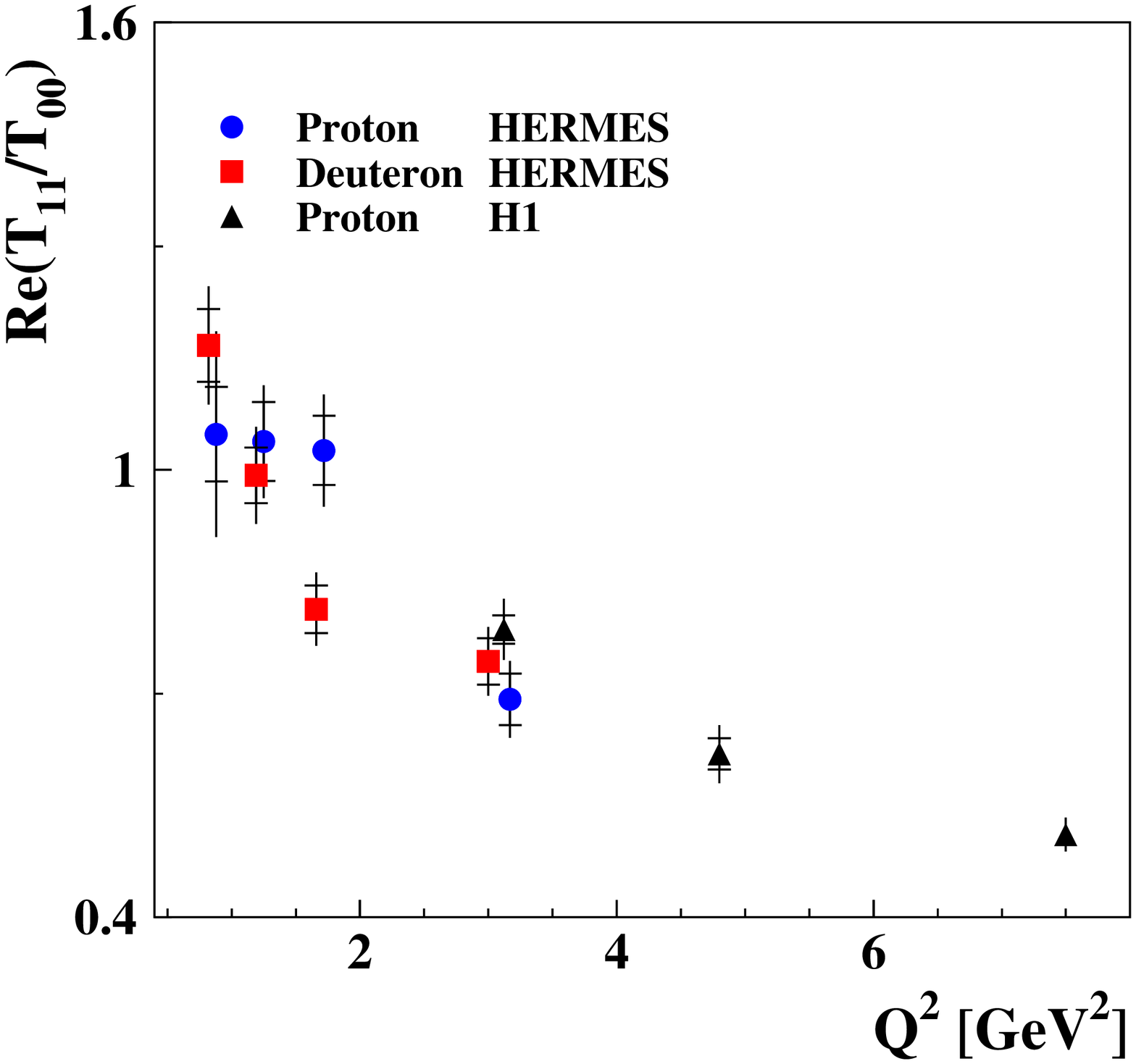}
\includegraphics[width=0.53\textwidth]{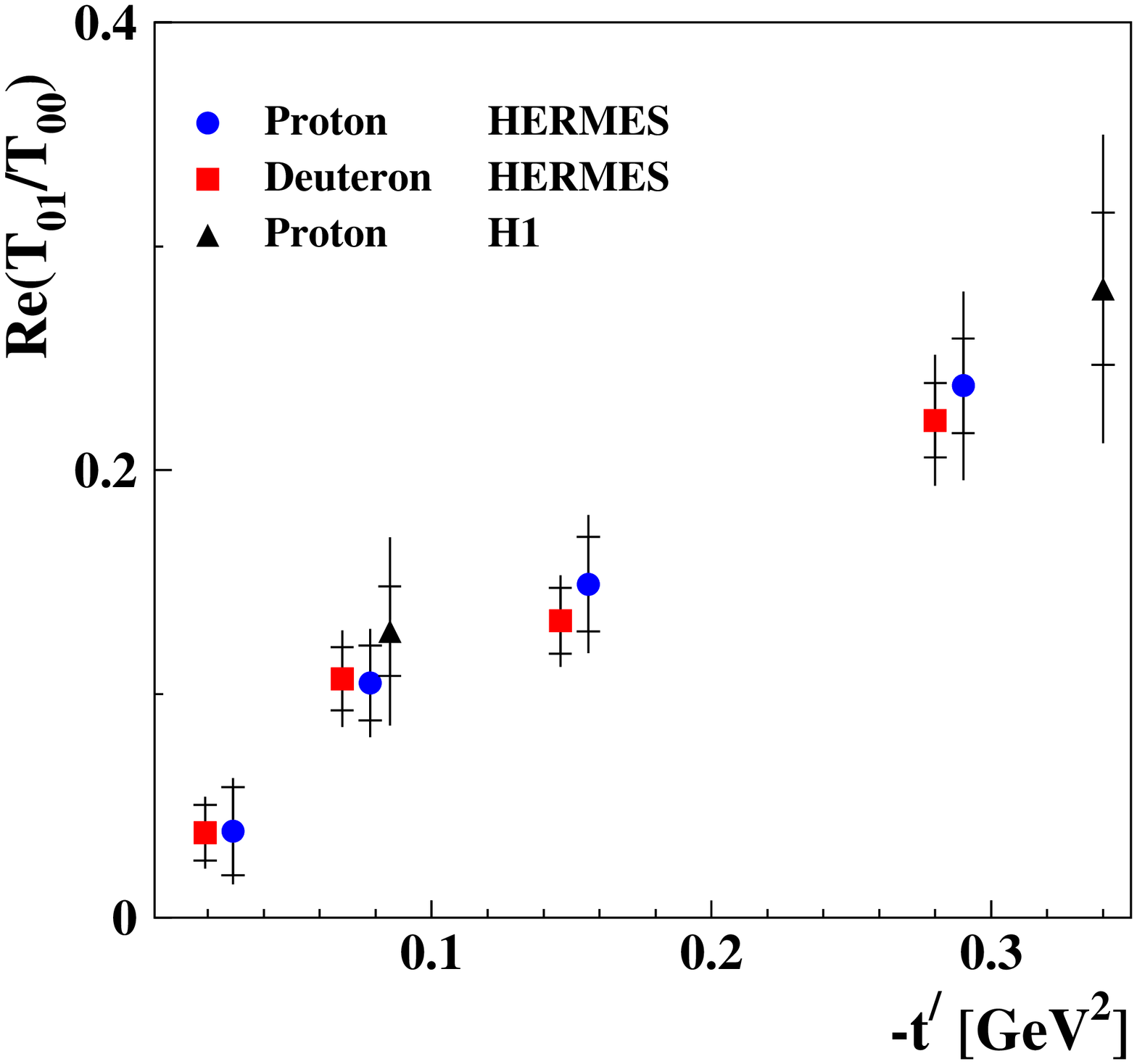}    
  \caption{The kinematic dependences of $\mathrm{Re}(T_{11}/T_{00})$ and
$\mathrm{Re}(T_{01}/T_{00})$ for proton and deuteron data.
Points of HERMES are the same as in Figs.~\ref{fig-t11} and \ref{fig-t01}.
The H1 results are from Ref.~\cite{H1-amp}.
Inner error bars show the statistical uncertainty and the outer ones indicate
statistical and systematic uncertainties added in quadrature.
}
\label{world}
\end{figure*}

The ratios $\mathrm{Re}(t_{1-1})/(-t^\prime)$
and $\mathrm{Im}(t_{1-1})/(-t^\prime)$ are presented versus $Q^2$ 
in Fig.~\ref{fig-t1m1}. Both ratios  appear to be compatible with zero
 within  $2.5$ $\sigma$ of the total
uncertainty  both for the proton and deuteron target and only the combined data 
provides a non-zero signal for $\mathrm{Im}(t_{1-1})$. The fit of $\mathrm{Re}(t_{1-1})$ with the function
\begin{eqnarray}
\mathrm{Re}(t_{1-1})=(-t^\prime)\frac{h}{Q}   
\label{fit-Q-t1-1}
\end{eqnarray}
 corresponds to the second term in Eq.~(\ref{Q-asympt1-1}).
 The result, presented in Tab.~\ref{t1m1} for the combined proton and
deuteron data, shows a signal of $2\sigma$ significance with respect to the total uncertainty.
The result of the fit of $\mathrm{Im}(t_{1-1})$ with the same function (\ref{fit-Q-t1-1})
is  also presented in Tab.~\ref{t1m1}.
The parameter $h$  deviates from zero by more than three standard deviations of the total uncertainty.
A fit of the combined proton and deuteron data with both terms in   Eq.~(\ref{Q-asympt1-1})
 provides  values of $C_1$ and $C_2$ that  are consistent with zero within $1.5$ $\sigma$ of the total 
uncertainty for both $\mathrm{Re}(t_{1-1})$ and $\mathrm{Im}(t_{1-1})$.
This means that the statistical precision of the HERMES data  
does not allow a reliable verification of Eq.~(\ref{Q-asympt1-1}) or an extraction of 
the gluon transversity GPD.

%%%%%%%%%%%%%%
\subsection{Comparison of HERMES and H1 Results  }
Here we compare the results of the present work with the analysis of the $\rho^0$-meson production
data on the proton by the H1 collaboration~\cite{H1-amp}, obtained for the CM
energy range $36$ GeV $\leq W \leq 180$ GeV,  photon virtuality
$2.5$~GeV$^2$ $\leq Q^2 \leq 60$ GeV$^2$, and $-t^\prime\leq 3$ GeV$^2$.
For the comparison of the $t^\prime$ dependence, the results at the mean value of $Q^2$ equal to 3.3  GeV$^2$
are chosen which are the closest ones to the $Q^2$ region of the HERMES data. In the H1 data analysis presented 
in Ref.~\cite{H1-amp}, the imaginary parts of the amplitude ratios were not extracted.  The additional hypothesis 
was used in Ref.~\cite{H1-amp} that for any ratio of the amplitudes, the approximate relation 
$|t_{\lambda_V\lambda_{\gamma}}|^2=[\mathrm{Re}(t_{\lambda_V\lambda_{\gamma}})]^2$ is valid within the 
 experimental accuracy. The comparison of the HERMES and H1 results is presented in Fig.~\ref{world}. 
 As seen from the figure the HERMES and H1 results agree  within their total uncertainties.
 No strong  dependence on $W$ is observed for the amplitude ratios $\mathrm{Re}(t_{11})$ and
 $\mathrm{Re}(t_{01})$. Considering the  differences in $Q^2$ between HERMES ($\langle Q^2 \rangle =2.0$~GeV$^2$) 
 and H1 ($\langle Q^2 \rangle=3.3$~GeV$^2$), no $Q^2$ dependence can be seen
 for $\mathrm{Re}(t_{01})$  at small $|t^\prime|$  even  for $Q^2\leq 3.3$ GeV$^2$ (see also the 
 discussion in Sec.~\ref{sec:kin-dep-t01}). The ratios  $\mathrm{Re}(t_{10})$ and $\mathrm{Re}(t_{1-1})$ obtained 
by H1~\cite{H1-amp} are compatible with zero at $Q^2=3.1$ GeV$^2$ within the experimental accuracy. This is 
consistent with the HERMES results for the proton shown in Figs.~\ref{fig-t10} and \ref{fig-t1m1}.

%%%%%%%%%%%%%%%%%%%%%%%%%%%%%%%%%%%%%%%%%%%%%%%%%%%%%%%%%
\section{Conclusions}
 Exclusive $\rho^0$-meson electroproduction is studied in the HERMES experiment, using a $27.6$~GeV
 longitudinally polarized electron/positron beam and unpolarized hydrogen and deuterium  targets in the
 kinematic region $0.5$~GeV$^2 <Q^2<7.0$~GeV$^2$, $3.0$~GeV $<W<6.3$~GeV, and $-t^\prime<0.4$~GeV$^2$.
These data are used  to determine the real and imaginary parts of the  ratios 
$T_{11}/T_{00}$, $T_{01}/T_{00}$, $T_{10}/T_{00}$, $T_{1-1}/T_{00}$, and 
$|U_{11}/T_{00}|$ for 16 bins in $Q^2$ and $-t^\prime$.
%% using a nine-parameter fit.
Systematic uncertainties due to the background contribution,  uncertainties in
the  Monte Carlo input  parameters, 
and the uncertainty of the applied amplitude method are given. Except for $T_{10}/T_{00}$, 
the amplitude ratios for the proton are compatible with those 
for the deuteron. The extracted amplitude ratios $T_{11}/T_{00}$, $|U_{11}/T_{00}|$ and $T_{01}/T_{00}$ are 
 found to be sizable. The ratios $T_{10}/T_{00}$ and $T_{1-1}/T_{00}$ for the proton are found to be
compatible with zero within experimental uncertainties. The ratio $T_{1-1}/T_{00}$ for the deuteron is 
also zero within experimental accuracy, while $\mathrm{Re}(T_{10}/T_{00})$ is slightly positive 
 and $\mathrm{Im}(T_{10}/T_{00})$ is slightly negative, except for the bin with the smallest $Q^2$ values.

The SDMEs calculated in terms of these helicity amplitude ratios  generally agree with the results  published in Ref.~\cite{DC-24}.
 The amplitude method was shown to provide more accurate polarized SDMEs than the previous analysis using the SDME method. 

 The statistical precision available in this analysis permits the parameterization of the kinematic dependences of amplitude ratios and the 
 extraction of the phase difference between various helicity amplitudes. The real part of $T_{11}/T_{00}$ is found to  follow the asymptotic 
 $1/Q$ behavior predicted by pQCD~\cite{IK,KNZ}. The imaginary part of $T_{11}/T_{00}$ grows with $Q^2$, in  contradiction to the 
large-$Q^2$ asymptotic  behavior expected from  pQCD. The phase difference 
$\delta_{11}$ between the amplitudes  $T_{11}$ and $T_{00}$ grows with $Q^2$ and 
has a mean value of about 30 degrees in the HERMES kinematic region. This 
is in  agreement with the published result of the SDME method~\cite{DC-24} and  in contradiction to  
calculations~\cite{golos1,golos2,golos3,IK,KNZ} based on pQCD. For the first time, 
the $Q^2$ dependence of $\delta_{11}$ observed 
in Ref.~\cite{DC-24} is shown to be related to the increase  with  $Q^2$ of the imaginary part of 
the ratio of the helicity amplitudes $T_{11}/T_{00}$.

The behavior of $\mathrm{Im}(T_{01}/T_{00})$ is found to be in  agreement with the asymptotic pQCD  
behavior $\sqrt{-t^\prime}/Q$, while the extracted value of $\mathrm{Re}(T_{01}/T_{00})$ is likely to be 
in disagreement with the pQCD prediction. The data indicate non-zero values of the  phase difference 
$\delta_{01}$ 
between the amplitudes $T_{01}$ and $T_{00}$ and  the decrease of $\delta_{01}$ with $Q^2$.

The ratio $|U_{11}/T_{00}|$ is found to be constant in the HERMES kinematic region, in disagreement with the
asymptotic pQCD behavior at  large $Q^2$. No pion-pole-like behavior is observed
at small $|t^\prime|$. The UPE signal is seen with a very high significance for  both  proton and
deuteron data, confirming the existence of unnatural-parity exchange contributions with a higher 
precision
than that obtained with the SDME method~\cite{DC-24}.

The $Q^2$ dependence of the amplitude ratio $\mathrm{Re}(T_{11}/T_{00})$ and $t^\prime$ dependence 
of the amplitude ratio $\mathrm{Re}(T_{01}/T_{00})$ are also compared to those extracted  
by the H1  collaboration at the center-of-mass 
energy range $36$~GeV $\leq W \leq 180$~GeV,  photon virtuality $2.5$~GeV$^2 \leq Q^2 \leq 60$~GeV$^2$, and 
 $-t^\prime\leq 3$~GeV$^2$. No strong dependence of the amplitude ratios $\mathrm{Re}(T_{11}/T_{00})$ 
and $\mathrm{Re}(T_{01}/T_{00})$ on $W$, and 
no $Q^2$ dependence of $\mathrm{Re}(t_{01})$ at small $|t^\prime|$ are observed. 
The HERMES and H1 results agree  within their total uncertainties.

\section*{Acknowledgments}
We would like to thank M.~Diehl, S.V.~Goloskokov, P.~Hoy\-er, and P.~Kroll for many useful
discussions.

We gratefully acknowledge the DESYmanagement for its support and the staff
at DESY and the collaborating institutions for their significant effort.
This work was supported by 
the Ministry of Economy and the Ministry of Education and Science of Armenia;
the FWO-Flanders and IWT, Belgium;
the Natural Sciences and Engineering Research Council of Canada;
the National Natural Science Foundation of China;
the Alexander von Humboldt Stiftung,
the German Bundesministerium f\"ur Bildung und Forschung (BMBF), and
the Deutsche Forschungsgemeinschaft (DFG);
the Italian Istituto Nazionale di Fisica Nucleare (INFN);
the MEXT, JSPS, and G-COE of Japan;
the Dutch Foundation for Fundamenteel Onderzoek der Materie (FOM);
the Russian Academy of Science and the Russian Federal Agency for 
Science and Innovations;
the U.K.~Engineering and Physical Sciences Research Council, 
the Science and Technology Facilities Council,
and the Scottish Universities Physics Alliance;
and the U.S.~Department of Energy (DOE) and the National Science Foundation (NSF).
%%%%%%%%%%%%%%%%%%%

%%%%%%%%%%%%%%%%%
\appendix
\section{Systematic Uncertainty of the  Amplitude Method for an Unpolarized Target}
\label{app:app2}

The  systematic uncertainty arising from  the neglect of amplitudes that make small contributions
to the angular distribution applies if only data on
unpolarized targets are considered.
If the set of observables were enlarged and data
on targets with transverse and longitudinal polarizations were added, then all the amplitude ratios
could be extracted and there would be no contributory systematic uncertainty of the amplitude method.

The SDME analysis~\cite{DC-24} of the HERMES data has shown the absence of any signal of the UPE amplitudes
$U_{01}$, $U_{10}$, and $U_{1-1}$, violating the SCHC approximation. The contribution of the greatest UPE
amplitude $U_{11}$ to any SDME changes it by a value which is less or about one standard deviation of
the statistical uncertainty. We neglect the contributions of $U_{01}$, $U_{10}$, and $U_{1-1}$ to the SDMEs
and do not consider that as a possible source of  systematic uncertainty. 
 The terms with the amplitudes $U_{01}$, $U_{10}$, and $U_{1-1}$ in all the formulas considered in this appendix 
are also ignored. As explained in Sec.~\ref{sec:schc} the fractional contribution to SDMEs of NPE 
helicity-flip amplitudes are suppressed by $\alpha^2$. The systematic uncertainty of the
extracted amplitude ratios due to the neglect of the NPE nucleon helicity-flip amplitudes is the only uncertainty
of the amplitude method which is considered below.

The  true ratio of the  NPE amplitude without helicity flip
$T_{\lambda_V \frac{1}{2}\lambda_{\gamma} \frac{1}{2}}$ to
$T_{0 \frac{1}{2} 0 \frac{1}{2}}$ is denoted by $t_{\lambda_V \lambda_{\gamma} }$.
The  true ratio of the nucleon-helicity-flip amplitude
$T^s_{\lambda_V \lambda_{\gamma}} \equiv
T_{\lambda_V -\frac{1}{2}\lambda_{\gamma} \frac{1}{2}}$
to $T_{0 \frac{1}{2} 0 \frac{1}{2}}$ is denoted by $t^s_{\lambda_V \lambda_{\gamma} }$, in
particular,
$t^s_{00}\equiv T_{0 -\frac{1}{2} 0 \frac{1}{2}}$\\$/T_{0 \frac{1}{2} 0 \frac{1}{2}}$,
while 
$|u_{11}|^2\equiv (|U_{1 \frac{1}{2}1\frac{1}{2} }|^2+
|U_{1 -\frac{1}{2}1 \frac{1}{2}}|^2)/|T_{0\frac{1}{2}0\frac{1}{2}}|^2$.
In the following, we estimate the effect of neglecting small contributions of the NPE nucleon spin-flip
amplitudes, $T^s_{\lambda_V \lambda_{\gamma} }$, which bias the fitted amplitude ratios by $\delta t_{\lambda_V \lambda_{\gamma} }$
and $\delta |u_{11}|$ respectively.
In order to avoid misunderstandings,
we  note that the notation $t_{\lambda_V \lambda_{\gamma}}$ and $|u_{11}|$ in the main
text was used for the fitted amplitude ratios which corresponds now to
$t_{\lambda_V \lambda_{\gamma}}+\delta t_{\lambda_V \lambda_{\gamma}}$ and $|u_{11}|+\delta |u_{11}|$, respectively.
 
 As  shown in Eqs.~(\ref{eqang1}-\ref{eqang3}), any SDME is multiplied by a function of the angles
$\theta$, $\Phi$, $\phi$ and these functions are linearly independent. This means that every SDME (rather
than  combinations of them) is determined in the fit individually. Now we assume that the true amplitude ratios are known, so that 
the true values of all SDMEs can be calculated. 

The exact expression for $r^{04}_{00}$, which was presented in Ref.~\cite{DC-24}
and rewritten here in terms of the ratios of the helicity amplitudes  $t_{\lambda_V \lambda_{\gamma} }$, 
$t^s_{\lambda_V \lambda_{\gamma} }$  and $|u_{11}|$, is
\begin{eqnarray}
r^{04}_{00}=\frac{\epsilon (1+|t^s_{00}|^2)+|t_{01}|^2+|t^s_{01}|^2}{N_{tot}}
\label{pr-r04-00}
\end{eqnarray}
where
\begin{eqnarray}
\nonumber
N_{tot}&=&\epsilon \Big(1+|t^s_{00}|^2+2|t_{10}|^2+2|t^s_{10}|^2\Big) \\
 \nonumber 
&+&|t_{11}|^2+|t^s_{11}|^2+|t_{01}|^2+|t^s_{01}|^2 \\
&+& |t_{1-1}|^2+|t^s_{1-1}|^2 + |u_{11}|^2.
\label{pr-Ntot}
\end{eqnarray}
It is more convenient technically to consider a logarithm of $r^{04}_{00}$. The contribution
to $\ln(r^{04}_{00})$ of the small NPE nucleon-helicity-flip amplitudes $|t^s_{00}|^2$, $|t^s_{10}|^2$, $|t^s_{11}|^2$, $|t^s_{01}|^2$ and
$|t^s_{1-1}|^2$, linear in  these small quantities, is
\begin{eqnarray}
&&\;\;\Delta \ln(r^{04}_{00}) = \Delta r^{04}_{00}/r^{04}_{00} = \frac{\epsilon |t^s_{00}|^2+|t^s_{01}|^2}{\epsilon +|t_{01}|^2}
 \nonumber \\
&-&   \Big[ \epsilon (|t^s_{00}|^2+2|t^s_{10}|^2)+|t^s_{11}|^2+|t^s_{01}|^2+|t^s_{1-1}|^2\Big] \nonumber \\
 \nonumber 
&/&  \Big[\epsilon (1+2|t_{10}|^2)+ |t_{11}|^2+|t_{01}|^2 \\
&+&|t_{1-1}|^2 + |u_{11}|^2\Big].
\label{lin-r04-00}
\end{eqnarray}

The formulas
for $r^{04}_{00}$ used in the fit of the angular distribution are
\begin{eqnarray}
\widetilde{r}^{04}_{00}=\frac{\epsilon +|t_{01}|^2 +\delta |t_{01}|^2
}{\widetilde{N}_{tot}}
\label{fit-r04-00}
\end{eqnarray}
where
\begin{eqnarray}
\widetilde{N}_{tot}&=&\epsilon \Big[1+2\Big(|t_{10}|^2+\delta |t_{10}|^2\Big)\Big] \nonumber \\
&+&\Big(|t_{11}|^2+\delta |t_{11}|^2\Big)+\Big(|t_{01}|^2+\delta |t_{01}|^2\Big) \nonumber \\
&+&\Big(|t_{1-1}|^2+\delta |t_{1-1}|^2\Big)+\Big(|u_{11}|^2+\delta |u_{11}|^2\Big)\;\;\;\;
\label{fit-Ntot}
\end{eqnarray}
with $\delta |t_{\lambda_V \lambda_{\gamma}}|^2=
t^*_{\lambda_V \lambda_{\gamma}}\delta t_{\lambda_V \lambda_{\gamma}}+
t_{\lambda_V \lambda_{\gamma}}\delta t^*_{\lambda_V \lambda_{\gamma}}$
 and $\delta |u_{11}|^2$
 $=2|u_{11}|\delta |u_{11}|$.

The  contribution to $\ln(\widetilde{r}^{04}_{00})$ linear in the small quantities $\delta |t_{\lambda_V \lambda_{\gamma}}|^2$ and 
$\delta |u_{\lambda_V \lambda_{\gamma}}|^2$  is given by the relation
\begin{eqnarray}
&& \delta \ln(\widetilde{r}^{04}_{00}) =\frac{\delta \widetilde{r}^{04}_{00}}{\widetilde{r}^{04}_{00}}
 =\frac{\delta |t_{01}|^2}{\epsilon +|t_{01}|^2 } \nonumber\\
&-&\Big[2\epsilon \delta |t_{10}|^2 +\delta |t_{11}|^2+\delta |t_{01}|^2 \nonumber \\
&& \; \;+ \; \delta |t_{1-1}|^2 + \delta |u_{11}|^2\Big]\nonumber \\
&/&\Big[\epsilon \Big(1+2|t_{10}|^2\Big) + |t_{11}|^2+|t_{01}|^2\nonumber \\
&& \; \;+ \; |t_{1-1}|^2 + |u_{11}|^2\Big].
\label{del-r04-00}
\end{eqnarray}
A comparison of Eqs.~(\ref{pr-r04-00}) and (\ref{pr-Ntot}) with Eqs.~(\ref{fit-r04-00}) and (\ref{fit-Ntot}) shows
that $\widetilde{r}^{04}_{00}=r^{04}_{00}$ when $\delta t_{\lambda_V \lambda_{\gamma}}=
\delta |u_{11}|=t^s_{\lambda_V \lambda_{\gamma}}=0$.
Since the fit has to reproduce the angular distribution, the contribution of the 
nucleon-helicity-flip
amplitudes to $r^{04}_{00}$ is  compensated by the deviation of the obtained amplitude ratios
from the true amplitude ratios. This assumes the validity of the relation
\begin{eqnarray}
\Delta \ln(r^{04}_{00})=\delta \ln(\widetilde{r}^{04}_{00})
\label{cond-r04-00}
\end{eqnarray}
which provides the linear relation between $|t^s_{\lambda_V \lambda_{\gamma}}|^2$
and $\delta |t_{\lambda_V \lambda_{\gamma}}|^2$,  $\delta |u_{\lambda_V \lambda_{\gamma}}|^2$.
Considering all other SDMEs $r^{\eta}_{\lambda_V \lambda'_V}$,  use of the same method provides
the set of relations for $\eta=1-3,\;5-8$
\begin{eqnarray}
\Delta \ln(r^{\eta}_{\lambda_V \lambda'_V})=\delta \ln(\widetilde{r}^{\eta}_{\lambda_V \lambda'_V})
\label{cond-r-all}
\end{eqnarray}
which can be used to determine $\delta t_{\lambda_V \lambda_{\gamma}}$ and 
$\delta |u_{11}|$.

In order to get an  approximate solution of Eq.~(\ref{cond-r04-00}), the contributions of 
the $s$-channel helicity violating amplitudes $T_{01}$, $T_{10}$, and $T_{1-1}$ 
($T^s_{01}$, $T^s_{10}$, and $T^s_{1-1}$) are 
neglected since they are small compared to $T_{00}$ and $T_{11}$ ($T^s_{00}$ and $T^s_{11}$). Putting 
Eqs.~(\ref{lin-r04-00}) and (\ref{del-r04-00}) into Eq.~(\ref{cond-r04-00}) yields the relation
\begin{eqnarray}
|t^s_{00}|^2
-\frac{\epsilon |t^s_{00}|^2+|t^s_{11}|^2}{\epsilon+|t_{11}|^2+|u_{11}|^2}
=-\frac{\delta |t_{11}|^2+\delta |u_{11}|^2}{\epsilon+|t_{11}|^2+|u_{11}|^2}. \nonumber \\
\label{eq-r04-00}
\end{eqnarray}
Considering in the same way the matrix element $r^1_{1-1}$ (or $\mathrm{Im}(r^2_{1-1})$)
and using Eq.~(\ref{cond-r-all}) we get the relation
\begin{eqnarray}
\nonumber
&& \frac{|t^s_{11}|^2}{|t_{11}|^2-|u_{11}|^2}
-\frac{\epsilon |t^s_{00}|^2+|t^s_{11}|^2}{\epsilon+|t_{11}|^2+|u_{11}|^2} \\ [0.2cm]
&=&\frac{\delta |t_{11}|^2-\delta |u_{11}|^2}{|t_{11}|^2-|u_{11}|^2}
-\frac{\delta |t_{11}|^2+\delta |u_{11}|^2}{\epsilon+|t_{11}|^2+|u_{11}|^2}.
\label{eq-r1-1m1}
\end{eqnarray}
The solution of the system of two equations (\ref{eq-r04-00}) and (\ref{eq-r1-1m1}) is
\begin{eqnarray}
\label{sol01-r04-r1}
\delta |t_{11}|^2&=&|t^s_{11}|^2-|t^s_{00}|^2|t_{11}|^2,\\
\delta |u_{11}|^2&=&-|t^s_{00}|^2|u_{11}|^2.
\label{sol02-r04-r1}
\end{eqnarray}
Dividing Eq.~(\ref{sol01-r04-r1}) by $|t_{11}|^2$ and Eq.~(\ref{sol02-r04-r1})  by $|u_{11}|^2$ these 
solutions can be rewritten for the fractional systematic uncertainties of
$|t_{11}|^2$ and $|u_{11}|^2$ in the form
\begin{eqnarray}
\label{frac-sol01-r04-r1}
\frac{\delta |t_{11}|^2}{|t_{11}|^2}&=& \Bigl |\frac{T^s_{11}}{T_{11}}\Bigr |^2-\Bigl |\frac{T^s_{00}}{T_{00}}\Bigr |^2,\\
\frac{\delta |u_{11}|^2}{|u_{11}|^2}&=& 2\frac{\delta |u_{11}|}{|u_{11}|}=-|t^s_{00}|^2\equiv -\Bigl |\frac{T^s_{00}}{T_{00}}\Bigr |^2.
\label{frac-sol02-r04-r1}
\end{eqnarray}

In order  to obtain estimates for the real and imaginary parts of $\delta t_{11}$, we  consider the SDMEs
$\mathrm{Re}(r^5_{10})$ and $\mathrm{Re}(r^8_{10})$, respectively. The exact expression for $\mathrm{Re}(r^5_{10})$
taken from Ref.~\cite{DC-24} and rewritten in terms of $t_{\lambda_V \lambda_{\gamma}}$,
$t^s_{\lambda_V \lambda_{\gamma}}$, and $|u_{11}|$ is
\begin{eqnarray}
\nonumber
\mathrm{Re}(r^5_{10})&=&\frac{1}{\sqrt{8}}\mathrm{Re}\Big[t_{11}-t_{1-1}+(t^s_{11}-t^s_{1-1})(t^s_{00})^*\\
&& \; \; \;\; + \; 2t_{10}(t_{01})^*+2t^s_{10}(t^s_{01})^*\Big]/N_{tot}\;\;
\label{exact-r5}
\end{eqnarray}
where $N_{tot}$ is defined by Eq.~(\ref{pr-Ntot}). Using Eq.~(\ref{cond-r-all})  
an approximate equation analogous to Eqs.~(\ref{eq-r04-00}) and (\ref{eq-r1-1m1})
is obtained:
\begin{eqnarray}
\nonumber
&&\frac{\mathrm{Re}[t^s_{11}(t^s_{00})^*]}{\mathrm{Re}(t_{11})}
-\frac{\epsilon |t^s_{00}|^2+|t^s_{11}|^2}{\epsilon+|t_{11}|^2+|u_{11}|^2}\\
&=&\frac{\mathrm{Re}(\delta t_{11})}{\mathrm{Re}(t_{11})}
-\frac{\delta |t_{11}|^2+\delta |u_{11}|^2}{\epsilon+|t_{11}|^2+|u_{11}|^2}.
\label{eq-r5-10}
\end{eqnarray}
 Substituting into this equation the solutions for $\delta |t_{11}|^2$ and $\delta |u_{11}|^2$ given in 
Eqs.~(\ref{sol01-r04-r1}) and (\ref{sol02-r04-r1}) leads to:
\begin{eqnarray}
\mathrm{Re}(\delta t_{11})=\mathrm{Re}[t^s_{11}(t^s_{00})^*]-|t^s_{00}|^2\mathrm{Re}(t_{11}).
\label{sol-r5-10}
\end{eqnarray}
Considering $\mathrm{Re}(r^8_{10})$  analogously  we get 
\begin{eqnarray}
\mathrm{Im}(\delta t_{11})=\mathrm{Im}[t^s_{11}(t^s_{00})^*]-|t^s_{00}|^2\mathrm{Im}(t_{11}).
\label{sol-r6-10}
\end{eqnarray}
Combining $\mathrm{Re}(\delta t_{11})$ and $\mathrm{Im}(\delta t_{11})$ into the complex number 
$\delta t_{11}=\mathrm{Re}(\delta t_{11})+i\cdot\mathrm{Im}(\delta t_{11})$, the equation
\begin{eqnarray}
\delta t_{11}=t^s_{11}(t^s_{00})^*-|t^s_{00}|^2t_{11}
\label{com-sol-r5-r6}
\end{eqnarray}
is obtained which is equivalent to  Eqs.~(\ref{sol-r5-10}) and (\ref{sol-r6-10}). From Eq.~(\ref{com-sol-r5-r6}), 
it follows that
\begin{eqnarray}
\frac{\delta t_{11}}{t_{11}}=\Bigl(\frac{T^s_{11}}{T_{11}}-\frac{T^s_{00}}{T_{00}}\Bigr)
\Bigl (\frac{T^s_{00}}{T_{00}}\Bigr )^* .
\label{frac-unc-t11}
\end{eqnarray}

An analogous consideration of $r^5_{00}$ and $r^8_{00}$ provides the expression 
\begin{eqnarray}
\frac{\delta t_{01}}{t_{01}}=\Bigl(\frac{T^s_{01}}{T_{01}}-\frac{T^s_{00}}{T_{00}}\Bigr)
\Bigl (\frac{T^s_{00}}{T_{00}}\Bigr )^*,
\label{frac-unc-t01}
\end{eqnarray}
while considering  $r^1_{11}$ and $\mathrm{Im}(r^3_{1-1})$ leads to:
\begin{eqnarray}
\frac{\delta t_{1-1}}{t_{1-1}}=\Bigl(\frac{T^s_{1-1}}{T_{1-1}}-\frac{T^s_{00}}{T_{00}}\Bigr)
\Bigl (\frac{T^s_{11}}{T_{11}}\Bigr )^*.
\label{frac-unc-t1m1}
\end{eqnarray}
Consideration of $\mathrm{Im}(r^6_{1-1})$ and $\mathrm{Im}(r^7_{1-1})$ leads to the relation
\begin{eqnarray}
\frac{\delta t_{10}}{t_{10}}=\Bigl(\frac{T^s_{10}}{T_{10}}-\frac{T^s_{00}}{T_{00}}\Bigr)
\Bigl (\frac{T^s_{11}}{T_{11}}\Bigr )^*.
\label{frac-unc-t10}
\end{eqnarray}

As shown in Ref.~\cite{MSI} the ratios of $T^s_{\lambda_V \lambda_{\gamma}}$ and
$T_{\lambda_V \lambda_{\gamma}}$ obey the inequality
\begin{eqnarray}
|T^s_{\lambda_V \lambda_{\gamma}}/T_{\lambda_V \lambda_{\gamma}}| \le v_T/(2M),
\label{sim-in}
\end{eqnarray}
which gives the inequality
\begin{eqnarray}
\Bigl |\frac{T^s_{\lambda_V \lambda_{\gamma}}}{T_{\lambda_V \lambda_{\gamma}}}-
\frac{T^s_{00}}{T_{00}}\Bigr | \le \Bigl |\frac{T^s_{\lambda_V \lambda_{\gamma}}}{T_{\lambda_V \lambda_{\gamma}}} \Bigr |+
\Bigl |\frac{T^s_{00}}{T_{00}}\Bigr | 
 \le \frac{v_T}{M}.
\label{sim-in2}
\end{eqnarray}
Inserting Eqs.~(\ref{sim-in}) and (\ref{sim-in2}) into expressions (\ref{frac-unc-t11}-\ref{frac-unc-t10})
yields the formulas of interest for the systematic uncertainty of $t_{\lambda_V \lambda_{\gamma}}$
\begin{eqnarray}
|\delta t_{\lambda_V \lambda_{\gamma}}| \le \frac{v_T^2}{2M^2} |t_{\lambda_V \lambda_{\gamma}}|.
\label{syst-unc}
\end{eqnarray}
Since
$|\mathrm{Re}(\delta t_{\lambda_V \lambda_{\gamma}})| \le |\delta t_{\lambda_V \lambda_{\gamma}}|$ and
$|\mathrm{Im}(\delta t_{\lambda_V \lambda_{\gamma}})| $
\linebreak
$\le |\delta t_{\lambda_V \lambda_{\gamma}}|$,
the right hand side of relation (\ref{syst-unc}) can be used both for the real and imaginary parts
of $\delta t_{\lambda_V \lambda_{\gamma}}$ yielding  Eqs.~(\ref{syst-amp-meth01}) and (\ref{syst-amp-meth02}).
Relation (\ref{syst-amp-meth03})
follows immediately from  Eqs.~(\ref{frac-sol02-r04-r1}) and (\ref{sim-in}).
\vspace{5.0cm}

\end{document}